\documentstyle[12pt,aaspp4]{article}
\lefthead{JENKINS \& TRIPP}
\righthead{THERMAL PRESSURES FROM C~I EXCITATION}
\begin{document}
\slugcomment{Note: Figs 4, 6, 7a-j and 8 are slightly degraded to
make the total file size not too large.  A gzipped postscript file of this paper
that shows better versions of these figures can be found at
ftp://astro.princeton.edu/library/preprints/pop843.ps.gz}
\title{The Distribution of Thermal Pressures in the Interstellar Medium
from a Survey of C~I Fine-Structure Excitation\footnote{
Based on observations with the NASA/ESA Hubble
Space Telescope obtained at the Space Telescope Science Institute,
which is operated by the Association of Universities for
Research in Astronomy, Incorporated, under NASA contrtact
NAS5-26555.}}
\author{Edward B. Jenkins and Todd M. Tripp}
\affil{Princeton University Observatory\\
Princeton, NJ 08544-1001; ebj@astro.princeton.edu,
tripp@astro.princeton.edu}
\begin{abstract}
We used the {\it Space Telescope Imaging Spectrograph\/} (STIS) with its
smallest entrance aperture (0\farcs 03 wide slit) and highest resolution
echelle gratings (E140H and E230H) to record the interstellar absorption
features for 10 different multiplets of neutral carbon at a resolving
power of $\lambda/\Delta\lambda=200,000$ in the UV spectra of 21
early-type stars.  Our objective was to measure the amount of C~I in
each of its three fine-structure levels of the ground electronic state,
so that we could determine the thermal pressures in the absorbing gas
and how much they vary in different regions.  Our observations are
principally along directions out to several kpc in the Galactic plane
near longitudes $\ell = 120\arcdeg$ and $300\arcdeg$, with the more
distant stars penetrating nearby portions of the Perseus and
Sagittarius-Carina arms of the Galaxy.  We devised a special analysis
technique to decipher the overlapping absorption features in the
different multiplets, each with different arrangements of the closely
spaced transitions.  In order to derive internally consistent results
for all multiplets, we found that we had to modify the relative
transition $f-$values in a way that made generally weak transitions
stronger than amounts indicated in the current literature.

We compared our measured relative populations of the excited
fine-structure levels to those expected from equilibria calculated with
collisional rate constants for various densities, temperatures and
compositions.  The median thermal pressure for our entire sample was
$p/k=2240\,{\rm cm}^{-3}\,$K, or slightly higher if the representative
temperatures of the material are much above or below a most favorable
temperature of 40K for the excitation of the first excited level at a
given pressure.  For gas that is moving outside the range of radial
velocities permitted by differential Galactic rotation between us and
the targets, about 15\% of the C~I indicates a thermal pressure $p/k >
5000\,{\rm cm}^{-3}\,$K.  For gas within the allowed velocities, this
fraction is only 1.5\%.  This contrast reveals a relationship between
pressure enhancements and the kinematics of the gas.

Regardless of velocity, we usually can register the presence of a very
small proportion of the gas that seems to be at $p/k \gtrsim 10^5\,{\rm
cm}^{-3}\,$K.  We interpret these ubiquitous wisps of high pressure
material to arise either from small-scale density enhancements created
by converging flows in a turbulent medium or warm turbulent boundary
layers on the surfaces of dense clouds moving through an intercloud
medium.  For turbulent compression, our C~I excitations indicate that
the barytropic index $\gamma_{\rm eff}\gtrsim 0.90$ must apply if the
unperturbed gas starts out with representative densities and
temperatures $n=10\,{\rm cm}^{-3}$ and $T=100\,$K.  This value for
$\gamma_{\rm eff}$ is larger than that expected for interstellar
material that remains in thermal equilibrium after it is compressed from
the same initial $n$ and $T$.  However, if regions of enhanced pressure
reach a size smaller than $\sim 0.01\,$pc, the dynamical time starts to
become shorter than the cooling time, and $\gamma_{\rm eff}$ should
start to approach the adiabatic value $c_p/c_v=5/3$.

Some of the excited C~I may arise from the target stars' H~II regions or
by the effects of optical pumping from the sub-millimeter line radiation
emitted by them.  We argue that these contributions are small, and our
comparisons of the velocities of strongly excited C~I to those of
excited Si~II seem to support this outlook.

For 6 stars in the survey, absorption features from interstellar excited
O~I could be detected at velocities slightly shifted from the persistent
features of telluric origin.  These O~I* and O~I** features were
especially strong in the spectra of HD~93843 and HD~210839, the same
stars that show exceptionally large C~I excitations.

In appendices of this paper, we present evidence that (1) the wavelength
resolving power of STIS in the mode we used is indeed about 200,000, and
(2) the telluric O~I* and O~I** features exhibit some evidence for
macroscopic motions, since their broadenings are in excess of that
expected for thermal Doppler broadening at an exospheric temperature
$T=1000\,$K.

\end{abstract}

\keywords{ISM: atoms --- ISM: kinematics and dynamics --- ISM: lines and
bands --- techniques: spectroscopic --- ultraviolet: ISM}

\section{Introduction}\label{intro}

The many diverse ways now available to observe the interstellar medium
(ISM) in the Galactic disk has brought about an awareness that the gas
exhibits a vast range of densities and temperatures, along with large
contrasts in the molecular fractions and degrees of ionization.  In
large part, the extremes in these quantities arise from the gas being
driven by injections of energy and momentum from the impulsive,
far-reaching disturbances created by shock waves arising from supernova
explosions  (McKee \& Ostriker 1977; McCray \& Snow 1979; Mac Low,
McCray, \& Norman 1989) whose energy inputs are sufficient to maintain a
network of very hot gases in the disk of the Galaxy  (Cox \& Smith 1974;
Smith, 1977; Jones et al. 1979).

Supplementing the disturbances created by explosions are the much more
gradual effects from certain classes of stars  (Abbott 1982; McKee
1986).  For instance, stars that have bipolar outflows  (Bally 1982;
Genzel \& Downes 1982; K\"onigl 1996) or have large rates of mass loss
while they are on the main sequence  (Lamers 1981) have a strong
influence on the dynamics and physical state of the gas that surrounds
them.  Spherically symmetric outflows should create volumes of hot, low
density gas surrounded by dense shells of cooler gas  (Castor, McCray,
\& Weaver 1975; Weaver et al. 1977).  Adding to the effects of direct
mechanical energy are pressurizations which arise from the ultraviolet
radiations from hot stars that rapidly create zones of fully ionized gas
whose temperatures are much higher than that of the ambient, initially
neutral material.  For a brief time, such regions can have substantially
elevated thermal pressures that are eventually relieved by a dynamical
response.

Taken together, supernovae and their accompanying young stellar
associations represent sources that are largely responsible for creating
some of the vivid contrasts in conditions, along with distinctive
morphological signatures manifested in the forms of shells and loops 
(Heiles 1979).   The outcomes from these disturbances are made even more
conspicuous by the tendency for many of these sources to act together,
since they are clustered in space and time.  Their direct effects also
bring about various secondary processes that influence the state of the
gas, such as interactions from Alfv\'en waves  (Elmegreen 1997), density
fluctuations caused by turbulence  (V\'azquez-Semadeni, Passot, \&
Pouquet 1995; Ballesteros-Paredes, V\'azquez-Semadeni, \& Scalo 1999),
and the nonlinear responses of gaseous material arising from thermal and
gravitational instabilities  (Wada, Spaans, \& Kim 2000).  In high
density clouds, these processes and their compound effects play an
important role in star formation.

From a superficial viewpoint, gases in the Galactic disk show an inverse
relationship between density and temperature, leading to an approximate
equality of thermal pressure $p=nkT$ for different locations.  While
this may be true for the simplest depiction of the states of widely
different phases of the gas in the Galactic disk, it is unlikely to hold
over small scales and at all times.  The weight of material within the
Galactic gravitational potential produces a substantial reservoir of
total pressure $p_{\rm total}/k\approx 3\times 10^4{\rm cm}^{-3}\,$K 
(Boulares \& Cox 1990), made up of several sources that supplement the
thermal one, such as turbulence, cosmic rays and magnetic fields.  Under
the right conditions, these other forms can be transformed into
temporary excesses or rarefactions in thermal pressure.  Our purpose in
studying thermal pressures is to examine the inevitable deviations that
must arise from the random, tempestuous character of the medium.  Any
information that we can gather about base pressure value together with
the size and character of fluctuations around this value represent a
valuable constraint for theories on the processes that shape the global
conditions of the ISM.

An excellent probe of densities and temperatures in diffuse H~I regions
is the relative population of neutral carbon atoms in excited
fine-structure levels of the ground electronic state.  The levels'
energy separations, collisional excitation rate constants, and
spontaneous radiative decay rates are ideal for differentiating
different regimes of density and temperature that are expected in this
particular phase of the ISM.  The first large scale survey of C~I
absorption features in the uv spectra of hot stars was carried out by
Jenkins \& Shaya  (1979), who used the spectrometer on the {\it
Copernicus\/} satellite  (Rogerson et al. 1973).  The sample size was
doubled by a subsequent survey, again with {\it Copernicus}, carried out
by Jenkins, Jura \& Loewenstein   (1983).  In a discussion of the
outcomes from both surveys, Jenkins et al.  (1983) were not able to
recover the full distribution of pressures from their data because a
substantial fraction of the cases had large uncertainties. 
Nevertheless, they were able to express quantitatively the relative
fraction of material at either extremely low or high pressures.  These
conclusions gave some guidance on the applications of general theories
of the ISM that had just been developed, such as the calculations of
random pressure fluctuations caused by supernova explosions  (McKee \&
Ostriker 1977) and the crushing effects on clouds overtaken by blast
waves  (Cox 1979, 1981).

A significant investment in space astronomy technology and facilities
since the launch of {\it Copernicus\/} almost 30 years ago now enables
us to greatly surpass the earlier achievements in studying C~I
absorption features.  Over the last decade, a number of studies have
provided interesting new insights on physical conditions based on
interpretations of the C~I fine-structure excitation.  In an early
observation using the {\it Goddard High Resolution Spectrograph\/}
(GHRS) on the {\it Hubble Space Telescope\/} (HST), Smith et al.  (1991)
studied the C~I absorptions along line of sight to $\xi$~Per and found
evidence for large contrasts in conditions for components at different
radial velocities.  Jenkins \& Wallerstein  (1995) discovered an
unusually large fine-structure excitation of C~I for high velocity gas
in the Vela supernova remnant, using a spectrum recorded by GHRS at
intermediate resolution (with the G160M grating) for the star HD~72089. 
This result was confirmed and recorded in more detail by Jenkins et al. 
(1998)  using the {\it Space Telescope Imaging Spectrograph\/} (STIS) 
(Kimble et al. 1998; Woodgate et al. 1998), a later generation
spectrograph that replaced GHRS.  Wannier et al.  (1999) explored the
physical conditions in the outer envelope of the B5 molecular cloud in
Perseus by recording 3 C~I multiplets (plus a CO absorption band) in the
spectra of 3 stars at strategic locations behind the cloud's periphery.

In this paper, we report on our use of STIS to record the spectra of 21
early-type stars.  The power of STIS to record many wavelength elements
simultaneously allowed us encompass ten different C~I multiplets with
only a modest amount of effort.  To achieve the best possible wavelength
resolution to differentiate parcels of gas with small separations in
radial velocity, we employed the highest resolution modes of the echelle
spectrograph (E140H and E230H) and used its narrowest entrance slit
(0\farcs 03 wide).  As shown by the results in
Appendix~\ref{resolving_pwr}, the spectrograph in this configuration
yields $\lambda/\Delta\lambda=200,000$, a resolving power ten times
greater than that of {\it Copernicus\/} and at least twice as good as
the highest resolution echelle grating on GHRS.  This is especially
beneficial for studying C~I, since its features probably behave very
much like those of other neutral species that are normally expected to
be in the singly-ionized state in H~I regions, such as Na~I and K~I.  It
is well documented that Na~I and K~I can have widths narrower than
$1\,{\rm km~s}^{-1}$  (Welty, Hobbs, \& Kulkarni 1994; Welty \& Hobbs
2001).

The observing program was organized to yield a good sampling of
directions within the limits of HST observing time allocated for the
project.  The details of target selection and STIS observing
configurations are discussed in \S\ref{tgt_select} and \S\ref{obs},
respectively.  In \S\ref{reduction} we discuss various phases of the
data reduction, including the correction for scattered light for
obtaining a proper zero level for intensities (\S\ref{scat_light}), the
sensing and elimination of a high-frequency modulation of the spectrum
caused by uneven responses to half-pixel photoevents by the detector's
electronics (\S\ref{rebalance}), special precautions to co-align the
wavelengths of separate exposures before they were combined
(\S\ref{comb_exp}), our method of defining the continuum levels above
the absorption features (\S\ref{cont_def}), and the means we used to
register accurately the wavelengths from one multiplet to another
(\S\ref{vel_reg}).

The C~I multiplets consist of complex blends of individual lines from
different fine-structure levels.  The patterns vary from one multiplet
to the next, and in \S\ref{principles}$-$\ref{specific_appl} we describe
a special analysis technique that takes advantage of these changes to
unravel the structures of the velocity profiles in the absence of such
interference.  However, to obtain self-consistent results for all of the
multiplets, we found it necessary to adjust their relative $f-$values. 
This was done by a special analysis of all of the data in the survey, as
described in \S\ref{fval}, followed by various checks to show that
outcomes were not influenced by fine-scale structures in the C~I
velocity profiles that were saturated but still not resolved by the
instrument (\S\ref{substructures}).  Section~\ref{analysis} ends with
one subsection that presents an example of profile reconstructions to
show that the outcome of the analysis is consistent with the input data
(\S\ref{reconstructions}) and another that discusses various sources of
error in the results (\S\ref{errors}).

After some introductory remarks in \S\ref{genl_remarks}, we describe in
\S\ref{theoretical} our calculations for the expected fine-structure
population ratios for various interstellar conditions and compositions
(\S\ref{theoretical}).  A graphical means for comparing for each line of
sight the C~I population ratios with these expected values is presented
in \S\ref{presentations}, followed later by a general discussion on the
implications from the entire survey in \S\ref{interpretation} and how
the findings relate to the kinematics of the gas in \S\ref{kinematics}. 
Accompanying these main arguments are some digressions on a few
detections of the absorptions by neutral oxygen atoms in excited
fine-structure states (\S\ref{OI*}), the unusual conditions seen toward
one star, HD~210839 (\S\ref{210839_comp}), and a simple check
(\S\ref{simple_check}) on some of the more surprising outcomes expressed
in \S\ref{pts} that bypasses the assertion that the $f-$values need
revision, as we claimed earlier in \S\ref{fval}.  In
\S\S\ref{hii_regions},\ref{other_HII} we cover the possibility that some 
small fraction of the C~I might arise from H~II regions instead of H~I
regions.  Finally, in the discussion section (\S\ref{discussion}) we
present our interpretation that the admixtures of small amounts of
high-pressure gas with the ordinary material may signify the presence of
density enhancements that arise from turbulence, and that perhaps this
is the same phenomenon that is responsible for small-scale structures in
the gas. 

\section{Selection of Targets}\label{tgt_select}

Our objective was to obtain moderately good signal-to-noise ratios
($S/N$) on as many targets as possible within our allocation of HST
orbits for this program.  To accomplish this, we chose targets that were
in the Continuous Viewing Zone (CVZ) of HST, thus eliminating the losses
of time caused by Earth occultations.  The advantage in integration time
is about a factor of two over non-CVZ targets.  A few targets were
outside the CVZ; these targets were chosen to satisfy the requirements
of another research program which had pooled its orbital allocations
with ours.  A consequence of our favoring the CVZ is that the galactic
longitudes of the targets are clustered around $\ell = 120\arcdeg$ and
$300\arcdeg$.

The target stars in our survey are listed in Table~\ref{tgt_stars}. 
They were selected such that their predicted ultraviolet fluxes would
yield far-UV MAMA global count rates just below the maximum permitted
value of $2\times 10^5\,{\rm s}^{-1}$.  Generally, stars that were
expected to have much lower count rates were disfavored, again in the
spirit of maximizing the $S/N$ in a short observing time.  We generally
tried to avoid stars that had visual binary companions that might
complicate the acquisition process, but there were several exceptions
that did not pose any problems.  These stars can be identified in
Table~\ref{tgt_stars} as ones with an ``A'' following the HD number.  To
make the interstellar features easy to separate from stellar ones, we
chose stars that had large projected rotational velocities (listed in
column~4 of Table~\ref{tgt_stars}).

\placetable{tgt_stars}

Noteworthy information about the targets is given in
Table~\ref{tgt_stars}, such as the identification by HD number (col.~1),
spectral classification (col.~2), radial velocity (col.~3), projected
rotational velocity $v \sin i$ (col.~4), $V$ magnitude (col.~5), and the
far-UV continuum flux $F_\lambda$ at 1300$\,$\AA\ (col.~6).  Readers are
cautioned against using flux values obtained from our observations in
the HST archive, since they do not consider the effect from the reduced
transmission of the narrow entrance slit.  The last two columns of the
table show important parameters for our observations.  Column~7 lists
the $S/N$ that we obtained at the most favorable wavelength
(1280$\,$\AA) for obtaining a strong signal with the E140H observing
mode.  Approximate values of $S/N$ for multiplets at wavelengths well
removed from this one can be calculated by multiplying $S/N_{1280}$ by
the numbers in parentheses below the multiplet identifiers in column~1
of Table~\ref{fval_results}.  The identification numbers for the
exposures that exist in the HST archive are given in column~8.  Readers
who wish to know the exact details about the exposures (observing dates,
durations, etc.) may construct the appropriate tables in this archive 
(these observations are listed under Program ID's 8043 and 8484).

Various parameters that pertain to the lines of sight and are relevant
to the interstellar conditions are given in Table~\ref{los}.  For each
line of sight identified by the star's HD number (col.~1) we give the
galactic coordinates (cols.~2$-$3), estimated distance (col.~4), the
difference between radial velocities in the heliocentric system and the
Local Standard of Rest (LSR) (col. 5), the expected radial velocity at
the position of the star arising from differential galactic rotation
(col.~6), $B-V$ color excess (col.~7), and H~I column density (col.~8).
\clearpage
\placetable{los}

\begin{deluxetable}{
r    
l    
c    
c    
c    
c    
c    
l    
}
\tablecolumns{8}
\tablewidth{0pt}
\tablecaption{Target Stars and Exposures\label{tgt_stars}}
\tablehead{
\colhead{HD~~} & \colhead{Sp. Type} &
\colhead{$v_*$\tablenotemark{b}} &
\colhead{$v \sin i$\tablenotemark{c}} &
\colhead{$V$\tablenotemark{d}} &
\colhead{$F_\lambda$\tablenotemark{e}} & \colhead{$S/N$} & \colhead{HST
Archive} \\
\colhead{} & \colhead{\& [Ref.]\tablenotemark{a}} &
\colhead{(${\rm km~s}^{-1}$)} & \colhead{(${\rm km~s}^{-1}$)} &
\colhead{} & \colhead{(1300$\,$\AA)} &
\colhead{(1280$\,$\AA)} & \colhead{Rootname}\\
\colhead{(1)} & \colhead{(2)} & \colhead{(3)} & \colhead{(4)} &
\colhead{(5)} & \colhead{(6)} & \colhead{(7)} & \colhead{(8)}
}
\startdata
108\phantom{A}&O6:f?pe~[5]\tablenotemark{f}&
$-63$\tablenotemark{g}&115\tablenotemark{h}&7.40&
2.0&25&O5LH01010$-$1080 \nl
3827\phantom{A}&B0.7~V((n))~[4]&
$-22$\tablenotemark{g}&125\tablenotemark{h}&8.00&
13.&36&O54359010$-$9030 \nl
15137\phantom{A}&O9.5~II$-$III(n)~[4]&
$-35$\tablenotemark{g}&135\tablenotemark{h}&7.87&
1.8&25&O5LH02020$-$2050 \nl
69106\phantom{A}&B0.5~IVnn~[7]&
+6\tablenotemark{g}&320\tablenotemark{h}&7.14&
13.&42&O5LH03010$-$3050 \nl
88115\phantom{A}&B1.5~IIn~[7]&
$-18$\tablenotemark{g}&245\tablenotemark{h}&8.30&
2.2&23&O54305010$-$5060 \nl
\nl
93843\phantom{A}&O5.5~III(f)~[7]&
$-9$\tablenotemark{g}&90\tablenotemark{h}&7.33&
9.\tablenotemark{i}&35&O5LH04010$-$4040 \nl
94493\phantom{A}&B1~Ib~[7]&
$-1$\tablenotemark{g}&145\tablenotemark{h}&7.27&
5.&26&O54306010, 6020 \nl
99857A&B0.5~Ib~[7]&
$-10$\tablenotemark{g}&180\tablenotemark{h}&7.45&
3.0&31&O54301010$-$1060 \nl
103779\phantom{A}&B0.5~Iab~[7]&
$-9$\tablenotemark{j}&78\tablenotemark{k}&
7.21&7.&25&O54302010, 2020 \nl
106343\phantom{A}&B1.5~Ia~[7]&
$-7$\tablenotemark{g}&85\tablenotemark{h}&6.22&
3.8\tablenotemark{l}&28&O54310010, 0020 \nl
\nl
109399\phantom{A}&B0.7~II~[7]&
$-50$\tablenotemark{g}&125\tablenotemark{h}&7.62&
5.&24&O54303010, 3020 \nl
116781A&B0~IIIne~[7]&\nodata&
\nodata&7.60&
4.&22&O5LH05010$-$5040 \nl
120086\phantom{A}&B3~III~[3]&
$-1$\tablenotemark{g}&120\tablenotemark{h}&7.87&
8.&30&O5LH06010$-$6050 \nl
122879\phantom{A}&B0~Ia~[7]&
2\tablenotemark{g}&92\tablenotemark{k}&6.41&
9.&35&O5LH07010$-$7040 \nl
124314A&O6~V(n)((f))~[6]&
$-13$\tablenotemark{g}&270\tablenotemark{h}&6.64&
5.&23&O54307010, 7020 \nl
\nl
203374A&B0~IVpe~[2]&
$-7$\tablenotemark{m}&350:\tablenotemark{h}&6.68&
2.5&30&O5LH08010$-$8060 \nl
206267A&O6.5~V~[5]\tablenotemark{n}&
$-8$\tablenotemark{m}&155\tablenotemark{h}&5.62&
5.&23&O5LH09010$-$9040 \nl
208947\phantom{A}&B2~V~[1]\tablenotemark{o}&
+2\tablenotemark{m}&250\tablenotemark{h}&6.43&
\nodata&40&O5LH0A010$-$A040 \nl
209339A&B0~IV~[2]&
$-20$\tablenotemark{m}&220\tablenotemark{h}&6.65&
\nodata&35&O5LH0B010$-$B040 \nl
210839\tablenotemark{p}\phantom{A}&O6~I(n)f~[5]&
$-74$\tablenotemark{m}&275\tablenotemark{h}&5.04&
10.&18&O54304010, 4020 \nl
\tablebreak
224151\phantom{A}&B0.5~II~[2]\tablenotemark{q}&
$-26$\tablenotemark{m}&115\tablenotemark{k}&6.00&
5.&26&O54308010, 8020 \nl
\enddata
\tablenotetext{a}{Sources of spectral types: [1] =  (Johnson, H. L. \&
Morgan 1953), [2] =  (Morgan, Code, \& Whitford 1955), [3] =  (Hill, P.
W. 1970), [4] =  (Walborn 1971). [5] =  (Walborn 1972), [6] =  (Walborn
1973), [7] =  (Garrison, Hiltner, \& Schild 1977).}
\tablenotetext{b}{Star's heliocentric radial velocity, taken from the
indicated sources.}
\tablenotetext{c}{Star's projected rotational velocity, taken from the
indicated sources.}
\tablenotetext{d}{$V$ magnitudes from Mermilliod  (1987).}
\tablenotetext{e}{Star's continuum flux at 1300$\,$\AA, measured from
large-aperture IUE spectra (except where noted), expressed in units of
$10^{-11}{\rm erg~cm}^{-2}{\rm s}^{-1}{\rm \AA}^{-1}$.}
\tablenotetext{f}{Identified as a spectroscopic binary by Hutchings 
(1975), with the secondary being much fainter than the primary.  However
this duplicity is questioned by Vreux \& Conti  (1979).}
\tablenotetext{g}{ (Evans 1967).}
\tablenotetext{h}{Values from the catalog of Uesugi \& Fukuda  (1981)
provided by the Vizzier web site (Strasbourg Data Center).}
\tablenotetext{i}{Only a small-aperture IUE exposure is available. 
Fluxes scaled using the TD1 flux at 1565$\,$\AA\ reported by Thompson et
al.  (1978).}
\tablenotetext{j}{ (Thackeray, Tritton, \& Walker 1973)}
\tablenotetext{k}{ (Howarth et al. 1997).}
\tablenotetext{l}{Only a small-aperture IUE exposure is available. 
Fluxes scaled using the ANS flux at 1500$\,$\AA\ reported by Wesselius
et al.  (1982)}
\tablenotetext{m}{ (Wilson 1953).}
\tablenotetext{n}{HD~206267 is a strong x-ray source  (Schulz,
Bergh"fer, \& Zinnecker 1997).  Its A component is a triple system 
(Stickland 1995), with the primary being brightest by $\Delta m = 0.8$ 
(1971).}
\tablenotetext{o}{Identified as a spectroscopic binary by Petrie \&
Petrie  (1967), with $\Delta m = 0.6$.}
\tablenotetext{p}{$\lambda$~Cep}
\tablenotetext{q}{Identified as a spectroscopic binary by Hill \& Fisher 
(1987), with $\Delta m = 2.0$.}
\end{deluxetable}
\clearpage
\begin{deluxetable}{
r    
r    
r    
c    
c    
c    
c    
c    
}
\tablecolumns{8}
\tablewidth{0pt}
\tablecaption{Lines of Sight\label{los}}
\tablehead{
\colhead{HD~~} & \colhead{$\ell$} & \colhead{$b$} &
\colhead{Dist.\tablenotemark{a}} &
\colhead{$v_{\rm LSR}-v_{\sun}$\tablenotemark{b}} &
\colhead{$v_{\rm gal.~rot.}$\tablenotemark{c}} &
\colhead{$E(B-V)$\tablenotemark{d}} & \colhead{$\log N({\rm
H~I})$\tablenotemark{e}} \\
\colhead{} & \colhead{(deg.)} & \colhead{(deg.)} &
\colhead{(pc)} & \colhead{(${\rm km~s}^{-1}$)} & \colhead{(${\rm
km~s}^{-1}$)} & \colhead{} & \colhead{(${\rm cm}^{-2}$)}\\
\colhead{(1)} & \colhead{(2)} & \colhead{(3)} & \colhead{(4)} &
\colhead{(5)} & \colhead{(6)} & \colhead{(7)} & \colhead{(8)}
}
\startdata
108\phantom{A}&117.93&+1.25&2000&\phs 8.6&$-29.3$&0.50&21.53\nl
3827\phantom{A}&120.79&$-23.23$&2100&\phs 4.0&$-26.7$&0.04&20.56\nl
15137\phantom{A}&137.46&$-7.58$&2800&\phs 1.6&$-34.7$&0.33&21.11\nl
69106\phantom{A}&254.52&$-1.33$&1400&$-17.2$&\phs 15.3&0.18&21.08\nl
88115\phantom{A}&285.32&$-5.53$&3700&$-12.4$&~$-6.1$&0.18&21.02\nl
\nl
93843\phantom{A}&288.24&$-0.90$&2900&$-11.1$&$-14.8$&0.28&21.33\nl
94493\phantom{A}&289.01&$-1.18$&3500&$-11.0$&$-14.9$&0.20&21.11\nl
99857A&294.78&$-4.94$&3200&$-9.9$&$-26.3$&0.34&21.31\nl
103779\phantom{A}&296.85&$-1.02$&3700&$-8.9$&$-30.9$&0.22&21.16\nl
106343\phantom{A}&298.93&$-1.83$&3000&$-8.4$&$-32.4$&0.29&\nodata\nl
\nl
109399\phantom{A}&301.71&$-9.88$&2800&$-8.6$&$-34.1$&0.25&21.11\nl
116781A&307.05&$-0.07$&1900&$-5.9$&$-30.0$&0.43&\nodata\nl
120086\phantom{A}&329.61&+57.50&1000&\phs 7.0&~$-4.6$&0.01&20.41\nl
122879\phantom{A}&312.26&+1.79&2400&$-4.0$&$-39.5$&0.37&\nodata\nl
124314A&312.67&$-0.42$&1100&$-4.2$&$-19.3$&0.53&21.34\nl
\tablebreak
203374A&100.51&+8.62&730&\phs
13.8&~$-5.4$&0.60&21.11\tablenotemark{f}\nl
206267A\tablenotemark{g}&99.29&+3.74&790&\phs
13.5&~$-5.6$&0.53&\nodata\nl
208947\phantom{A}&106.55&+9.00&510&\phs 12.5&~$-5.2$&0.19&\nodata\nl
209339A&104.58&+5.87&1000&\phs 12.6&$-10.0$&0.37&\nodata\nl
210839\tablenotemark{h}\phantom{A}&103.83&+2.61&880&\phs
12.4&~$-8.4$&0.57&21.15\nl
\nl
224151\phantom{A}&115.44&$-4.64$&1100&\phs 8.5&$-15.7$&0.49&21.32\nl
\enddata
\tablenotetext{a}{Spectroscopic distances computed from the spectral
types and $V$ magnitudes given in columns 2 and 5 of
Table~\protect\ref{tgt_stars}, the color excesses $E(B-V)$ given in
column 7 of this table, and using the absolute magnitudes of Vacca et al 
(1996) for the O stars and Lesh  (1968) for the B stars and assuming
that $A_V=3.1E(B-V)$.  Known spectroscopic binaries, as noted in
Table~\protect\ref{tgt_stars}, had their $M_V$ values adjusted as noted. 
Since we avoided stars with low values of $v \sin i$ (column 4 of
Table~\protect\ref{tgt_stars}), special corrections in $M_V$ outlined by
Lamers et al.  (1997)  are not needed.}
\tablenotetext{b}{Correction to be added to heliocentric velocity scales
shown in Figs.~\protect\ref{210839_f1f2}$-$\protect\ref{210839_si_ii} to
obtain velocities in the ``standard definition'' of Local Standard of
Rest (LSR)  (Kerr \& Lynden-Bell 1986).  While more rigorous derivations 
(Mihalas \& Binney 1981), pp. 389$-$409) may be more physically accurate
and better suited for the analysis presented in \S\ref{kinematics}, we
retained the standard definition to make it more easy to compare our
radial velocities with those reported by radio astronomers.}
\tablenotetext{c}{Calculated LSR velocity arising from differential
galactic rotation for a point defined by the coordinates given in
columns 2 and 3 and the distance in column 4, using the Galactic
rotation curve of Clemens  (1985) and assuming the Sun is 8.5~kpc from
the center of the Galaxy.}
\tablenotetext{d}{$B$ and $V$ magnitudes from Mermilliod  (1987).  Color
excess computed using the Johnson's  (1963) intrinsic colors for the
spectral types listed in column 2 of Table~\ref{tgt_stars}.}
\tablenotetext{e}{Derived from Ly$\alpha$ absorption in IUE spectra by
Diplas \& Savage  (1994)}.
\tablenotetext{f}{From Table~2 of Diplas \& Savage  (1994), hence of
reduced accuracy due to uncertain stellar parameters.}
\tablenotetext{g}{Situated within IC~1396, an H~II region with
bright-rimmed globules  (Weikard et al. 1996), similar to the Orion
Trapezium.}
\tablenotetext{h}{$\lambda$~Cep}
\end{deluxetable}
\clearpage
\section{Observations}\label{obs}

All of the targets were observered with both the E140H and E230H echelle
modes of STIS  (Woodgate et al. 1998) between 1999 February and 2000
August. Since the C~I lines were expected to be quite narrow and our
analysis benefits from the highest possible resolution, spectra with
both gratings were obtained with the $0\farcs 1 \times 0\farcs 03$ slit,
the narrowest slit available. Similarly, to maximize the resolution of
the data, the spectra were extracted with ``Hi-Res'' half-pixel
centroiding (see \S\ref{rebalance} below). We estimate that this STIS
configuration and analysis option provides a resolving power of $R =
\lambda /\Delta \lambda \approx 200,000$ or ${\rm FWHM} \approx
1.5\,{\rm km~s}^{-1}$ (see Appendix~\ref{resolving_pwr}).  The grating
tilts were selected to cover as many of the C~I multiplets as possible:
the E140H grating was positioned to provide spectra extending from $\sim
1160 - 1361\,$\AA , while the E230H observations were set up to cover
$\sim 1630 - 1902\,$\AA. While this second setup covers only one C~I
multiplet -- the strongest one at 1657$\,$\AA\ -- it has an additional
advantage of covering a weak Si~II line at 1808.013$\,$\AA, which is
useful to compare with the strong transition from its excited
fine-structure level at 1264.738$\,$\AA.

\begin{figure}[ht]
\plotone{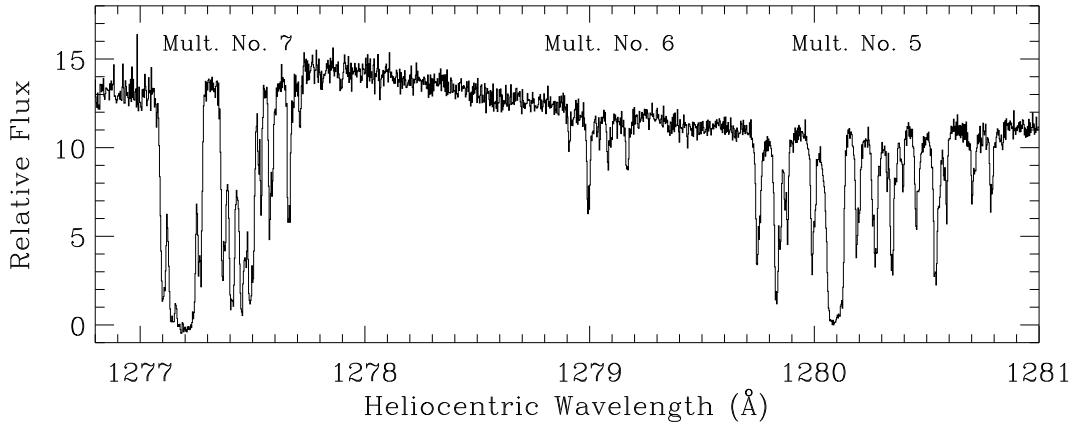}
\caption{A part of the spectrum of HD~210839.  This wavelength interval
appears in a single diffraction order of the E140H grating on STIS.  The
three multiplets of C~I are identified in
Table~\protect\ref{fval_results}.\label{spec210839}}
\end{figure}
Figure~\ref{spec210839} shows a single diffraction order of the E140H
grating.  Three C~I multiplets are visible over the wavelength range
covered by this order.

\placefigure{spec210839}

All of the stars are too bright for standard acquisition procedures. 
Thus the targets had to be initially acquired with a neutral-density
filter and then peaked-up in the narrow slit with a low-dispersion
grating before switching to the high-dispersion modes for the science
exposures.

\section{Data Reduction}\label{reduction}
\subsection{Scattered Light Correction}\label{scat_light}

The data were reduced with the CALSTIS routines developed by the STIS
Investigation Definition Team. In a number of respects, these routines
perform functions that are similar to those of the standard pipeline
reductions, so that the calibrated data benefit from the usual
flat-fielding, wavelength calibrations, and so forth.  Notable
exceptions, however, are the extractions with Hi-Res sampling mentioned
earlier and a more elegant means of estimating scattered light developed
by Lindler \& Bowers  (2001). This method explicitly accounts for
grating scatter from the echelle and cross-disperser elements in STIS
(using a model for these effects derived from preflight test exposures),
as well as for the point spread functions of the telescope,
spectrograph, and MAMA detectors.  By comparing spectra with and without
these corrections, we verified that small-scale artifacts are not being
introduced with an amplitude that is large enough to matter for the
objectives of our program.

\subsection{Rebalancing the Intensities in Adjacent Hi-Res
Pixels}\label{rebalance}

To achieve the highest wavelength resolution possible with the output
from the STIS MAMA detector (and avoid aliasing with our narrow-slit
spectra), we made use of the separate Hi-Res intensity output signals 
(Woodgate et al. 1998), rather than binning them together in pairs (the
usual default with the STIS data products).  At times, the raw signals
showed a significant, high-frequency modulation that arose from a small
imbalance in the responses to the ``even-fold'' and ``odd-fold'' events,
i.e., ones that contribute to adjacent Hi-Res pixels, as sensed by the
outputs from the fine-array elements in the MAMA detector's anode 
(Kasle \& Morgan 1991).  To overcome this effect, we measured the
relative intensity amplitudes of even-numbered pixels with respect to
their odd-numbered neighbors and then performed a least-squares fit to
the outcomes to a create a low-order polynomial representation along
each spectral order.  (We could not just evaluate a global average for
the imbalance, because it changed slowly with position on the detector
format.)  In performing the fit, the relative weights for these even-odd
signal ratios were scaled to the signal strength so that the best-fit
polynomial would not be influenced by nonsense information in the
middles of strong absorption features with a zero-intensity signal
level.  After measuring the effect in this manner, we corrected
alternate Hi-Res pixels so that their responses would be matched to
their neighbors.

\subsection{Combining Exposures}\label{comb_exp}

Some of the target stars required more than one exposure over a given
spectral interval to achieve the desired $S/N$.  Before these exposures
were added together to obtain a composite spectrum, we checked for small
shifts in the positions of the strongest, sharpest features.  All of the
spectra that exhibited shifts relative to the first exposure in a
sequence were realigned to eliminate the degradation in resolution that
would have occurred if the spectra were simply added together in their
original form.  A typical accuracy in registering one spectrum with
respect to another was about 0.2 Hi-Res pixels = $0.14\,{\rm
km~s}^{-1}$, and the largest shifts between spectra were of order
1~Hi-Res pixel = $0.7\,{\rm km~s}^{-1}$.  The intensities for spectra
that had non-integral numbers for the magnitudes of their shifts were
computed using midpoint interpolations represented by sinc functions 
(Bracewell 1965, pp. 194$-$195). 

When the spectra were added together, they were assigned relative
weights in proportion to the inverse squares of their respective errors.
These errors were taken from the error vectors supplied by the CALSTIS
routines but subsequently smoothed with a five-point running average. 
The smoothing reduces the effect that measurements with large positive
or negative random noise excursions in individual samples are
systematically assigned an inappropriate weight, either too large or too
small.  New errors $\epsilon_c$ for the composite spectrum were derived
from the relation $\epsilon_c=(\sum_i \epsilon^{-2})^{-0.5}$.

Some multiplets appear simultaneously at opposite ends of two adjacent
echelle orders.  These multiplets are identified in
Table~\ref{fval_results} -- see note $m$.  We did not combine these
separate recordings, since their agreement (or lack thereof) could be
used to validate our conclusions about random and systematic errors
(\S\ref{errors}).

\subsection{Definition of the Continuum Level}\label{cont_def}

The spectral segments spanning the C~I multiplets were normalized to
continuum levels defined by the fits of Legendre polynomials to fluxes
at wavelengths on either sides of (or between) the lines.  A detailed
description of the method and its merits has been presented by Sembach
\& Savage  (1992).  To be sure that we were not defining continuum
levels in places where there were very weak lines, we determined the
velocity intervals over which C~I could be seen in the strongest lines. 
These intervals were subsequently marked off for the weaker lines to
serve as a guide for regions to avoid when the continuum levels were
defined, so that the fitting regions were not contaminated by very weak
absorption features.  In some cases, we had to violate these guides to
constrain the continuum levels over broad expanses of wavelengths that
were technically forbidden.  When these circumstances arose, we
exercised our best judgement on a choice of regions that avoided a
downward bias in the outcome.

\subsection{Velocity Registration of Multiplets}\label{vel_reg}

When we analyzed the results from different multiplets to create a
composite picture of the C~I, C~I* and C~I**, we recognized the
importance of having an accurate registration of the velocity scales. 
Otherwise, the velocity resolution of the final result would be inferior
to the original resolution of the individual spectra.  The wavelength
scales supplied by CALSTIS are not accurate enough for this purpose.

To align the spectra, we selected in each multiplet a velocity marker
consisting of a feature that was not blended with other transitions.  We
then used pairs of these markers to measure the relative offsets from
one multiplet to the next.  In some cases, these shifts were as large as
$1.3\,{\rm km~s}^{-1}$, but usually they were less than $0.4\,{\rm
km~s}^{-1}$.  In a few cases where the absorptions from isolated excited
fine-structure levels were too weak, we had to use the absorptions out
of just the unexcited level, recognizing that, in principle, some were
contaminated by nearby lines from excited levels.  The alignments
started with the weakest multiplets and progressed to successively
stronger ones so that the comparisons had the smallest disparities in
strength (and hence profile shapes, if they had any asymmetries).

For all but one of the multiplets, the wavelength scales supplied by the
STIS data reduction routine were excellent (except for the small offsets
described above).  The one exception was the echelle order containing
the 1657$\,$\AA\ multiplet.  The individual transitions in this
multiplet span a large range in wavelength, and it was apparent from the
spacing of the absorption features that scale factor for wavelength vs.
linear distance on the detector was slightly incorrect.  This is
probably caused by the paucity of emission lines supplied by the
on-board calibration lamp in this region.  We compensated for the error
by adding an artificial correction term $\Delta\lambda$ to the published
laboratory wavelengths $\lambda$ in this multiplet, with
$\Delta\lambda=0.002615(\lambda-1656.15\,{\rm \AA})$.

\section{Analysis of the C~I Multiplets}\label{analysis}
\subsection{General Principles}\label{principles}

Figure~\ref{spec210839} shows that the complex patterns of velocity
components for C~I, C~I* and C~I** overlaid on the closely spaced line
configurations within each multiplet often create a strongly overlapping
arrangement of absorption that is usually difficult to interpret by
simple inspection.  If one were to try to avoid this confusion by
inspecting only lines with large separations from others or lines at the
extreme ends of the multiplet, a very large fraction of the data would
be overlooked.  This is not a palatable notion, since our objective is
to obtain the most accurate results given the resolution and S/N of the
data.  The challenge, then, is to untangle the overlapping lines and
their multiple velocity components, and to do so without sacrificing
information where there is overlap.  With the analysis of any single
multiplet the solution is likely to be ambiguous, since Doppler velocity
shifts cannot be distinguished uniquely from the shifts in going from
one line to the next.  However, the line patterns differ from one
multiplet to the next, and this change allows one to resolve the
ambiguities if many multiplets are interpreted collectively.

One approach to analyzing the C~I multiplets is to create hypothetical
absorption models that consist of superpositions of various Gaussian
components, compare them to the data, and then modify them so that they
converge on the best solution by minimizing the $\chi^2$ of the fit. 
This is a popular analysis scheme among interstellar line investigators
[e.g., Vidal-Madjar et al  (1977), Ferlet et al.  (1980), Lemoine et al. 
(1995), Welsh et al.,  (1990, 1991), Hobbs \& Welty  (1991), Welty et
al.  (1991, 1994), Vallerga et al.,  (1993), Spitzer \& Fitzpatrick, 
(1993, 1995), Fitzpatrick \& Spitzer  (1994), Crane et al.,  (1995) and
H\'ebrard et al.  (1999)].  However, there is the requirement that the
initial definition of the model's properties (such as the number of
components and their approximate positions, amplitudes and widths) be
based on a subjective initial guess or, alternatively, on evidence from
the investigations of lines from
other species in the same line of sight.  Also, one may end up with the
lingering concern that another solution set, created from a different
initial trial model or method of convergence, could provide a better
fit.  Even so, component fitting yields useful insights and has even
been employed with good success to C~I multiplets with up to 7
identified velocity components  (Jenkins \& Wallerstein 1995; Jenkins et
al. 1998).

For our survey of C~I and its excited fine structure states, we wish to
avoid the labor-intensive task of model fitting and invoke a purely
mechanical analysis method that presents a more objectively derived
output product.  One such procedure that can be applied to single
transitions is the apparent optical depth method  (Savage \& Sembach
1991), where we first derive an apparent optical depth $\tau_a$ as a
function of radial velocity,
\begin{equation}\label{tau_a}
\tau_a(v) = \ln \left( {I_0(v)\over I(v)}\right)~.
\end{equation}
from a recorded intensity $I(v)$ and an estimate for the continuum level
$I_0(v)$ that would be present in the absence of any absorption.  When
$\tau_a(v)$ is not so large that noise and uncertainties in the
background level create substantial errors, we obtain a differential
column density per unit velocity through the relation
\begin{equation}\label{N_a}
N_a(v) = 3.768\times 10^{14}{\tau_a(v)\over f\lambda}{\rm cm}^{-2}({\rm
km~s}^{-1})^{-1}~,
\end{equation}
where $f$ is the transition's oscillator strength and $\lambda$ is
expressed in \AA.  If the detailed velocity structure is unresolved by
the instrument, then the derived $N_a(v)$ is a smoothed version of the
true distribution $N(v)$.  If there are unresolved narrow structures
{\it and\/} the lines are strong, then $N_a(v)$ underrepresents $N(v)$
and one must apply a correction at each velocity based on a comparison
of $N_a(v)$ for two or more lines of differing strength, much as one
would invoke a curve-of-growth analysis for the equivalent widths of
saturated lines  (Jenkins 1996).  We must watch for this effect when we
analyze strong absorptions.

\subsection{Specific Applications to C~I
Multiplets}\label{specific_appl}

\subsubsection{Equations that Unravel Blended Features}\label{unravel}

As we stated earlier, the features within the multiplets strongly
overlap each other, and this precluded our being able to derive $N_a(v)$
for either C~I, C~I* or C~I** by a straightforward application of
Eq.~\ref{N_a} to the observed continuum-normalized intensities. 
Consequently, we had to develop a new method of analysis that could
eliminate the profile confusion and present simple representations of
the separate $N_a(v)$'s for the three species.
 
Within a single multiplet that contains a mixture of lines from the
three excitation levels, we re-expressed their separate $N_a(v)$ in
terms of a collection of discrete unknowns $X_{k,\epsilon}$, with the
index $k$ specifying any of $L$ velocity bins and $\epsilon$ specifying
the level of excitation ($\epsilon = 0,1,2$ for C~I, C~I* and C~I**,
respectively).  For the survey described here, we found it adequate to
have $L=300$ for a contiguous set of bins, each $0.5\,{\rm km~s}^{-1}$
wide, covering the interval from $-100$ to $+50\,{\rm km~s}^{-1}$.  No
C~I lines are apparent outside of this velocity range, except for the
star HD~69106, where we had to redefine the velocities to run from $-50$
to $+150\,{\rm km~s}^{-1}$.

As with the unknown variables $X_{k,\epsilon}$, we considered the
observed optical depths $\tau_a(\lambda)$ over the relevant wavelength
interval in terms of a large collection of discrete measurements
$\tau_a(i)$ sampled at successive wavelengths $[\lambda_0(i-200)/6\times
10^5] + \lambda_0$.  ($\lambda_0$ is an arbitrary reference wavelength
within or near the multiplet.)  Associated with each $\tau_a(i)$ is an
expectation for the combined contribution from all transitions (each
designated with the subscript $\ell$) within a multiplet,
\begin{equation}\label{tau_sum}
\tau(i)=C\sum_\ell (f\lambda)_\ell X_{k,\epsilon}
\end{equation}
where $C=(3.768\times 10^{14})^{-1}$, i.e., the inverse of the constant
that appears in Eq.~\ref{N_a}.  For each case $\ell$, the index $k$ is
set to the nearest integer in the expression
\begin{equation}\label{k}
k=i-6\times 10^5{\lambda_\ell-\lambda_0\over \lambda_0}
\end{equation}
and $\epsilon$ is defined according to the line's originating excitation
level.  Our objective is to find a solution set for all $X_{k,\epsilon}$
that minimizes
\begin{equation}\label{chisq}
\chi^2=\sum_i\left\{ \left[
\tau_a(i)-\tau(i)\right]/\sigma_{\tau(i)}\right\}^2
\end{equation}
where $\tau(i)$ is the true, composite optical depth given in
Eq.~\ref{tau_sum}, but sampled as an apparent optical depth $\tau_a(i)$
with an estimated uncertainty $\sigma_{\tau(i)}$.  Noise, systematic
uncertainties arising from improper definitions of the continuum level
$I_0$ and the zero-intensity baseline, and the incorrect interpretations
of unresolved saturated structures can, in principle, all conspire to
make $\tau_a(i)$ slightly different from the real $\tau(i)$.

The derivative of $\chi^2$ with respect to any one of the variables
$X_{j,\epsilon}$ is given by
\begin{equation}\label{partial_chisq_X}
{\partial \chi^2\over\partial X_{j,\epsilon}}=-2A\sum_i\sum_\ell\left\{
\delta\left[ i,\left( j+6\times 10^5{\lambda_\ell-\lambda_0\over
\lambda_0}\right) \right] (f\lambda)_\ell\left[ {\tau_a(i)-\tau(i)\over
\sigma_{\tau(i)}^2}\right] \right\}~.
\end{equation}
The $\delta$ expression is a Kronecker delta that activates the terms
only when there is a close match in wavelength between a portion of an
individual line's absorption profile with the relevant $X_{j,\epsilon}$. 
From the argument of $\delta$, it is clear that the meaning of the index
$j$ in terms of wavelength offsets is the same as that for $k$ given in
Eq.~\ref{k}.  Nevertheless, $j$ is distinct from $k$ that appears when
Eq.~\ref{tau_sum} is substituted into the expression for $\tau(i)$.

For each of the 900 possible combinations of $(j,\epsilon$), we have
different forms of Eq.~\ref{partial_chisq_X}.  The best solutions for
$X_{j,\epsilon}$ are found by setting these expressions equal to zero
for a minimum in $\chi^2$ and then solving all of the equations
simultaneously. In each case, we have an equation that involves a sum of
a constant number that includes $\tau_a(i)$, a term that is composed of
$X_{j,\epsilon}$ times a coefficient, and then numerous additional terms
that include other $X_{k,\epsilon}$ variables that overlap in wavelength
(multiplied by other coefficients).  By expanding the sums to cover
results for many different multiplets, we unravel the ambiguity in the
solutions for $X_{j,\epsilon}$ arising from the overlap of the lines in
any single multiplet. That is, we can obtain unique answers for all
$X_{j,\epsilon}$ by virtue of the different arrangements of the
transitions within each multiplet.

A large fraction of the coefficients in the $900\times 900$ matrix for
the entire system of equations are zero.  The only non-zero ones are
those along diagonals that satisfied the identity required by the
Kroneker delta of Eq.~\ref{partial_chisq_X}.  This condition allows us
to take advantage of numerical techniques for solving sparse matrices
[e.g., Press et al.  (1992)], which makes the problem tractable.

The most elementary way to express the error $\sigma_{\tau(\lambda)}$ is
simply to multiply the error in intensity $\sigma_{I(\lambda)}$ by $|
d\tau/dI |$ to get $\sigma_{I(\lambda)}/I(\lambda)$.  However, this is a
poor approximation when $\sigma_{I(\lambda)}$ is not very much less than
$I(\lambda)$ or when there are errors in the estimate for the zero level
that can have a large relative effect on the derived value of
$I(\lambda)$.  For this reason, we felt that it was prudent to declare
much larger errors for low intensities by creating a modification (the
second term in the equation) to yield
\begin{equation}\label{sigma_tau}
\sigma_{\tau(\lambda)} = {\sigma_I(\lambda)\over I(\lambda)}~\times ~ {2
\over 1 + \tanh [\pi (I-I_t)/0.1] } 
\end{equation}
that doubled the assumed error in the vicinity of transition intensity
$I_t$ and made it virtually infinite for intensities that were much
lower.  The effect of this representation for $\sigma_{\tau(\lambda)}$
is that we essentially discard all intensity measurements below a level
of about $I_t-0.1$ and then provide a smooth transition to the simple
form $\sigma_{I(\lambda)}/I(\lambda)$ for much larger intensities.

To arrive at reasonable values for $I_t$, we studied the departures
between a Gaussian distribution in $\tau$ with a standard deviation
equal to $\sigma_{I(\lambda)}/I(\lambda)$ (i.e., our adopted
approximation for the probability density) and the actual distribution
that arises from a transformation of a Gaussian distribution with a
standard deviation $\sigma_{I(\lambda)}$ through the nonlinear operation
of taking a logarithm.  This comparison was done for various values of
$I/I_0$ and the signal-to-noise ratio ($S/N$) at the continuum.  For a
spectrum with $S/N=10$, the true probability distribution begins to show
moderate differences from the approximation when $I/I_0=0.20$, and the
deviations become appreciable for lower intensities. For $S/N=20$ the
departures become noticeable only when the intensity approaches
$I/I_0=0.15$.

For all multiplets in Table~\ref{fval_results} except the ones near
1657$\,$\AA\ (Mult. nr.~2), 1192$\,$\AA\ (Mult. nr.~12) and 1189$\,$\AA\
(Mult. nr.~14), we adopted $I_t=0.1$ because their $S/N$ generally
ranged between 20 and 30.  The signal quality for the 3 exceptions was
lower, generally $8 < S/N < 12$, so we set $I_t=0.2$ when their
$\sigma_{\tau(\lambda)}$ were evaluated.  We believe that these 3
multiplets also have a greater chance of having an incorrect assignment
of the zero intensity level, so the absolute insensitivity to
intensities lower than $0.1I_0$ seemed warranted.  To check that our
derivations of $N_a$ were not sensitive to the choice of $I_t$, we
performed a duplicate, experimental analysis where each of the values of
$I_t$ were increased by a factor of two.  There were only very small
changes in the outcomes.  The rederivations of $f$-values described in
\S\ref{comparisons} are likewise not sensitive to the adopted values of
$I_t$.

Since unwarranted changes in $\sigma_{I(\lambda)}$ may be caused by
noise fluctuations in $I(\lambda)$, we needed to take a special
precaution that the correlation between the two did not give negative
noise spikes an unjustifiably higher weight.  To guard against this from
happening, we smoothed the $\sigma_{I(\lambda)}$ vector with a 5-point
running mean to reduce the correlation. 

\subsubsection{A Validation of the Method}\label{validation}

To demonstrate that the analysis described above gives results that are
equivalent to the more traditional techniques of profile fitting, we
compare the two for one particular example and show that they yield
results that are nearly identical.  Jenkins et al.  (1998) interpreted
the complex pattern of absorption by C~I, C~I* and C~I** toward HD~72089
by defining a set of Gaussian velocity components and then using a
component-fitting program developed by one of the authors
[E.~Fitzpatrick, see e.g., Spitzer \& Fitzpatrick  (1993) or Fitzpatrick
\& Spitzer  (1997)] to minimize the $\chi^2$ between the data and the
components' column densities, central velocities, and velocity
dispersions.  We now analyze once again the same observations, but this
time by using the equations in \S\ref{unravel} that unravel the apparent
optical depths.

\begin{figure}
\plotone{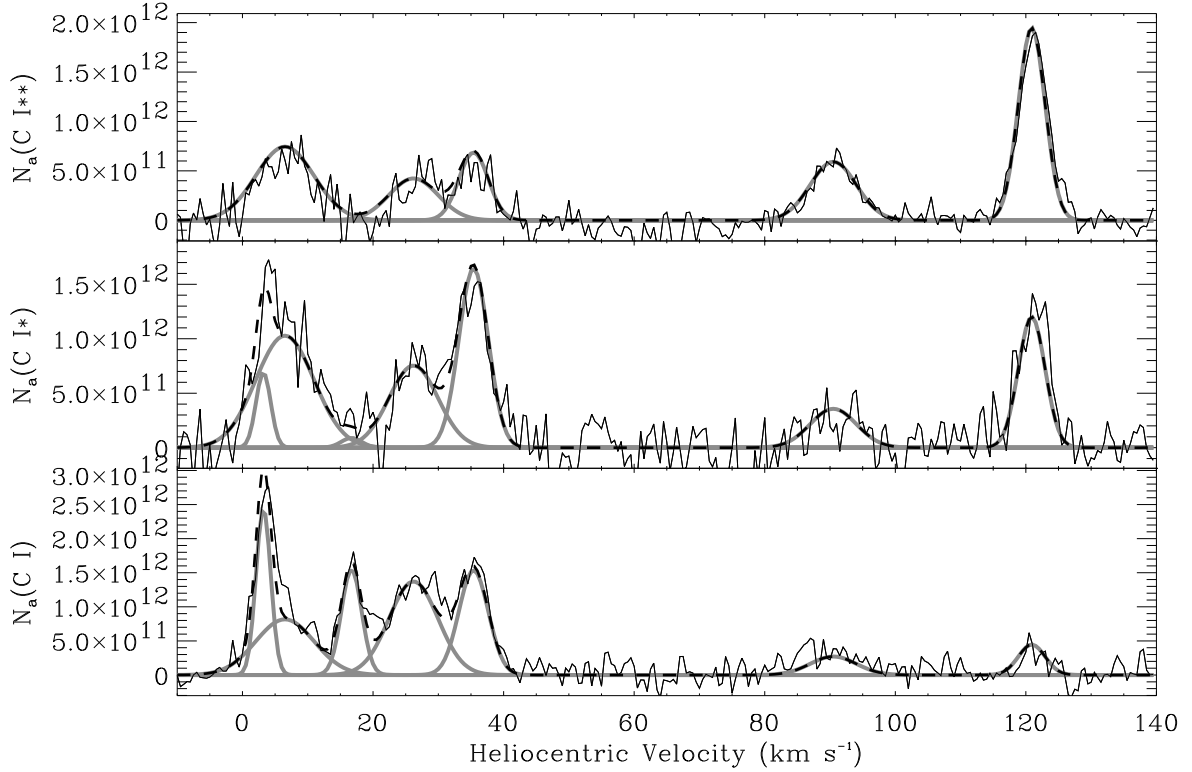}
\caption{A comparison between a re-analysis done here for the STIS
observations of C~I multiplets in the spectrum of HD~72089 and the
original analysis of the same data by Jenkins et al  (1998) using a
component-fitting technique.  Column densities per unit velocity were
derived according to the method described in \S\protect\ref{unravel}
(top panel: C~I**, middle panel: C~I*, and bottom panel: C~I) and are
plotted (thin, solid line) as a function of heliocentric radial
velocity.  Overplotted on our determinations are the original,
best-fitting Gaussian components defined by Jenkins et al (thick, gray
lines), and their sum is depicted by a heavy, dashed
line.\label{72089chk}}
\end{figure}

Figure~\ref{72089chk} shows the outcome of the comparison between the
two methods.  To within the uncertainties that probably arise from
differences in defining continuum levels and backgrounds, the outcomes
are virtually identical.  To make the comparison more meaningful, we
used the $f$-values given by Morton  (1991), as done by Jenkins et al 
(1998), rather than the revised $f$-values described in \S\ref{fval}
below.

\placefigure{72089chk}

\subsection{Self Consistent $f-$values}\label{fval}
\subsubsection{Optical Depth Comparisons}\label{comparisons}

After deriving $X_{j,\epsilon}$ and comparing the observed $\tau_a(i)$
with reconstructions based on Eq.~\ref{tau_sum}, we were not surprised
to find systematic differences in the strengths of features compared to
their expectations as we went from one multiplet to the next.  In the
face of inconsistencies, the system of equations discussed above struck
a compromise, with appropriate weight factors $\sigma_\tau(i)^{-2}$,
between the preferred solutions for lines within the various multiplets. 
We felt that it was most likely that the differences arise from errors
in the published $f$-values, although we could not rule out additional,
probably much smaller contributions from errors in the estimates for the
zero intensity levels associated with the observations.  On this
premise, we chose to solve for revisions in the $f$-values that gave
mutually consistent results.  After doing so, we substituted the new
$f$-values into the equations of \S\ref{specific_appl} to give modified
values for $X_{j,\epsilon}$, ones that we expect are better than those
that came from the relative $f$-values that gave disparate results.

Ultimately, we must trust that the $f$-values in one of the multiplets
are correct, so that we can establish an absolute standard against which
the other findings must be found to be in agreement.\footnote{One might
question why we bothered to derive the $f$-values of multiplets other
than the one we trusted if we were simply adjusting them so that they
were consistent with the C~I column densities derived from the primary
multiplet.  There are two answers to this question.  First, the lines in
a single multiplet can define $N_a(v)$ only under the condition that the
appropriate $\tau_a(v)$'s are much larger than their respective noise
levels or continuum level errors, or, in the case of very strong lines, 
not so large that the absorptions are nearly saturated.  When these
conditions are not satisfied, results from the other multiplets must
fill in the gaps.  Second, we combined the results for many stars to
obtain a more accurate set of relative $f$-values.  These generalized
results will not automatically give trivial agreements for certain lines
with the outcomes for the primary multiplet in any single, given star.} 
Of all the multiplets included in our study, the $^3P-$$^3P^0$ multiplet
at 1657.2\,\AA\ is assigned the highest accuracy (of order 3\%) in the
compilation by Wiese et al.  (1996).  Since the energy of this
multiplet's upper state is well isolated from others, deviations from LS
coupling are likely to be small.  This gave us some confidence about the
accuracy of the relative populations of C~I in the three levels of
excitation, as derived from this multiplet and others that are coupled
to it by our comparative analysis.

To solve for the best (relative) $f$-values, we repeated the general
strategy of minimizing the $\chi^2$ that we employed for determining the
the best set of $X_{j,\epsilon}$ outlined in \S\ref{specific_appl},
except that this time we differentiate with respect to the $f\lambda$
under question, which we designate as $(f\lambda)_{\ell^\prime}$.  This
gives the equation
\begin{equation}\label{partial_chisq_fl}
{\partial \chi^2\over\partial
(f\lambda)_{\ell^\prime}}=-2A\sum_*\sum_i\left\{ X_{j,\epsilon}\left[
{\tau_a(i)-\tau(i)\over
\sigma_{\tau(i)}^2}\right] \right\}
\end{equation}
which may be written for every line in a given multiplet.  As before,
Eq.~\ref{tau_sum} gives the expression for $\tau(i)$ and Eq.~\ref{k}
gives the transformations from $i$ to $k$.  The transformation from $i$
to $j$ is similar, except that the appropriate $\lambda_\ell$ is
$\lambda_{\ell^\prime}$, i.e., the one for the value of $f\lambda$ that
is being considered.  The sum with index ``$*$'' indicates that we
included data for all of the target stars.  When we derived solutions to
the equations, we needed to consider only one multiplet at a time, since
there is no mixing of signals across multiplets.\footnote{While the
third line in Multiplet~7.01 at 1277.1901$\,$\AA\ has a wavelength very
close to that of the first line in Multiplet~7 at 1277.2454$\,$\AA, it
is so weak that its contribution should be negligible (see
Table~\protect\ref{fval_results}).}

After deriving revised $f$-values, we repeated the determinations of
$X_{j,\epsilon}$ since small changes arose from the adjustments to
$(f\lambda)_{\ell^\prime}$.  We repeated the cycles that alternately
implemented Eqs.~\ref{partial_chisq_X} and \ref{partial_chisq_fl} until
the results converged.  Column 7 of Table~\ref{fval_results} shows the
final outcomes for $\log(f\lambda)$.  To give an indication on the
consistency of the results from one star to another (by using
Eq.~\ref{partial_chisq_fl} without the $\sum_*$ in each case), we
computed the median absolute deviation  (Mosteller \& Tukey 1977) of the
individual results from the overall result for all stars.  These results
are shown in the last column of the table.  For the purposes of
comparison, we show some previously listed transition strengths in the
recent literature (columns 4$-$6), again in terms of $\log (f\lambda)$.

Generally, from the median absolute deviations from star to star and
also the differences in the outcomes for multiplet nrs. 7 and 8.01 that
were recorded twice (in different orders) and evaluated separately, we
judge that random errors in $\log (f\lambda)$ are of order 0.025~dex. 
(Exceptions are a few weak lines where the median absolute deviations
are extraordinarily large.)  This is probably a good measure for the
errors in the relative $f$-values for multiplets whose strengths are
about the same order of magnitude.  However, in addition to these random
errors, we recognize that there could be a gradual buildup in a
systematic error as we progress from the strongest multiplets to the
weakest ones.  This effect can arise from the fact that each multiplet
is constrained by ones that are only mildly stronger or weaker.  There
is no direct coupling between the very strong and the very weak
multiplets because at places where the latter are strong enough to
register above the noise, the former are saturated and thus of no use. 
By design, the solutions reflect this by making the $\tau_a(i)-\tau(i)$
terms in Eq.~\ref{partial_chisq_X} very small compared to their
accompanying $\sigma_{\tau(i)}$.  As a result, there is no way restrain
a gradual, progressive creep in the systematic errors as we go from the
strong to the weak multiplets.  Unfortunately, the magnitude of this
error is difficult to assess.

\placetable{fval_results}
\clearpage
\begin{deluxetable}{
c     
c     
c     
c     
c     
c     
c     
c     
c     
c     
}
\tablecolumns{10}
\tablewidth{525pt}
\footnotesize
\tablecaption{New Line Strengths [$\log(f\lambda)$]\label{fval_results}}
\tablehead{
\colhead{} & \colhead{} & \colhead{} & \colhead{} &
\multicolumn{3}{c}{Previously Published $\log (f\lambda)$} & \colhead{}
& \multicolumn{2}{c}{This Paper} \\
\cline{5-7} \cline{9-10}\\
\colhead{Mult.} \\
\colhead{nr.\tablenotemark{a}} & \colhead{$\lambda$
(\AA)\tablenotemark{b}} & \colhead{$\epsilon$\tablenotemark{c}} &
\colhead{} & \colhead{M91\tablenotemark{d}} &
\colhead{WFD96\tablenotemark{e}} & \colhead{ZFC97\tablenotemark{f}} &
\colhead{} & \colhead{$\log (f\lambda)$} &
\colhead{MAD\tablenotemark{g}} \\
}
\startdata
2&1656.2672&1&&\phs 1.987&\phs 1.989&\nodata&&\tablenotemark{h}&\nodata\nl
(0.35)& 1656.9283&0&&\phs 2.367&\phs 2.362&\nodata&&\tablenotemark{h}&\nodata\nl
& 1657.0082&2\tablenotemark{i}&&\phs 2.242&\phs
2.236&\nodata&&\tablenotemark{h}&\nodata\nl
& 1657.3792&1&&\phs 1.765&\phs 1.771&\nodata&&\tablenotemark{h}&\nodata\nl
& 1657.9068&1&&\phs 1.890&\phs 1.893&\nodata&&\tablenotemark{h}&\nodata\nl
& 1658.1212&2&&\phs 1.765&\phs 1.771&\nodata&&\tablenotemark{h}&\nodata\nl
\nl
4&1328.8333&0&&\phs 1.887&\phs 1.924&\nodata&&\phs 2.077&0.010~($n=21$)\nl
(0.95)& 1329.0853&1&&\phs 1.410&\phs 1.452&\nodata&&\phs 1.680&0.025~($n=19$)\nl
& 1329.1004&1&&\phs 1.507&\phs 1.539&\nodata&&\phs 1.762&0.029~($n=19$)\nl
& 1329.1233&1&&\phs 1.285&\phs 1.328&\nodata&&\phs 1.613&0.024~($n=19$)\nl
& 1329.5775&2&&\phs 1.762&\phs 1.801&\nodata&&\phs 1.886&0.023~($n=17$)\nl
& 1329.6005&2&&\phs 1.285&\phs 1.325&\nodata&&\phs 1.610&0.046~($n=13$)\tablebreak
5&1279.8904&1&&\phs 1.243&\phs 1.208&\tablenotemark{j}&&\phs 
1.557&0.022~($n=19$)\nl
(1.0)& 1280.1353&0&&\phs 1.493&\phs 1.467&\tablenotemark{j}&&\phs 
1.789&0.024~($n=21$)\nl
& 1280.3328&2&&\phs 1.273&\phs 1.260&\tablenotemark{j}&&\phs 
1.548&0.036~($n=13$)\nl
& 1280.4042&1&&\phs 0.749&\phs 0.737&\tablenotemark{j}&&\phs 
1.228&0.083~($n=17$)\nl
& 1280.5970&1&&\phs 0.942&\phs 0.936&\tablenotemark{j}&&\phs 
1.357&0.032~($n=17$)\nl
& 1280.8470&2&&\phs 0.811&\phs 0.798&\tablenotemark{j}&&\phs
1.231&0.058~($n= 6$)\nl
\nl                                                                 
6&1279.0558&1&&$-0.026$\tablenotemark{k}&$-0.026$&\phs 0.412&&\phs 
0.892&0.078~($n=11$)\nl
(1.0)& 1279.2286&2&&\phs 0.684\tablenotemark{k}&\phs 0.684&\phs 0.617&&\phs 
1.061&0.035~($n= 5$)\nl
& 1279.4977&2&&$-0.600$\tablenotemark{k}&$-0.594$&\phs 0.063&&\phs
0.593\tablenotemark{l}&\nodata\nl
\nl                                                                 
7&1277.2454&0&&\phs 2.091&\phs 2.077&\tablenotemark{j}&&\phs 
2.225\tablenotemark{m}&0.019~($n=21$)\nl
(0.75)& 1277.2823&1&&\phs 1.967&\phs 1.955&\tablenotemark{j}&&\phs 
2.017\tablenotemark{m}&0.049~($n=21$)\nl
& 1277.5130&1&&\phs 1.489&\phs 1.455&\tablenotemark{j}&&\phs 
1.703\tablenotemark{m}&0.021~($n=19$)\nl
& 1277.5496&2&&\phs 2.016&\phs 2.004&\tablenotemark{j}&&\phs 
2.042\tablenotemark{m}&0.042~($n=19$)\nl
& 1277.7229&2&&\phs 1.268&\phs 1.297&\tablenotemark{j}&&\phs 
1.548\tablenotemark{m}&0.028~($n=13$)\nl
& 1277.9538&2&&\phs 0.092&\phs 0.019&\tablenotemark{j}&&\phs
0.593\tablenotemark{m}&0.154~($n= 1$)\tablebreak
7.01&1276.4825&0&&\phs 0.759&\phs 0.758&\phs 0.331&&\phs 1.178&0.041~($n=19$)\nl
(1.0)& 1276.7498&1&&\phs 0.564&\phs 0.508&\phs 0.485&&\phs 1.022&0.072~($n=12$)\nl
& 1277.1901\tablenotemark{n}&2&&$-0.362$&$-0.468$&\nodata&&\nodata&\nodata\nl      
\nl                                                                 
8.01&1270.1434&0&&$-0.307$&$-0.455$&$-0.265$&&\phs 0.439\tablenotemark{m}&0.061~($n=12$)\nl
(0.85)& 1270.4080&1&&$-1.080$&\nodata&
$-0.582$&&$-0.285\tablenotemark{m}$&\nodata\nl
& 1270.8439\tablenotemark{n}&2&&$-2.869$&\nodata&\nodata&&\nodata&\nodata\nl
\nl                                                                 
9&1260.7355&0&&\phs 1.696&\phs 1.681&\tablenotemark{j}&&\phs
1.870&0.025~($n=21$)\nl
(0.90)& 1260.9266&1&&\phs 1.219&\phs 1.231&\tablenotemark{j}&&\phs 
1.517&0.031~($n=18$)\nl
& 1260.9962&1&&\phs 1.094&\phs 1.122&\tablenotemark{j}&&\phs 
1.444&0.039~($n=18$)\nl
& 1261.1223&1&&\phs 1.316&\phs 1.268&\tablenotemark{j}&&\phs 
1.537&0.020~($n=18$)\nl
& 1261.4257&2&&\phs 1.094&\phs 1.105&\tablenotemark{j}&&\phs 
1.417&0.099~($n=12$)\nl
& 1261.5519&2&&\phs 1.571&\phs 1.582&\tablenotemark{j}&&\phs 
1.746&0.015~($n=16$)\nl
\nl                                                                 
12&1192.2175&0&&\phs 0.496&$-0.004$&\phs 0.019&&\phs 0.827&0.065~($n=17$)\nl
(0.40)& 1192.4507&1&&\phs 0.399&\phs 0.226&\nodata&&\phs 0.885&0.078~($n=11$)\nl
& 1192.8347\tablenotemark{n}&2&&\nodata&
$-0.476$&\nodata&&\nodata&\nodata\tablebreak
14&1188.8332\tablenotemark{o}&0&&\phs 1.299&\phs 1.105&\nodata&&\nodata&\nodata\nl 
(0.40)& 1188.9925&1&&\phs 0.822&\phs 0.736&\nodata&&\phs 1.156&0.116~($n=14$)\nl
& 1189.0650&1&&\phs 0.697&\phs 0.664&\nodata&&\phs 1.138&0.080~($n=14$)\nl
& 1189.2487&1&&\phs 0.919&\phs 0.529&\nodata&&\phs 1.041&0.079~($n=11$)\nl
& 1189.4469&2&&\phs 0.697&\phs 0.602&\phs 0.737&&\phs
1.18\tablenotemark{l}&\nodata\nl
& 1189.6307&2&&\phs 1.174&\phs 1.056&\phs 1.215&&\phs
1.42\tablenotemark{l}&\nodata\nl   
\enddata
\tablenotetext{a}{Numbering system of Moore  (1998), also adopted by
Morton  (1991).  Numbers in parentheses show the average multipliers
that should be applied to the S/N values listed in column 7 of
Table~\protect\ref{tgt_stars}.}
\tablenotetext{b}{Values taken from Morton  (1991)}
\tablenotetext{c}{Fine-structure excitation: 0~=~ground state ($^3P_0$),
1~=~first excited level ($^3P_1$), and 2~=~second excited level
($^3P_2$).}
\tablenotetext{d}{ (Morton 1991).}
\tablenotetext{e}{ (Wiese, Fuhr, \& Deters 1996).}
\tablenotetext{f}{ (Zsarg\'o, Federman, \& Cardelli 1997).}
\tablenotetext{g}{MAD = Median Absolute Deviation [ (Mosteller \& Tukey
1977), Chap.~10] for results from individual stars that are expected to
show a sustained absorption feature with a depth of at least 10\%.  The
number of stars used is given in parentheses.  If this number is less
than 4, no value for the MAD is given.}
\tablenotetext{h}{Results given by Wiese et al  (1996) for the
$f$-values of this multiplet were adopted as a standard against which
other $f$-values were compared.}
\tablenotetext{i}{In  (Morton 1991) there is an incorrect assignment of
the excitation energy for this particular transition.}
\tablenotetext{j}{In the paper by Zsarg\'o et al.  (1997), the results
given by Morton  (1991) for the $f$-values of this multiplet were
adopted as a standard against which other $f$-values were compared.}
\tablenotetext{k}{Values taken from erratum  (Morton 1992).}
\tablenotetext{l}{Only the spectrum of HD~206267A could be used for this
determination.}
\tablenotetext{m}{This transition is seen twice, in different orders. 
This determination is based on a consolidation of information from both
orders.}
\tablenotetext{n}{This line is too weak to show up in our spectra.}
\tablenotetext{o}{A Cl~I transition at 1188.7742 is likely to interfere
with this C~I line.  For this reason, we exclude this line from our
analysis.}
\end{deluxetable}
\clearpage

One could imagine that lines that are from the same level and very close
to each other in wavelength, such as those from the excited levels in
Multiplet nr. 4, might have larger than usual errors, especially if they
are not well resolved from each other because the C~I* and C~I**
features are intrinsically broad.  The errors that could arise would be
in the relative strengths of the lines within a group, while the sum of
$f$-values for the individual groups should remain accurate.

\subsubsection{Misleading Trends Arising from Unresolved, Saturated
Substructures?}\label{substructures}

\begin{figure}
\plotone{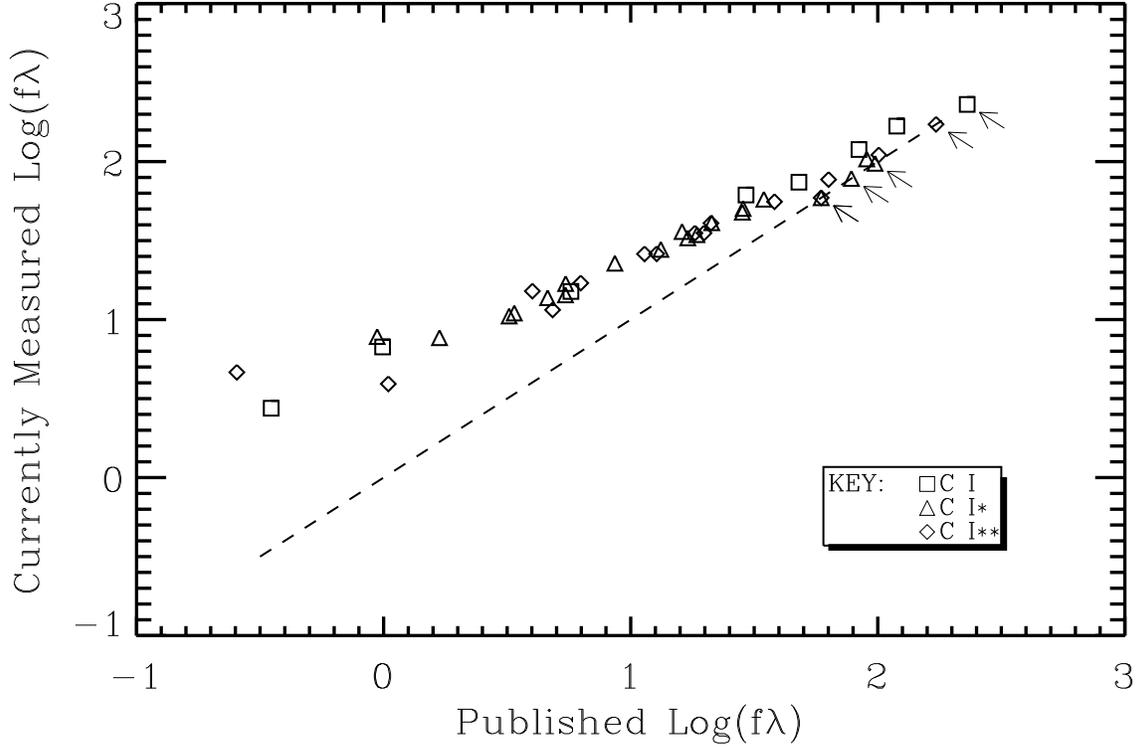}
\caption{A comparison of $f$-values: each symbol [squares are out of the
$^3P_0$ unexcited level (C~I), triangles from the $^3P_1$ first excited
level (C~I*), and diamonds from the $^3P_2$ second excited level
(C~I**)] has an abscissa corresponding to the value of $\log (f\lambda)$
given by Wiese et al.  (1996) and an ordinate equal to our values
reported in Table~\protect\ref{fval_results}.  Arrows indicate
transitions in Multiplet~2 at 1657$\,$\AA, which served as our reference
standard and fall on the line of equality (dashed line) by definition. 
For the probable errors of our determinations, see the discussions at
the ends of \S\S~\protect\ref{comparisons} and
\protect\ref{sys_errors}\label{fvalcomp}}
\end{figure}

A striking conclusion that emerges from a comparison of the previous
determinations of $\log (f\lambda)$ and our new derivations is that
there is a progressive increase in the differences between the two when
going from strong to weak transitions.  This phenomenon is illustrated
in Fig.~\ref{fvalcomp}.  In view of the smooth and progressive nature of
this trend, we must ask the question, ``Is this effect real and a
consequence of some generalized problem with the previous
determinations, or are we being systematically deceived by distortions
of the apparent optical depths from one transition to the next?''

\placefigure{fvalcomp}

In principle, an effect similar to that exhibited in Fig.~\ref{fvalcomp}
could arise if the profiles that we observed are really composed of very
narrow velocity components that do not overlap each other and are badly
saturated for the strong transitions.  While the spectrograph has a high
wavelength resolving power ($\lambda/\Delta\lambda = 200,000$; see
Appendix~\ref{resolving_pwr}), we still expect that many components are
well short of being resolved, if we consider the profiles of Na~I
recorded by Welty, Hobbs \& Kulkarni  (1994) as representative.  As we
pointed out earlier (\S\ref{principles}), when very narrow profiles are
badly saturated and unresolved, their apparent optical depths may not
accurately reflect the real optical depths, or even a smoothed
representation of them.  When we compare different transitions, the
resulting error will always work in the direction of making strong lines
appear to be not as strong as they should be, relative to the weaker
ones where the distortions virtually disappear.

Several tests indicate that it is unlikely that we are being deceived by
this effect.  First, nearly all of the values for the median average
deviations in our derived values of $\log (f\lambda)$, when measurable,
are below 0.1.  Many are well below this value.  Variations in the
special circumstances for saturated profiles from one star to the next
would probably create a more erratic outcome.  It is difficult to be
quantitatively precise in this conclusion however.

Second, in any given multiplet, the line out of the unexcited level is
much stronger than other lines that come from the excited levels (for
the most part, simply because there are more carbon atoms in the
unexcited level).  Since the saturation distortion becomes worse when
the lines are stronger, we would expect that the difference trend would
be more pronounced for the lines out of the unexcited level than for
those out of the excited levels.  This appears not to be happening,
since the points in Fig.~\ref{fvalcomp} all seem to lie on a common
track (as indicated in the caption, different symbols represent
transitions out of the three different levels).

Third, as an experiment to probe how sensitive the outcomes for the
$f$-values were to the wavelength resolution of the spectra, we repeated
the analyses which derived $N_a(v)$ and the transition strengths, but
with the same spectra after they had been artificially degraded by a
smoothing operation.  The degradation was carried out by convolving the
spectra with a Gaussian function that had a width equal to $\sqrt{3}$
times that of the instrumental profile, thus decreasing the net
resolution by a factor of 2 from the original recording.  Some random
differences were seen between the two sets of outcomes for
$\log(f\lambda)$ (degraded vs. original), with magnitudes generally of
order 0.05~dex or less.  In addition, there was a small shift (of order
0.025~dex) of the weaker transitions to even lower values, an effect
{\it opposite\/} to the one that we would have expected under the
hypothesis that the smoothing should make the weak lines look less
different from the strong ones.  That is, the trend was the reverse of
that shown in Fig.~\ref{fvalcomp} (but very much smaller in magnitude),
if we plotted the outcomes from the degraded spectrum on the $y$-axis
against the results at full resolution on the $x$-axis.

A fourth way to sense whether or not our revised strengths for weak C~I
lines are misguided is to compare one of the lines to a strong line from
another, similar element that has a much lower cosmic abundance.  A good
comparison element is sulfur, since it has a similar first ionization
potential to that of carbon (in H~I regions, both elements are
predominately singly ionized) and, like carbon, sulfur is not heavily
depleted.  One good pair of lines to consider is made up of the C~I line
at 1192.218$\,$\AA\ and the S~I line at 1295.653$\,$\AA.  As close as we
can determine, both features have nearly identical strengths (and
shapes) in the spectra of all stars, except for HD's 3827 and 120086
which do not have enough foreground material to make the lines visible. 
Since C~I and S~I are expected to have very similar velocity structures,
lines of equal strength should have identical systematic offsets, if
they exist, produced by the effects of saturated, unresolved components. 
As a result, we have confidence that the ratio of column densities
should be reflected by the inverse ratio of the lines' values of
$f\lambda$.

In the interstellar medium, the abundance of carbon relative to hydrogen
is about $1.4\times 10^{-4}$ over a broad range of average densities 
(Hobbs, York, \& Oegerle 1982; Cardelli et al. 1991, 1993, 1996; Sofia
et al. 1997; Sofia, Fitzpatrick, \& Meyer 1998).  While one could be
critical of these results because they rely on the calculated $f$-value
for one semi-forbidden line at 2325.403$\,$\AA, it is reassuring that
Sofia et al.  (1994) found similar relative abundances of carbon toward
$\xi$~Per and $\zeta$~Oph by measuring the damping wings on the allowed
transition of C~II at 1334.532$\,$\AA.  The amount of sulfur in the
interstellar medium is generally very close to its solar abundance ratio
relative to hydrogen, ${\rm S/H} = 1.9\times 10^{-5}$  (Federman et al.
1993; Spitzer \& Fitzpatrick 1993; Fitzpatrick \& Spitzer 1994, 1997;
Howk, Savage, \& Fabian 1999).

From the above, we conclude that free carbon atoms and ions in the
interstellar medium are about 7.5 times as abundant as the sulfur ones. 
If we multiply this ratio by a correction for the effects of
steady-state photoionization and recombination,
\begin{equation}\label{ioniz_corr}
{n({\rm C~I})n({\rm S~II})\over n({\rm C~II})n({\rm S~I})}={\alpha ({\rm
C})\Gamma ({\rm S})\over \Gamma ({\rm C})\alpha ({\rm S})}
\end{equation}
we find the expected $n({\rm C~I})/n({\rm S~I})$ should range between 16
and 24 for $A_V=0$ and 1.0, respectively (the variation is caused by
alterations in the spectral distribution of the ionizing radiation for
different depths inside a cloud).  In arriving at this conclusion, we
used the photoionization rates calculated by van Dishoeck  (1988) and
the fitting equations of Shull \& van Steenberg  (1982) for the
recombination rates.  We expect that the outcome for
Eq.~\ref{ioniz_corr} will not change much if we allow for additional
recombinations that could arise from collisions with grains or large
molecules, since the ratios of the reaction rate coefficients of C and S
differ little from $\alpha({\rm C})/\alpha({\rm S})$  (Weingartner \&
Draine 2001).  Dissociative recombination of CH$^+$ is another channel
for producing C~I; it is unclear whether or not this might have a
stronger influence than its counterpart involving SH$^+$ or if, for our
stars, these channels are at all important relative to the simple
recombinations of C and S with free electrons.  For the ratio of our
value for $f\lambda$ for the C~I transition at 1192.218$\,$\AA\ to that
of the S~I line at 1295.653$\,$\AA\ determined by Federman \& Cardelli 
(1995) [$\log (f\lambda)=2.052$] we obtain a value of 22, which is
approximately midway between the two extremes for the expected ratio of
C~I to S~I densities.  Had we instead used the values listed by either
Morton  (1991), Wiese et al.  (1996), or Zsarg\'o et al.  (1997), we
would have obtained 36, 114 and 108, respectively.

Our fifth and final test makes use of a large survey of K~I absorption.
Welty \& Hobbs  (2001) observed features toward HD~206267A and HD~210839
($\lambda$~Cep) at a resolution of $0.56\,{\rm km~s}^{-1}$, i.e., a
value about three times better than our STIS observations.  Both stars
are in our target list and their C~I lines resulted in $f$-values
similar to the general determinations listed in
Table~\ref{fval_results}.  Of the two, HD~210839 seems to have the
narrowest, well-separated components and is thus most likely to manifest
any bias that might be caused by our not resolving saturated features.
We believe that this star serves as a good test case to examine the
behavior of line profiles when they are not fully resolved. 

Welty \& Hobbs used a profile-fitting analysis, one that takes into
account the small amount of their instrumental smearing, to derive
component column densities, central velocities, and velocity
dispersions. On theoretical grounds, we expect the profiles of K~I and
C~I to be similar.  This expectation is validated by the good tracking
of $N$(K~I) to $N$(C~I) over many lines of sight studied by Welty \&
Hobbs.  With these considerations, we felt that it would be instructive
to perform a mock evaluation of $f$-values that duplicated our method
discussed in \S\ref{comparisons}, but using the K~I components acting as
surrogates for C~I.  The aim of our mock analysis was to find out what
would happen if the previously determined $f$-values, such as those
listed by WFD96, were indeed correct.  Would these $f$-values arise out
of the analysis intact, or would the saturation effects mislead us into
deriving progressively larger corrections in an upward direction as the
lines became weaker, as depicted in Fig.~\ref{fvalcomp}?

To answer this question, we undertook the following steps.  First, we
defined our synthetic C~I profiles on a fine velocity grid ($0.02\,{\rm
km~s}^{-1}$) such that they adequately sampled the K~I components, with
their very small $b$-values (generally ranging from 0.5 to $1.0\,{\rm
km~s}^{-1}$), as listed by Welty \& Hobbs.  For each component, we
estimated the relative proportions of atomic carbon in the three
different levels of fine-structure excitation, so that the behavior of
each $N_a$ approximately duplicated what we found for HD~210839.  While
these operations specified the detailed {\it shapes\/} of the C~I
features,\footnote{To obtain a good match with the strongest C~I lines
observed with STIS, we had to add some low-level, broad components that
added some very shallow wings to the main profiles.  These wings are
below the detection threshold in the K~I observations, but they show up
well in the much stronger Na~I D1 features recorded by Hobbs  (1976).}
they did not tell us {\it how much\/} C~I is really present.  For the
latter, we took the conservative assumption that the {\it weakest\/}
lines that showed measurable absorption were the only ones that could be
trusted (and their real $f$-values are lower than our determinations),
and this forced us to raise our logarithms of the overall outcomes for
$N$(C~I), $N$(C~I*) and $N$(C~I**) by the amounts 0.80, 0.65 and
0.40~dex, respectively.  These enhancements corresponded to the
disparities in $\log f\lambda$ between our determinations and those
listed by WFD96.  With this assumption, we effectively synthesized a
case where the weak lines duplicated what we saw with STIS, but strong
lines were driven much harder into saturation than what we surmised to
be true (and we are testing the proposition that this stronger
saturation might not be apparent at our resolution).

Following the definition of the C~I, C~I* and C~I** profiles, we
evaluated their detailed optical depths in each multiplet and then
smoothed the resulting intensities with the STIS instrumental profile
function (a Gaussian function with a width of $1.5\,{\rm km~s}^{-1}$
FWHM was adopted -- the smoothed synthetic profiles for weak lines gave
an excellent match with the STIS observations).  After adding random
noise of approximately the same magnitude as in the observations, we
analyzed collectively the synthesized profiles from all of the
multiplets using the same methods as with the real data.  In the end, we
obtained values of $\log f\lambda$ that were very close to those that we
specified initially (from WFD96), and the results did not show any
tendency to drift upward for the weakest lines.  The root mean squared
deviations from real to calculated values of $\log f\lambda$ equaled
0.06.

To summarize, a broad selection of different tests appears to support
our premise that the apparent optical depth comparisons are valid for
establishing our upward revisions in weak $f$-values.  This diminishes
our concern that we are being systematically misled by unresolved,
saturated velocity components.

In their introductory comments that recognize the complexity of the
interacting electronic states of C~I, Wiese et al.  (1996) acknowledge
that `` \dots for some multiplets of neutral carbon and nitrogen,
sizeable differences remain even between these multiconfiguration
results and accurate experimental data.''  While this may be true, it is
still surprising that for the weakest multiplets there are differences
between our findings and the published results that approach nearly one
order of magnitude.  However for such multiplets there are few
opportunities to compare theoretical calculations with experimental
determinations; their strengths are usually reported on the basis of
theory only.

\subsection{Reconstructions of the Profiles}\label{reconstructions}

\begin{figure}
\plotfiddle{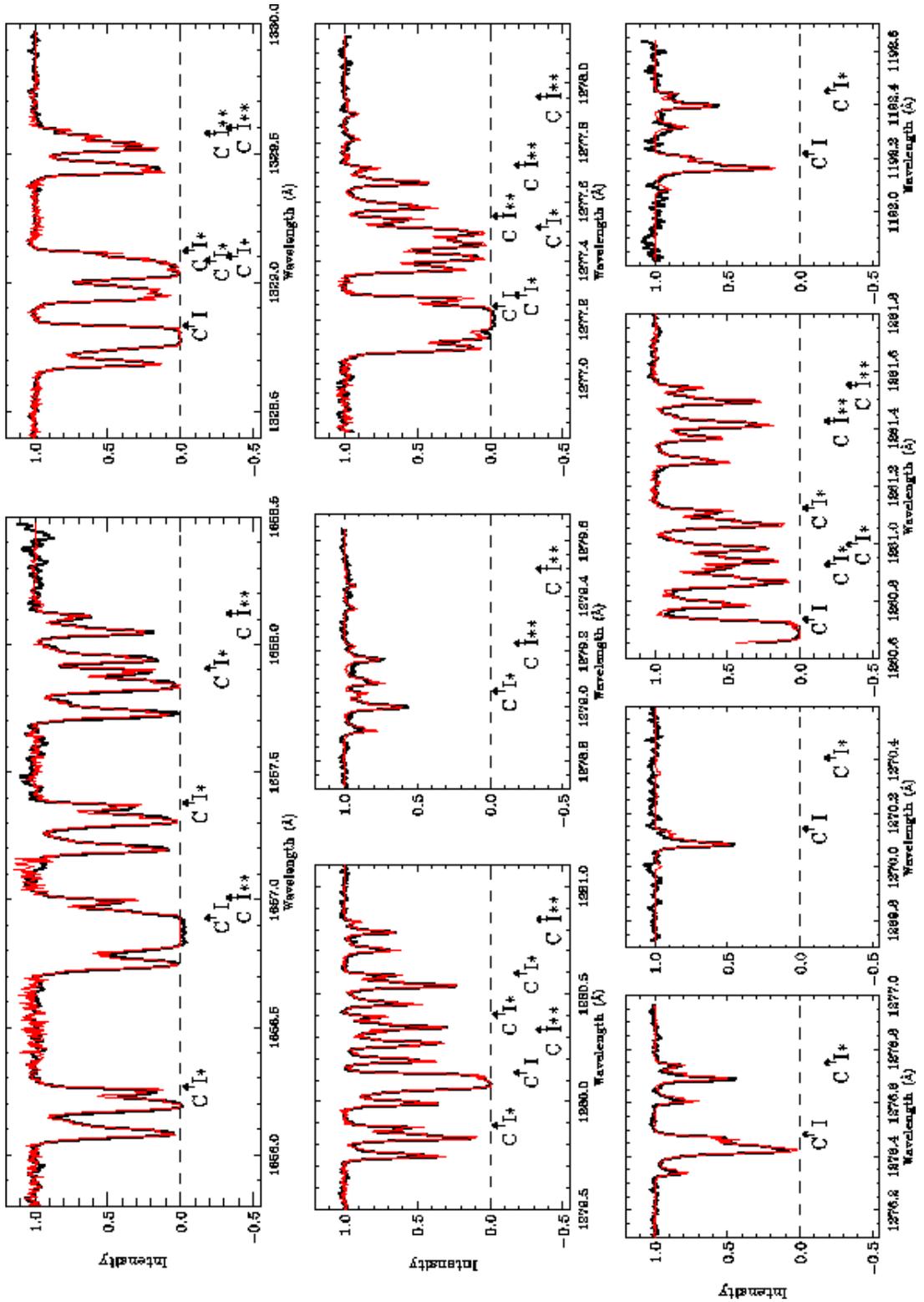}{18cm}{90}{70}{70}{300}{70}
\caption{A comparison of the original data normalized to a continuum
(black traces) for all multiplets in the spectrum of $\lambda$~Cep
(HD~210839) (except for the one at 1189$\,$\AA) against reconstructions
(overlaid red traces) generated from the derived $X_{k,\epsilon}$ being
substituted into Eq.~\protect\ref{tau_sum} to obtain $\tau(i)$, which in
turn gives the intensity through the relation
$I(v)=\exp[-\tau(i)]$.\label{210839_resid}}
\end{figure}

After deriving the three forms of $N_a(v)$ for a given star, we checked
that reconstructions of the intensity profiles $\exp[-\tau(i)]$ agreed
with the original data $\exp[-\tau_a(i)]$ recorded for each multiplet
(see Eqs.~\ref{tau_a}, \ref{tau_sum} and \ref{k}). 
Figure~\ref{210839_resid} shows such a series of comparisons for
$\lambda$~Cep (HD~210839).  This star has enough C~I to show absorptions
in all of the multiplets covered in the program.  One can see that the
reconstructions (red traces) agree well with the original data (black
traces).

\placefigure{210839_resid}

All stars except HD~108 showed excellent correspondences between the
original data and the reconstructions.  For HD~108, the tracking between
the the two was inferior to those of the other stars; some limited
portions of a few multiplets showed observed absorptions that were
stronger than those predicted by the analysis.
\clearpage
\subsection{Errors}\label{errors}
\subsubsection{Random Errors}\label{ran_errors}

The complexities of the calculations and weight factors discussed in
\S\ref{unravel} make it difficult to estimate the expected errors for a
given set of input spectra, which themselves have varying signal
qualities and degrees of influence.  Nevertheless, we can sense
empirically the random errors in the end results by studying the sizes
of the fluctuations in $N_a(v)$ over velocity ranges where we think
there is no C~I (if C~I is present at a level that is below what might
be obvious, our error estimate will be too large). 
Table~\ref{random_dev} shows the outcomes for the {\it rms\/} deviations
away from the zero level for blocks that are $5\,{\rm km~s}^{-1}$ wide. 
Since random errors may be correlated over small velocity differences,
it is probably not safe to conclude that the noise scales in proportion
to $\Delta v^{0.5}$ for smaller scales.  For scales greater than
$5\,{\rm km~s}^{-1}$, the {\it random\/} errors due to noise probably
scale in proportion to that expected for independent events, but one
must also consider the effects of possible systematic errors discussed
below.

Judging from the appearance of the residuals discussed in
\S\ref{reconstructions}, we estimate that in the vicinity of strongest
C~I features, the estimates for the errors shown in
Table~\ref{random_dev} should be doubled.

\placetable{random_dev}

\begin{deluxetable}{
r     
c     
c     
c     
c     
}
\tablecolumns{5}
\tablewidth{0pt}
\tablecaption{Observed Random Deviations\tablenotemark{a}~~in Column
Density\label{random_dev}}
\tablehead{
\colhead{HD~~} & \colhead{$\sigma [N({\rm C~I})]$} & \colhead{$\sigma
[N({\rm C~I}^*)]$} & \colhead{$\sigma [N({\rm C~I}^{**})]$} &
\colhead{Number}\\
\colhead{} & \colhead{($10^{12}{\rm cm}^{-2}$)} & \colhead{($10^{12}{\rm
cm}^{-2}$)} & \colhead{($10^{12}{\rm cm}^{-2}$)} & \colhead{of Samples}
}
\startdata
   108\phantom{A}&0.64&1.31&2.17&12\nl
  3827\phantom{A}&0.36&0.53&0.46&15\nl
 15137\phantom{A}&0.68&0.80&0.72&17\nl
 69106\phantom{A}&0.63&0.58&0.38&15\nl
 88115\phantom{A}&0.43&0.93&0.68&14\nl
\nl
 93843\phantom{A}&0.50&0.82&1.25&13\nl
 94493\phantom{A}&0.49&0.93&1.03&16\nl
 99857A&0.43&0.84&0.91&13\nl
103779\phantom{A}&0.49&0.70&0.70&14\nl
106343\phantom{A}&0.45&0.60&0.69&16\nl
\nl
109399\phantom{A}&0.50&1.34&0.49&13\nl
116781A&0.61&0.74&0.93&15\nl
120086\phantom{A}&0.32&0.39&0.38&22\nl
122879\phantom{A}&0.22&0.65&0.86&15\nl
124314A&0.42&0.85&0.65&16\nl
\nl
203374A&0.43&0.64&0.45&17\nl
206267A&0.58&0.68&0.52&17\nl
208947\phantom{A}&0.30&0.48&0.46&18\nl
209339A&0.39&0.69&0.81&18\nl
210839\phantom{A}&0.55&0.91&0.62&16\nl
\nl
224151\phantom{A}&0.27&0.70&0.45&11\nl
\enddata
\tablenotetext{a}{Evaluations of $\left\{\sum_n \left[ \int
N_a(v)dv\right]^2/n\right\}^{0.5}$, where the integration intervals are
over contiguous series of $5\,{\rm km~s}^{-1}$ wide blocks outside the
velocity ranges shown in Figs.~\protect\ref{210839_f1f2} and
7, i.e., where we expect there to be no C~I. 
Note that these are not {\it rms\/} deviations about a mean value of the
samples with a correction of $(n/n-1)^{0.5}$ to obtain a population
dispersion; they are simply the {\it rms\/} deviations about a zero
value.  The number $n$ of blocks is shown in the last column of the
table.}
\end{deluxetable}

\subsubsection{Systematic Errors}\label{sys_errors}

We recognize three principal sources of error that might lead to our
results being systematically offset from the true values of $N_a(v)$. 
First, there may be instrumental artifacts that are not accounted for in
the analysis, or perhaps done so incorrectly.  While errors in the
adopted background levels can lead to error, we found that the estimates
of these levels by the CALSTIS routines seem to be reasonably accurate. 
By inspecting the strongest interstellar features that are clearly
saturated with flat bottoms, such as C~II $\lambda 1334.532$, Si~II
$\lambda 1304.370$, O~I $\lambda 1302.169$, and the triplet of N~I at
1200$\,$\AA, we found that the apparent positions of the line bottoms
never deviated from zero by more than 3\% of the local continuum flux.

One instrumental effect for which we are not making explicit corrections
in our analysis is the existence of broad halos that are a part of the
point-spread functions of the Far- and Near-UV MAMA detectors in STIS 
(Leitherer 2000, pp. 128$-$129).  The most pronounced effects of the
halos are to alter the background levels and slightly degrade the
resolution.  However, since the bottoms of saturated lines appear to be
near zero, as we stated above, it is clear that the background
corrections of CALSTIS include the effects of the halos.  As for
degradation of the resolution, it is clear from our findings reported in
Appendix~\ref{resolving_pwr} that any such degradation must be very
small.

The second error source that could be coherent over large velocity
ranges is a deviation between our adopted continuum level and the
correct one.  Our choice of stars with large projected rotational
velocities (see col. 4 in Table~\ref{tgt_stars}) generally insured that
the continuum levels do not have sharp and unpredictable curves caused
by stellar features.  For most stars, the continuum curvatures were
inconsequential, as is evident in Fig.~\ref{spec210839} for HD~210839,
which made the continuum levels very easy to define. Nevertheless, in a
few cases for stars with lower values of $v\sin i$ the levels showed
undulations.  The worst example was HD~106343 in the vicinity of
1277$\,$\AA.  Here, stellar features had a depth of about 50\% and a
width of 0.5$\,$\AA\ FWHM.  This single star is unlikely to have much
impact on the final results.

Finally, for our third type of error we must consider the possibility
that some or most of our $f$-values are incorrect, despite our best
efforts to try to uncover evidence that we were being misled, as
discussed in \S\ref{substructures}.  While we had chosen the
1657$\,$\AA\ multiplet as our standard because we believed it to have
the most reliable $f$-value determinations, this multiplet is one of our
worst ones for giving clear, unambiguous signals -- see the relative
$S/N$ values in Table~\ref{fval_results} (number in parentheses below
the multiplet numbers in col. 1).  Upon inspection of the points in
Fig.~\ref{fvalcomp}, one might propose that other lines with strengths
that are about the same as the ones that we adopted as standards might
have been better choices.  If that is correct, then all values of $\log
f\lambda$ other than the ones belonging to the 1657$\,$\AA\ multiplet
should be reduced by about 0.1~dex.  The effect of this would be to
revise all of the values of $N_a(v)$ (for all three of the
fine-structure levels) upward by 0.1~dex, except for the weakest wings
of the lines that are primarily defined by the strongest multiplet.

\section{Results}\label{results}
\subsection{General Remarks}\label{genl_remarks}

The ultimate product from the analysis of all of the multiplets
described in \S\ref{specific_appl} is a synthesis of the functional
forms $N_a(v)$ for each of the three fine-structure levels of C~I.  The
upper two levels are primarily populated by collisions with other
particles, a process that competes with radiative de-excitations.  Our
main objective is to compare the populations in each level, so that we
can learn about the local temperature and density of the C~I-bearing
material.

Early studies of the C~I excitation from observations with the {\it
Copernicus\/} satellite indicated population ratios that were sometimes
inconsistent with the expected conditions for any given local density
and temperature  (Snow 1976; Drake \& Pottasch 1977; Frisch 1980).  It
was evident that the results had to be interpreted in terms of a
superposition of contributions from regions with different physical
conditions.  Much of this overlap will be overcome by the ability of the
STIS spectrum to show small differences in Doppler shifts. 
Nevertheless, we must still be prepared to handle cases where there are
superpositions of absorptions arising from regions with different
excitation conditions, but with nearly the same radial velocity.

\begin{figure}
\plotone{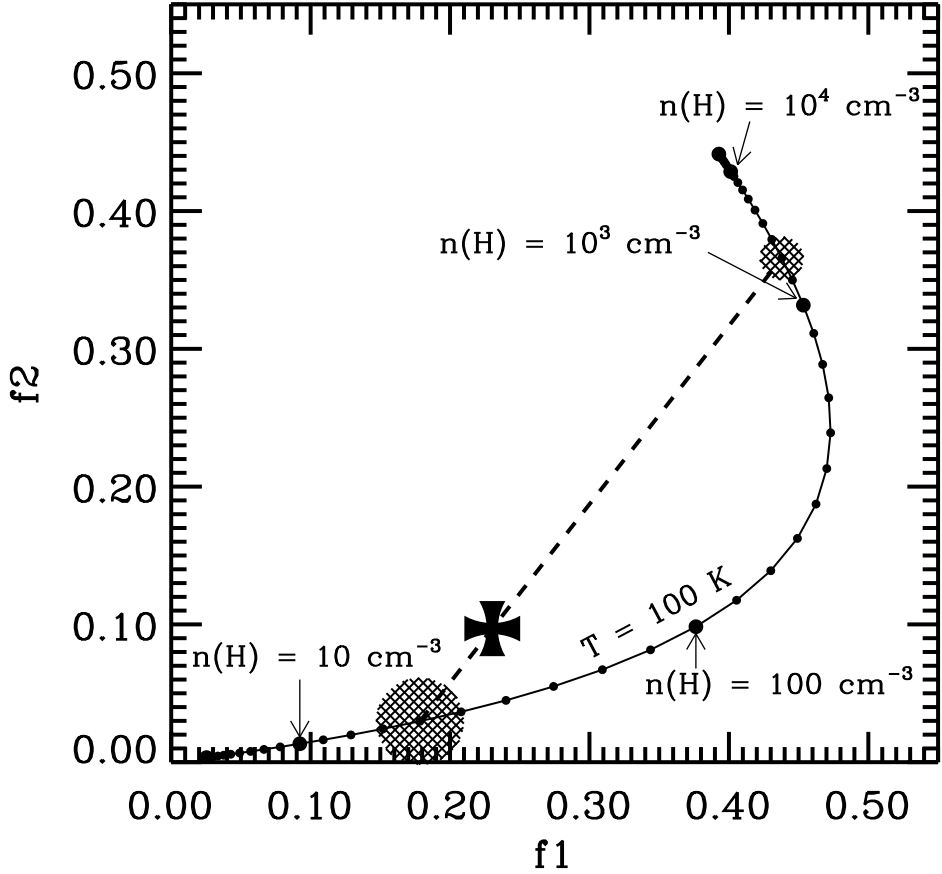}
\caption{Expected outcomes for $f1\equiv N({\rm C~I^*})/N({\rm C~I}_{\rm
total})$ and $f2\equiv N({\rm C~I^{**}})/N({\rm C~I}_{\rm total})$ for
absorption by C~I-bearing gases at various densities $n({\rm H})$ and a
single temperature $T=100\,$K.  An observed result at $f1=0.23$ and
$f2=0.10$ (at the Maltese cross) may arise from the superposition of
contributions from a thick cloud with a low density $n({\rm
H})=10^{1.4}\,{\rm cm}^{-3}$ (large, circular cross-hatched region) and
a thin cloud with $n({\rm H})=10^{3.2}\,{\rm cm}^{-3}$ (smaller
cross-hatched region).  This position in the diagram corresponds to the
``center of mass'' of the C~I-weighted combinations that lie on the
track representing homogeneous regions.\label{cartoon}}
\end{figure}

To understand what combinations of conditions can create these composite
level populations, it is useful to express the measurements at a given
velocity in the form of two quantities, $f1\equiv N({\rm C~I^*})/N({\rm
C~I}_{\rm total})$ and $f2\equiv N({\rm C~I^{**}})/N({\rm C~I}_{\rm
total})$.\footnote{In all quantitative statements and formulae in this
article, C~I always denotes atomic carbon in the lowest fine-structure
level {\it only}, while C~I$_{\rm total}$ refers to all of the atomic
carbon, irrespective of its fine-structure excitation.}  For a simple
introduction on how these quantities behave under different conditions,
we show in Figure~\ref{cartoon} the expected outcomes for $f1$ and $f2$
for C~I embedded within neutral hydrogen clouds having a single value
for the temperature, but with various internal densities $n({\rm H})$. 
As the density increases from a low value where collisional excitations
are infrequent to much higher densities where the level populations can
become appreciable, the ($f1$,$f2$) points advance along the curve from
the lower left to the upper right in the diagram.

\placefigure{cartoon}

With regard to superpositions, the $f1-f2$ representation has a useful
geometrical property that was exploited in an early study of C~I
excitation by Jenkins \& Shaya  (1979).  Specifically, they pointed out
that the outcome for a composite measurement on a diagram of $f2$ versus
$f1$ equals the ``center of mass'' of points that would be plotted for
the individual contributors, with ``weights'' assigned in proportion to
the respective local abundances of C~I$_{\rm total}$.  This concept is
illustrated in Fig.~\ref{cartoon}: an observed outcome at (0.23, 0.10)
may arise from C~I within a large amount of material having $n({\rm
H})=10^{1.4}\,{\rm cm}^{-3}$ which is then superposed on a contribution
from a gas that has a smaller column density of C~I but is considerably
more dense, i.e., $n({\rm H})=10^{3.2}\,{\rm cm}^{-3}$.  Of course, this
interpretation is not unique -- many other possible combinations of
contributions along the curve can produce the same result.  The
usefulness of this important theorem on composite measurements has been
reflected by the adoption of the representation of $f2$ vs. $f1$ in many
subsequent studies of C~I  (Jenkins et al. 1981, 1998; Jenkins, Jura, \&
Loewenstein 1983; Smith, et al. 1991; Jenkins \& Wallerstein 1995).

\subsection{Theoretical Expectations for $f1$ and
$f2$}\label{theoretical}
\subsubsection{Collisional Excitation}\label{collisional}

To interpret our measurements of $f1$ and $f2$ in terms of useful
properties of the gas which produces the C~I absorptions, we must
compute the loci of points on the diagram for a variety of physical and
chemical states.  In presentations that compare the expected outcomes
with the observations, to be discussed in \S\ref{presentations}, we
chose to exemplify three fundamental regimes: Case~1 -- neutral hydrogen
at various temperatures ($20 < T < 160\,$K) and densities, but with
negligible fractional concentrations of molecules, electrons and
protons, Case~2 -- gas with all of the hydrogen in molecular form at a
typical kinetic temperature $T=78\,$K, one that results in a rotational
excitation such that $N_{J=0}=N_{J=1}$ (again with negligible
ionization), and Case~3 -- fully ionized gas with $T=7000\,$K, as might
be found in a star's H~II region.  In all cases, we assumed an abundance
ratio He/H = 0.0975  (Anders \& Grevesse 1989), and that He is neutral
in all of the C~I-bearing zone of the H~II region that applies to Case
3.  Table~\ref{rel_pp} summarizes the relative partial pressures in the
three cases.

\placetable{rel_pp} 
\begin{table}[hb]
\caption{Relative Partial Pressures\label{rel_pp}}
\begin{tabular}{cccc}
\tableline\tableline
~~~Constituent~~~ & ~~Case 1~~ & ~~Case 2~~ & ~~Case 3~~ \\ \tableline
ortho-H$_2$&0.00&0.418&0.00\\
para-H$_2$&0.00&0.418&0.00\\
H~I&0.911&0.00&0.00\\
He~I&0.0888&0.163&0.0465\\
e&0.00&0.00&0.477\\
p&0.00&0.00&0.477\\
\hline
\end{tabular}
\end{table}

One can see from the results tabulated by Keenan  (1989) that for Case 1
(neutral atomic gas) excitations caused by the low concentrations of
electrons arising from the photoionization of carbon and other species
that can be ionized in H~I regions should be unimportant.  For
situations that might call for Case 2, we expect that regions that we
are investigating should not have all of the hydrogen in molecular form. 
Nevertheless, we find that we can regard the precise value of any
intermediate molecular fraction as an irrelevant parameter because the
outcomes for Case 2 are so similar to those from Case 1, as we will
demonstrate shortly.  The relevance of Case 3 will be addressed in a
later section (\S\ref{hii_regions}).  Generally, the three cases span a
useful range of expected conditions in C~I-bearing gas clouds, and one
can imagine what the consequences would be for intermediate states.

For the excitation and de-excitation of the C~I fine-structure levels,
we adopted rate coefficients from the following sources: Launay \&
Roueff  (1977) for atomic hydrogen, Schr\"oder et al  (1991)  for
molecular hydrogen in the ortho and para states, Staemmler \& Flower 
(1991) for neutral helium (with an extrapolation for $T > 150\,$K),
Johnson et al.  (1987) for electrons, and Roueff \& Le Bourlot  (1990)
for protons.  For the radiative decay rates of the upper levels, we used
the values given by Froese Fischer \& Saha  (1985).  The equilibrium
concentrations of the excited levels relative to the ground state are
given by
\begin{equation}\label{r10}
{n({\rm C~I^*})\over n({\rm
C~I})}={k_{0,1}(k_{2,1}+k_{2,0}+A_{2,1}+A_{2,0}) +
k_{0,2}(k_{2,1}+A_{2,1})\over k_{1,2}(k_{2,0}+A_{2,0})+(k_{1,0}+A_{1,0})
(k_{2,0}+k_{2,1}+A_{2,1}+A_{2,0})}
\end{equation}
and
\begin{equation}\label{r20}
{n({\rm C~I^{**}})\over n({\rm C~I})}={k_{0,2}(k_{1,0}+A_{1,0}+k_{1,2})
+ k_{1,2}k_{0,1}\over (k_{1,0}+A_{1,0})(k_{2,0}+k_{2,1}+A_{2,1}+A_{2,0})
+ k_{1,2}(k_{2,0}+A_{2,0})}
\end{equation}
where the indices 0, 1 and 2 refer to C~I, C~I* and C~I**, respectively,
and $k_{m,n}$ are the collision rates from state $m$ to state $n$ summed
in proportion to the concentrations of the projectile species that are
present.  The reverse reaction rates $k_{n,m}$ are related to the
forward ones $k_{m,n}$ through the principle of detailed balancing.  The
radiative decay rates $A_{2,1}$ and $A_{1,0}$ are the two quantities in
the equations that are independent of density and temperature, thereby
giving us the leverage to determine the physical conditions for the
C~I-bearing gas ($A_{2,0}$ is so small compared to the other two decay
rates that it can be neglected).

\subsubsection{Optical Pumping}\label{optical_pump}

In addition to the effects from collisions, C~I can be excited through
optical pumping, principally by ultraviolet photons from stars.  We have
adopted the rates computed by Jenkins \& Shaya  (1979) for the general
ISM, but used 10 times these rates for the H~II region, i.e., Case (3),
on the grounds that the stellar radiation field would be substantially
elevated.  For all cases, the pumping rates are simply added to the
appropriate collision rates.

Figure 6 of Jenkins \& Shaya  (1979) illustrates how the tracks in the
$f1-f2$ diagram are changed when the pumping rates are raised from one
to ten times the average rate.  For thermal pressures substantially less
than about $p/k=10^4\,{\rm cm}^{-3}$K, enhancements in the optical
pumping rate may create considerable elevations in $f1$, accompanied by
smaller changes in $f2$.  For example, when the pumping is at a level
equal to ten times the average rate and there are no collisional
excitations, $f1=0.13$ and $f2=0.02$.  The tracks for various
collisional excitation conditions all emanate from this displaced
origin, but the upward shifts in $f1$ and $f2$ become smaller as the
pressures increase.  When $p/k\gtrsim 10^4\,{\rm cm}^{-3}$K, the values
of $f1$ and $f2$ are virtually unchanged.

Pumping by the general background of line radiation emitted by other
excited C~I atoms in the Galaxy (see \S\ref{pts}) should be negligible. 
If the average emission intensity of $6.4\times 10^{-8}{\rm
erg~cm}^{-2}{\rm s}^{-1}{\rm sr}^{-1}$ at $609\,\mu$m recorded by Wright
et al.  (1991) for the Galactic plane is assumed to fill a solid angle
in the sky of about $2\pi\,{\rm sr}$ and is spread over a range of
Doppler velocities equal to $100\,{\rm km~s}^{-1}$ (making
$\Delta\nu=1.64\times 10^8{\rm Hz}$), we conclude that $U_\nu=8.2\times
10^{-26}{\rm erg~cm}^{-3}{\rm Hz}^{-1}$  and calculate an excitation
rate
\begin{equation}\label{ex_rate}
U_\nu B_{0,1}={g_1 U_\nu c^3 A_{1,0}\over 8\pi g_0 h \nu^3}=2.4\times
10^{-11}{\rm s}^{-1}
\end{equation}
which is smaller than $A_{1,0}$ by a factor of 3000.  A similar
conclusion holds for the $J=2\rightarrow 1$ radiation at $370\,\mu$m. 
The measured average radiation field at $370\,\mu$m is $1.16\times
10^{-7}{\rm erg~cm}^{-2}{\rm s}^{-1}{\rm sr}^{-1}$.\footnote{This value
is likely to be an overstatement of the true flux, since the COBE FIRAS
spectrometer is unable to resolve the $370\,\mu$m radiation from CO
7$-$6 emission at $369\,\mu$m  (Fixsen, Bennett, \& Mather 1999).}  With
the same assumptions as before, we find that $U_\nu B_{1,2}=1.1\times
10^{-11}{\rm s}^{-1}$ which is lower than $A_{2,1}$ by a factor of
$2.2\times 10^4$.

Next, we consider the consequences of the carbon atoms being in the
vicinity of, or in fact right next to, a bright H~II region that is
emitting the submillimeter line radiation by other carbon atoms inside
of it (or perhaps associated with its photon-dominated, dense, neutral
gas just outside).  If the brightness temperature of the $609\,\mu$m
$J=1\rightarrow 0$ line radiation is $1\,$K, then the corresponding
intensity $B_\nu=2kT\lambda^{-2}=7.4\times 10^{-14}{\rm erg~cm}^{-2}{\rm
s}^{-1}{\rm Hz}^{-1}{\rm sr}^{-1}$ should give an excitation rate $2\pi
B_\nu B_{0,1}/c=4.6\times 10^{-9}{\rm s}^{-1}$, i.e., about 6\% of the
value of $A_{2,1}$, if indeed the region that emits the C~I
fine-structure radiation subtends $2\pi\,{\rm sr}$ in the sky.  Peak
brightness temperatures of about $2.4\,$K (with a dispersion of
$1.8\,$K) have been recorded for bright, southern-hemisphere H~II
regions by Huang et al  (1999) using the AST/RO telescope.  Hence, the
pumping rate right next to one of these regions would be about
$1.1\times 10^{-8}{\rm s}^{-1}$.  We do not know the strength of the
corresponding radiation for $J=2\rightarrow 1$, but if we assume that
the C~I** associated with the H~II region has, at worst, a population
ratio equal to the ratio of statistical weights, $g_2/g_1=5/3$, then the
$1\rightarrow 2$ pumping rate just outside the region could be as high
as $2.4\times 10^{-8}{\rm s}^{-1}$ if $T_B=2.4\,$K for the peak emission
of the $609\,\mu$m line.  The two pumping rates discussed above are
about equivalent to 15 and 33 times the average optical pumping rates
for $0\rightarrow 1$ and $1\rightarrow 2$, respectively, estimated for 
the Galactic plane (away from bright, O-B associations). They do, of
course, assume the worst possible case, i.e., that the C~I that we are
examining is right next to the region and has a radial velocity that
exactly matches that of the emission profile's peak.  It should be clear
that even with some relaxation of these optimal conditions, the pumping
rates are not entirely trivial.  Nevertheless, we point out that most
locations on any line of sight will not be affected by this problem.

\subsection{Presentations}\label{presentations}

\begin{figure}
\epsscale{.7}
\plotone{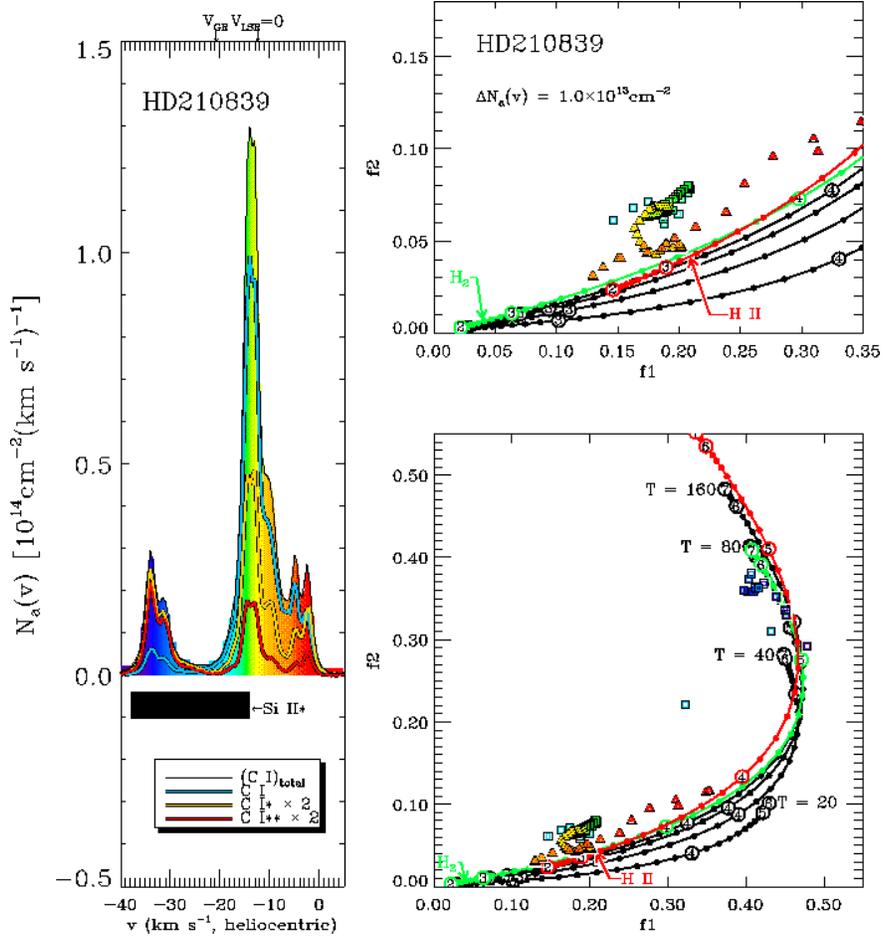}
\caption{{\it Left panel:\/}$N_a(v)$ for C~I, C~I* and C~I** (shown with
colored lines as indicated in the key) for the star HD~210839
($\lambda$~Cep), with the velocity region containing perceptible Si II*
absorption that might mark dense H~II regions (see
\S\protect\ref{hii_regions}) indicated by a solid, black, horizontal
bar.  The zero-point for the LSR velocity scale $v_{\rm LSR}=0$ and the
expected radial velocity from differential galactic rotation at the
star's position $v_{\rm GR}$ are indicated at the top.  {\it Right
panels:\/} The values of $f2$ plotted against $f1$, as defined in
\S\protect\ref{genl_remarks}, with colors that match the indicated
velocity regions in the plot of $N_a(v)$.  Triangular points come from
C~I outside the Si~II* velocity region, while box-like points come from
inside this interval.  The tracks with solid, round points on them
represent the theoretically expected loci for homogeneous regions at
various temperatures (as labeled) and thermal pressures (logarithmic
values for the units ${\rm cm}^{-3}$K as indicated by integers within
the large markers, with minor markers between them which are 0.1~dex
apart) for Cases 1 (neutral, atomic gas: black), 2 (neutral molecular
gas: green) and 3 (gas with all of the hydrogen fully ionized: red) --
see \S\protect\ref{theoretical}.\label{210839_f1f2}}
\end{figure}

The key results of our investigations are presented in
Figures~\ref{210839_f1f2} and 7.  The tall panel on the
left-hand side of Figure~\ref{210839_f1f2} depicts the outcome from our
derivations of $N_a(v)$ for C~I, C~I* and C~I** for the star HD~210839
($\lambda$~Cep).  The upper bound (thin black line) of the colored
region indicates $N_a(v)$ for C~I$_{\rm total}$, while the thick blue,
yellow and red traces apply to C~I, C~I* and C~I**, respectively.  (The
latter two quantities are magnified vertically by a factor of 2 to make
them easier to discern.)  The colors inside the profile for C~I$_{\rm
total}$ identify different zones in radial velocity which are keyed to
identical colors in the points which show $f1$ and $f2$ in the panels on
the right-hand side of the figure.  At the top of the panel showing the
profiles, we indicate the origin for the velocity scale for the Local
Standard of Rest (LSR) and the calculated heliocentric radial velocity
caused by differential galactic rotation at the position of the star
(see cols. 5 and 6 of Table~\ref{los}).  Quiescent gas along a line of
sight must be situated between these two markers.  Parts of the profiles
outside the markers must arise from gases with peculiar velocities for
their locations -- we will address this issue later
(\S\ref{kinematics}).

It is clear from our profiles for HD~210839, as well as nearly all of
the other stars in the survey, that we are able to derive $f1,f2$ pairs
for many different radial velocities.  This is best done by integrating
the relevant $N_a(v)$'s over finite velocity intervals, taking care that
the widths of these intervals are not so small that the results are
dominated by noise.  At the same time, we wish not to make the intervals
so large that we lose potentially useful distinctions of behavior at
different velocities.  As a good compromise, we chose to define the
boundaries for the velocity bins in terms of having a constant $N_a(v)$
for C~I$_{\rm total}$ inside them.  For all stars except HD's 3827,
120086 and 206267A (see labels in the $f1,f2$ panels of
Figs.~7$a$, $g$ and $i$), we chose these intervals to enclose $\Delta
N_a(v)=1.0\times 10^{13}{\rm cm}^{-2}$.  In essence, the intervals have
varying widths: they are broad when $N_a(v)$ is low, and they become
narrow within the tallest portions of the peaks in $N_a(v)$.  They are
indicated on the velocity plot as histogram-style bars for the colored
regions.  These bars are not easy to recognize for HD~210839 in
Fig.~\ref{210839_f1f2} because they are so narrow, but they are
prominent for stars with much lower column densities of C~I shown in
Figs.~7$a$, $c$ and $d$, such as HD's 3827, 88115, 93843, 94493, etc. 
The very small amount of C~I toward HD~120086 represents an extreme case
of coarse velocity intervals; we had to reduce $\Delta N_a(v)$ to
$2.5\times 10^{12}{\rm cm}^{-2}$ to give some level of differentiation
for different velocities.  As a consequence, the errors for these points
are relatively large.

The lower right panel of Fig.~\ref{210839_f1f2} shows the values of $f2$
plotted against their respective values of $f1$ with colors that match
the appropriate velocity-interval bands in the plot of $N_a(v)$. An
expanded version of this plot is shown on the upper right portion of the
figure.  We have refrained from showing error bars for the values of
$f1$ and $f2$, to avoid confusion.  One can judge the uncertainties from
random errors by examining the sizes of the fluctuations relative to the
signals in $N_a(v)$ at the appropriate matching velocities (indicated by
colors).  In the plots of $f2$ vs. $f1$, the black traces illustrate the
computed values of these variables for Case~1 (neutral atomic gas -- see
\S\ref{theoretical}) for the temperatures labeled at the end points. 
Different thermal gas pressures along these tracks are indicated by
points with a spacing $\Delta\log(p/k)=0.1$.  Numbers with circles
around them show the positions and values of points having integer
$\log(p/k)$.  The green and red traces represent Cases~2 (pure hydrogen
molecules) and 3 (fully ionized hydrogen with an enhanced pumping
field), respectively.

Plots of $N_a(v)$ and $f1$ vs. $f2$ for the remaining stars in the
survey are shown in Fig.~7.  The velocity scales and
ranges of $f1$ and $f2$ change from one star to the next, to accommodate
for the differences in the outcomes.  However to avoid confusing
distortions, the aspect ratio of the plots of $f1$ and $f2$ remain
unchanged: in all cases the range for $f2$ is half that for $f1$,
although the overall magnification may change from one plot to the next. 
Column densities integrated over the entire velocity ranges shown in the
plots are listed (in logarithmic form) in Table~\ref{col_dens}.

Some of the derived combinations of $f1$ and $f2$ yielded results that
fall outside the region in the diagram bounded by the outermost
theoretical curves on the right-hand side and a line running from the
origin to the point (0.33, 0.56) that corresponds to infinite
temperature and density (i.e., where the population ratios assume values
simply in proportion to the statistical weights of the levels).  These
combinations are clearly inadmissible within the framework of the
``center of mass'' construction for blended contributions discussed in
\S\ref{genl_remarks}.  In all instances where this happened, as shown by
the outlyers in the $f1,f2$ diagrams of Fig.~7, the
samples were drawn from velocity ranges where $N_a(v)$ was small.  Under
our scheme of always sampling a given value of $\Delta N_a(v)$, we had
to make the intervals of integration over velocity quite large, and this
led to our exposing the measurements to random and systematic errors
(\S\ref{errors}) that were significantly greater than the rest of the
samples.

\placefigure{210839_f1f2}

\clearpage

\placefigure{others_f1f2}

\begin{figure}
\epsscale{1.0}
\plotone{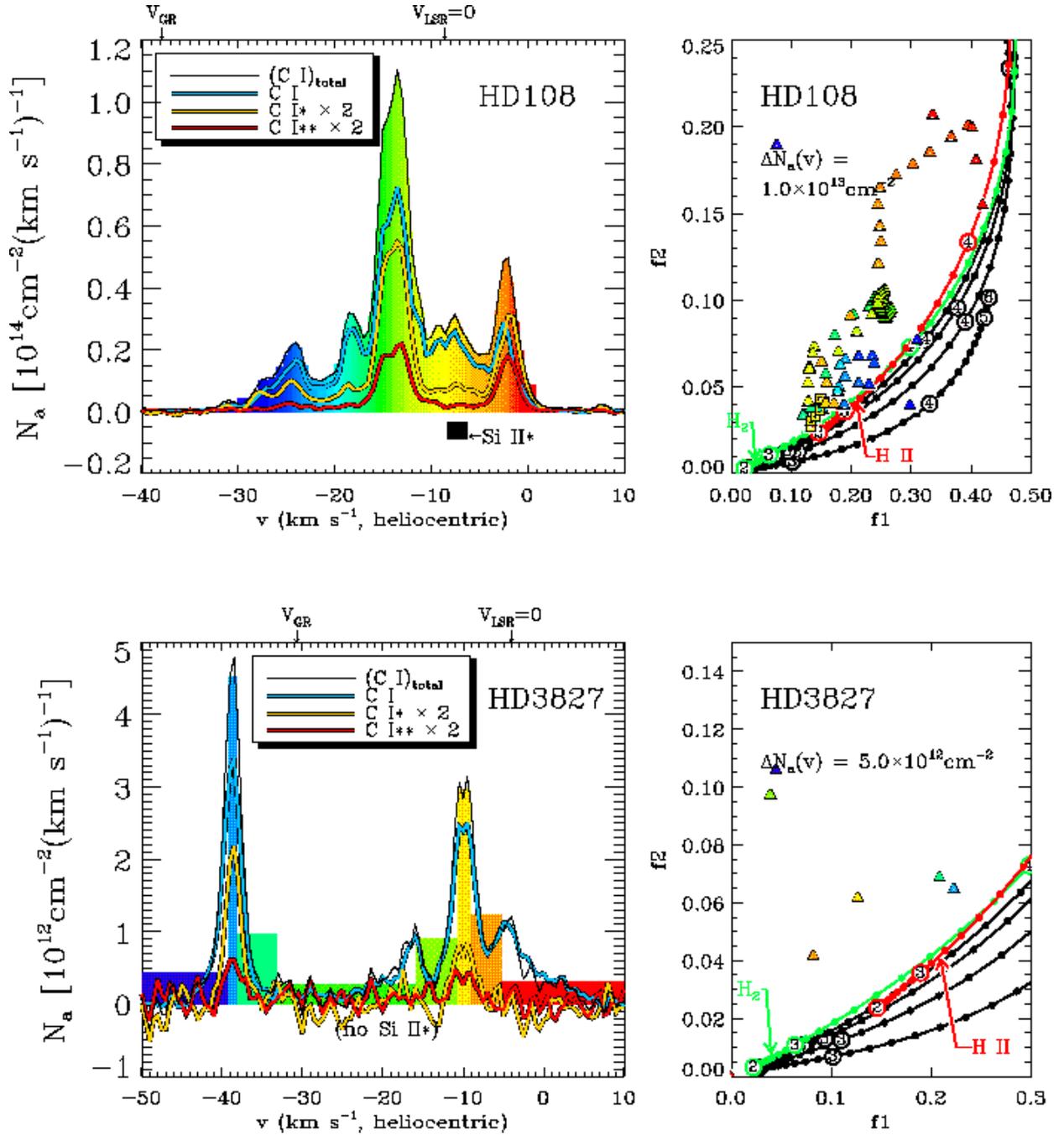}
\figurenum{7a}
\caption{Information on $N_a(v)$ (left panel) and $f2$ vs. $f1$ (right
panel) for stars in the survey, presented in the same style as for
HD~210839 in Fig.~\protect\ref{210839_f1f2}. See the caption for
Fig.~\protect\ref{210839_f1f2} and the text
(\S\protect\ref{presentations}) for details.\label{others_f1f2}}
\end{figure}
\clearpage
\begin{figure}
\plotone{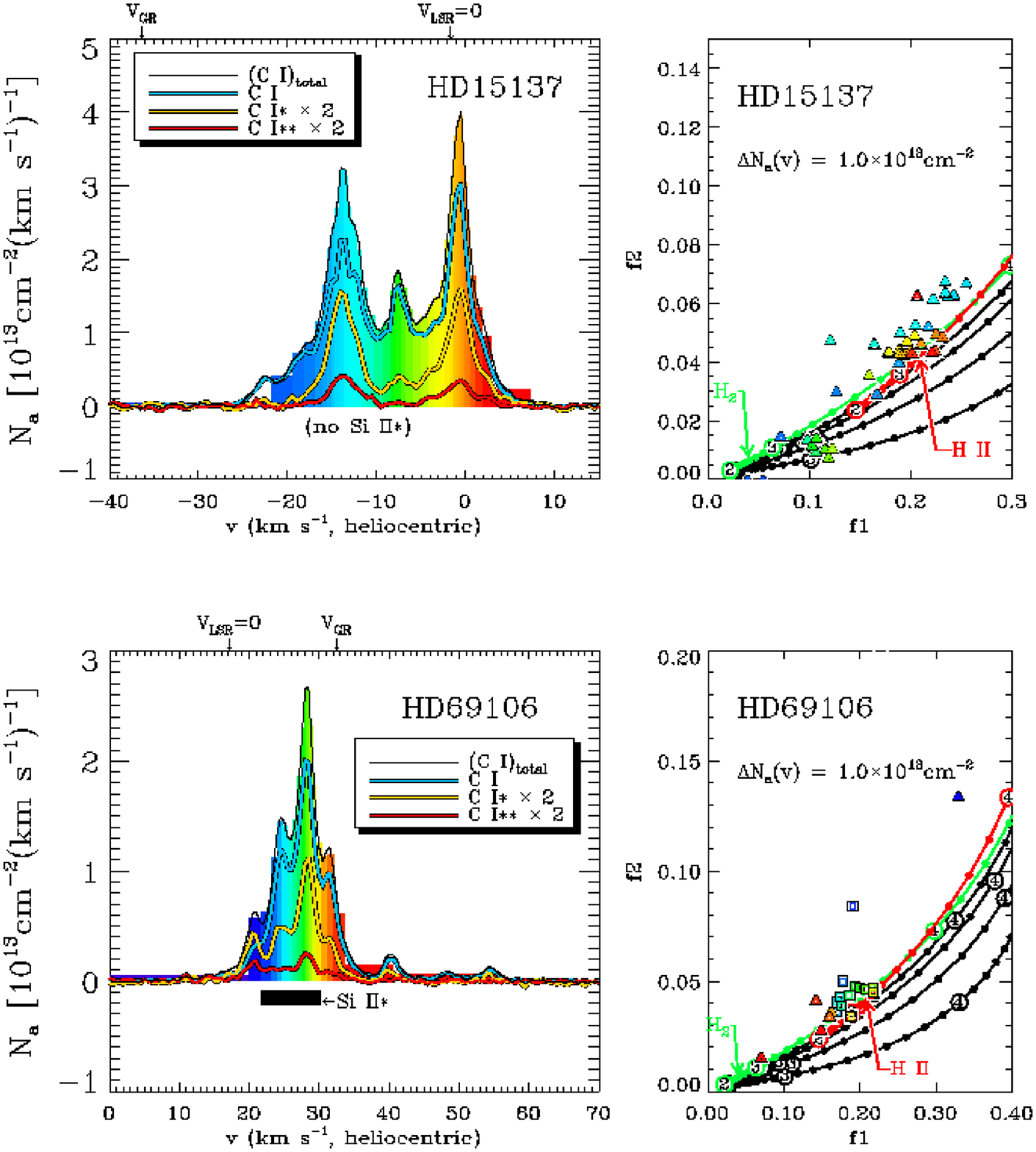}
\figurenum{7b}
\caption{continued}
\end{figure}
\clearpage
\begin{figure}
\plotone{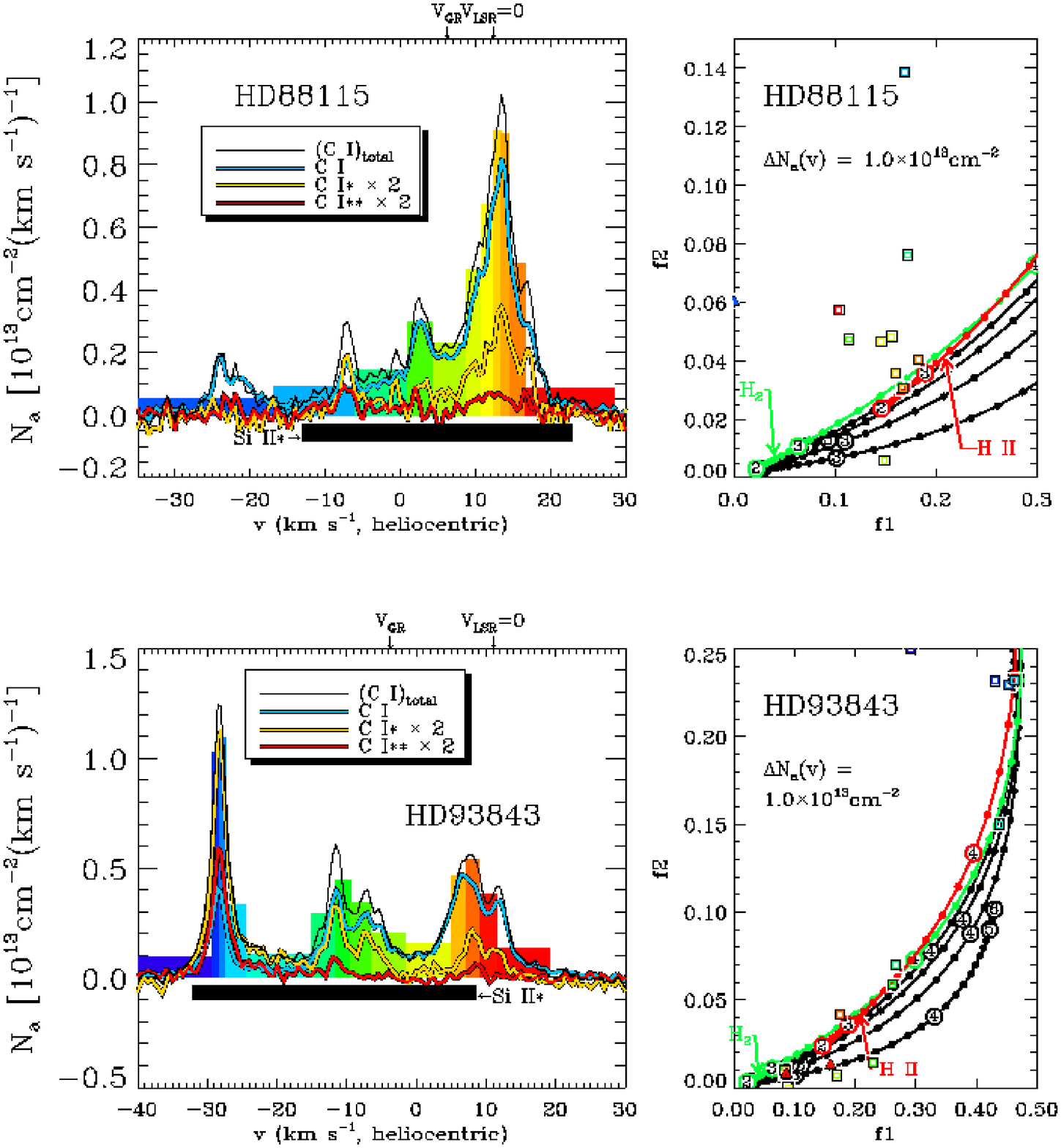}
\figurenum{7c}
\caption{continued}
\end{figure}
\clearpage
\begin{figure}
\plotone{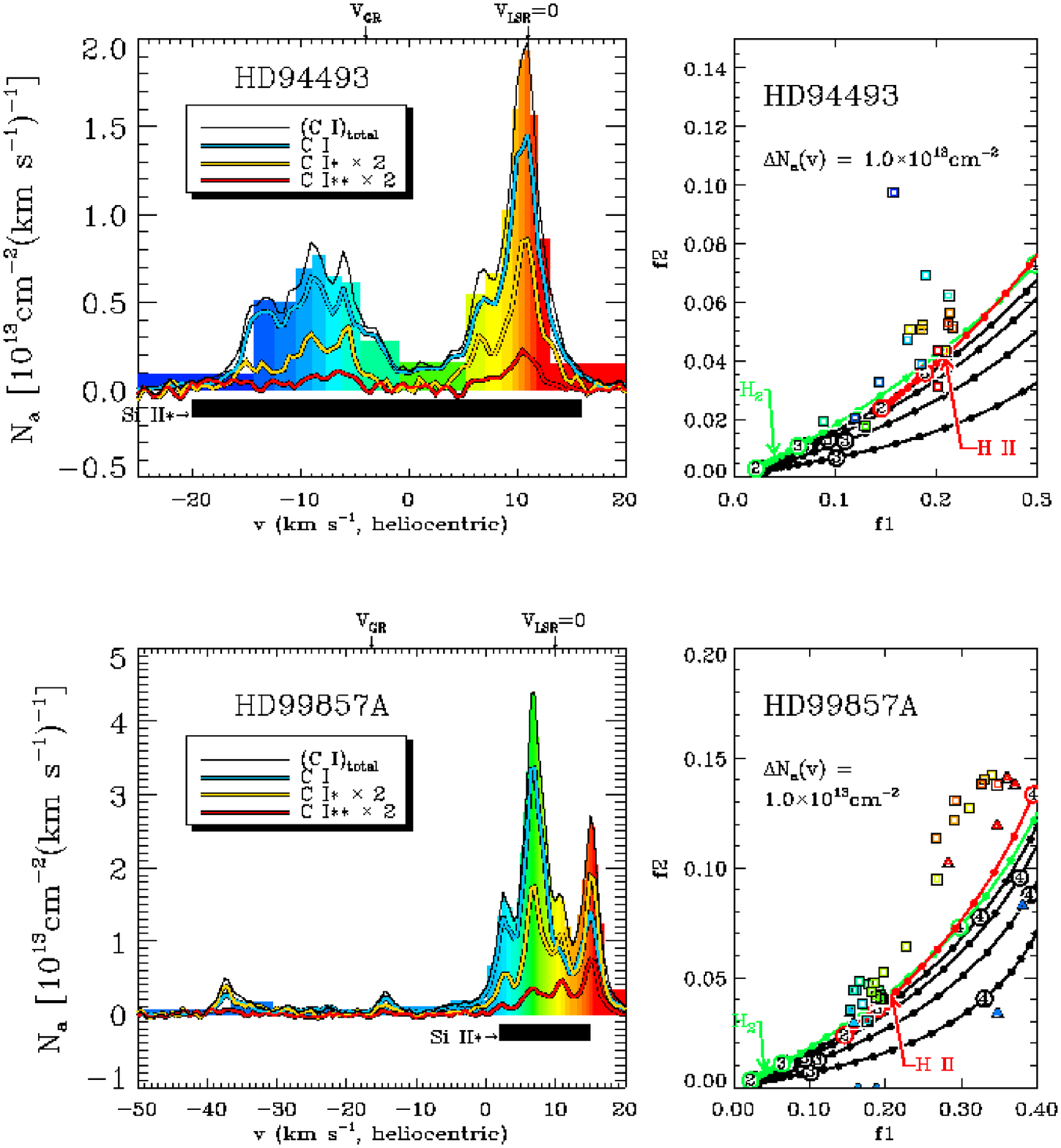}
\figurenum{7d}
\caption{continued}
\end{figure}
\clearpage
\begin{figure}
\plotone{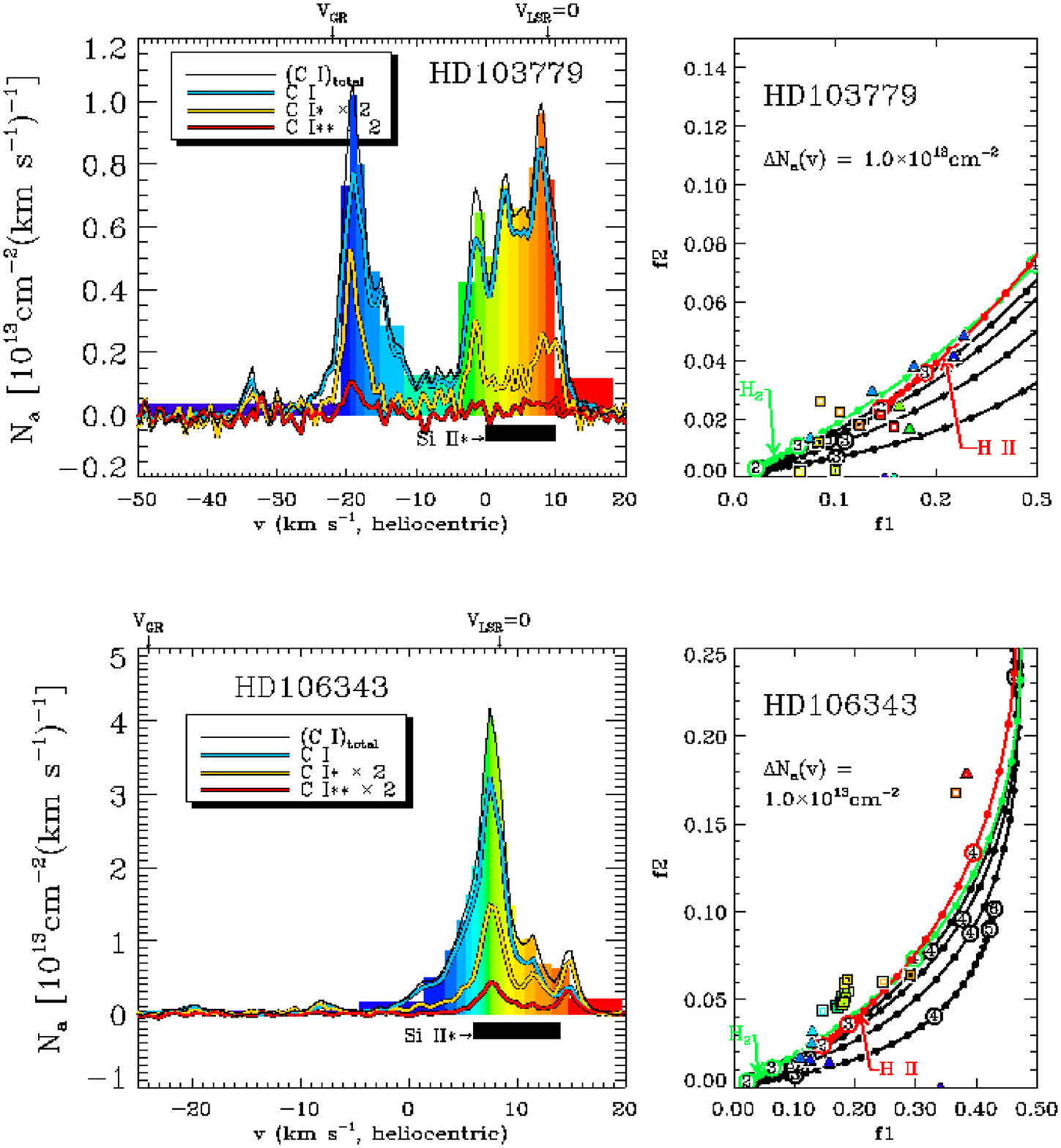}
\figurenum{7e}
\caption{continued}
\end{figure}
\clearpage
\begin{figure}
\plotone{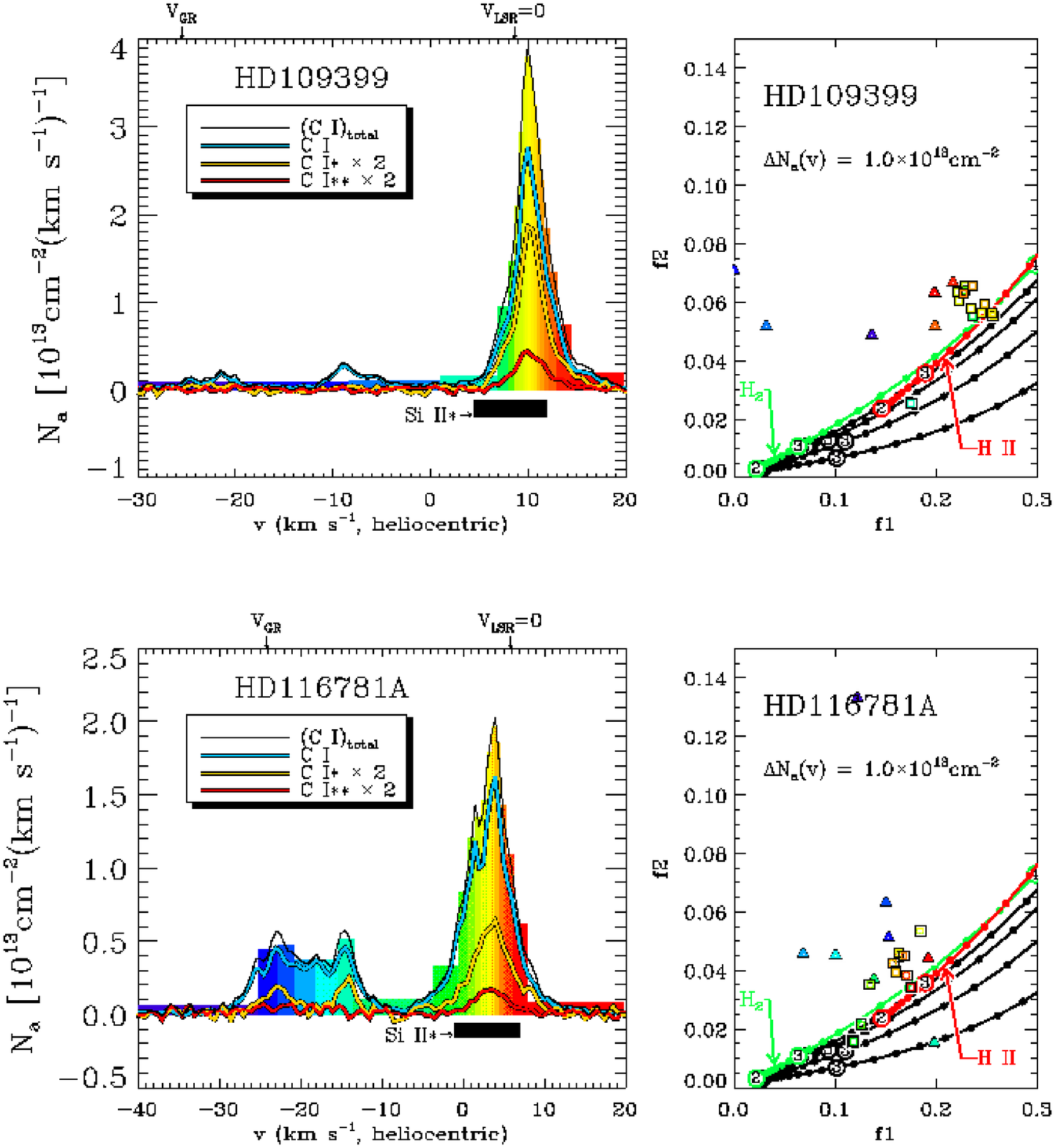}
\figurenum{7f}
\caption{continued}
\end{figure}
\clearpage
\begin{figure}
\plotone{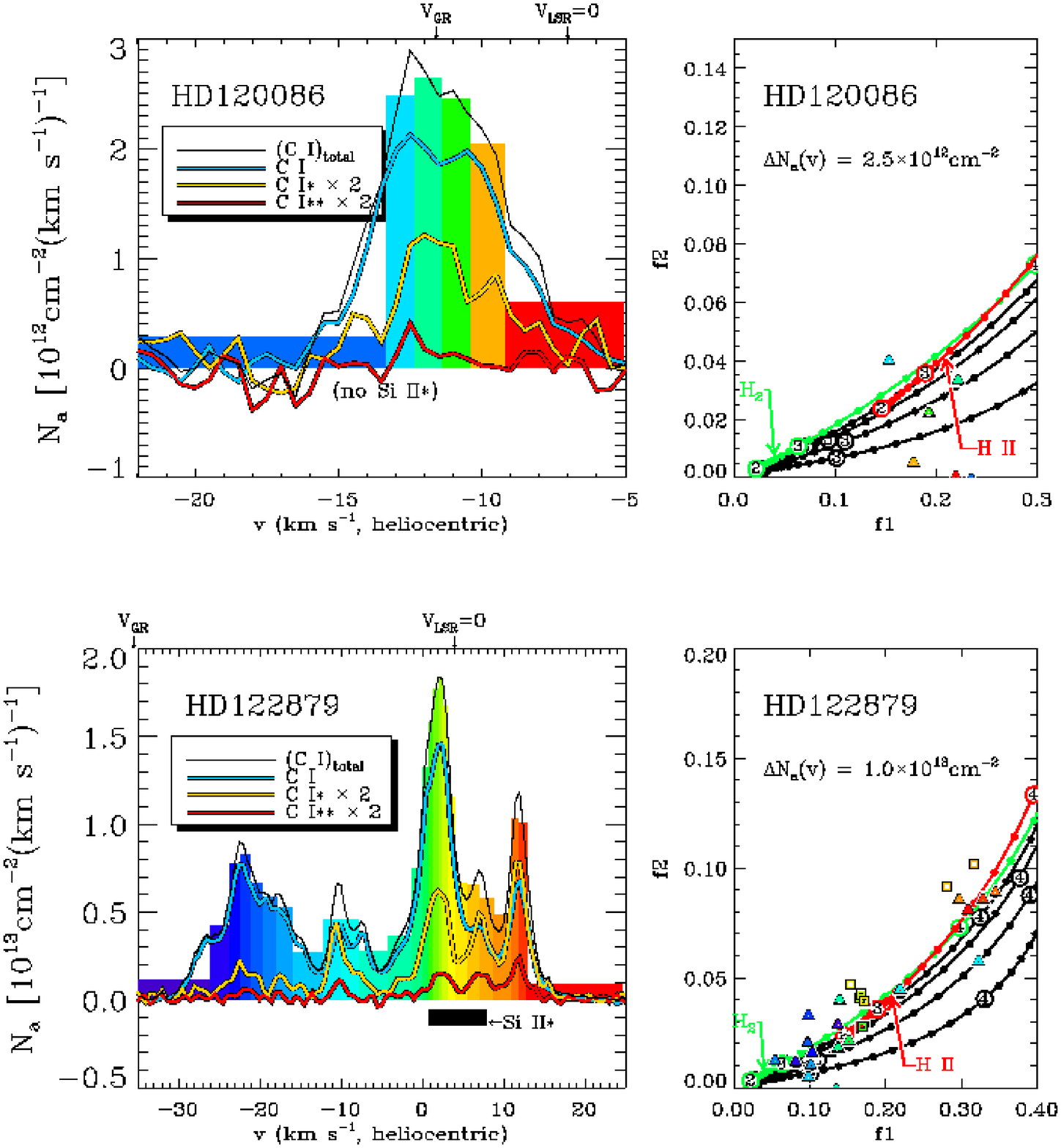}
\figurenum{7g}
\caption{continued}
\end{figure}
\clearpage
\begin{figure}
\plotone{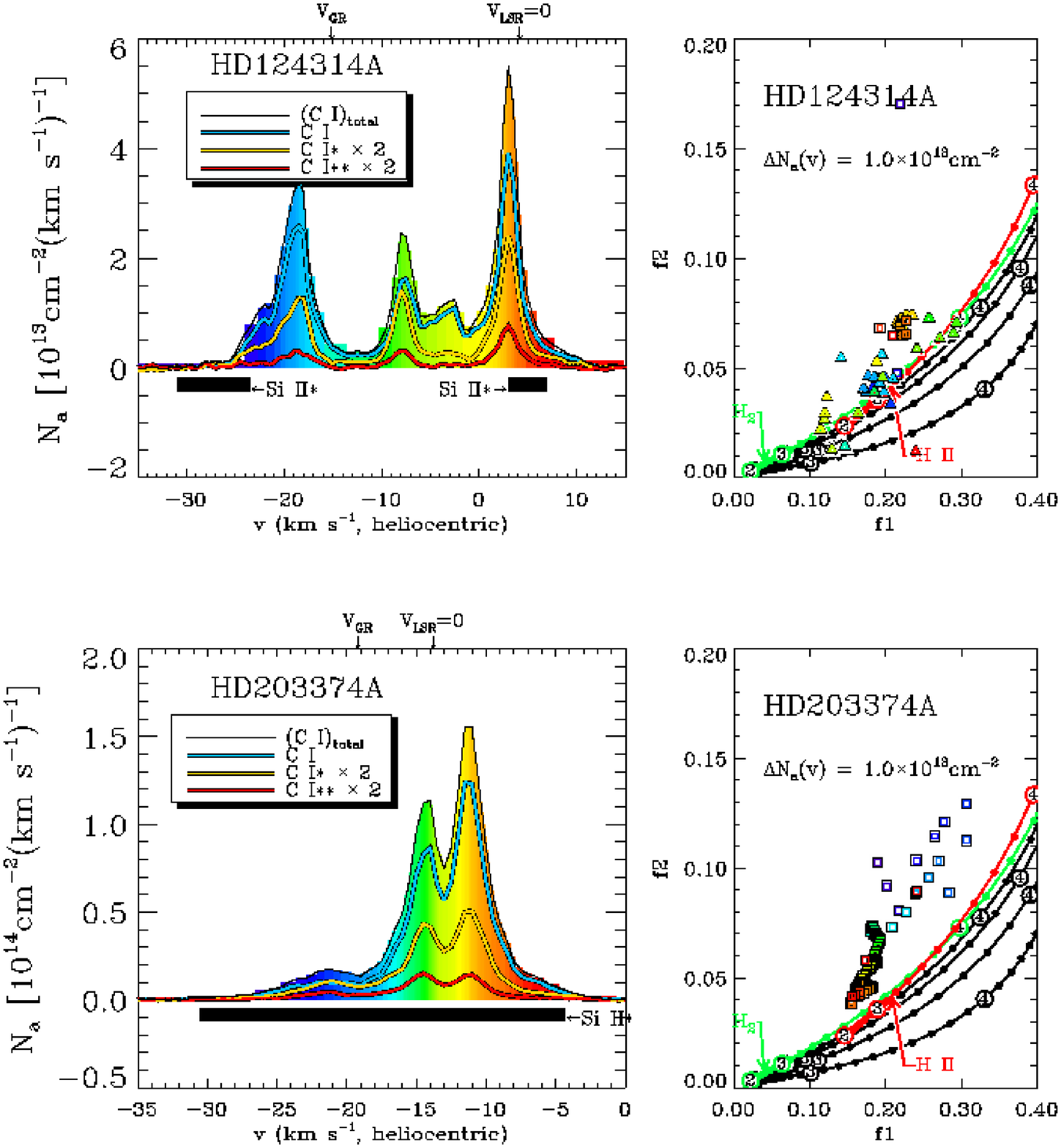}
\figurenum{7h}
\caption{continued}
\end{figure}
\clearpage
\begin{figure}
\plotone{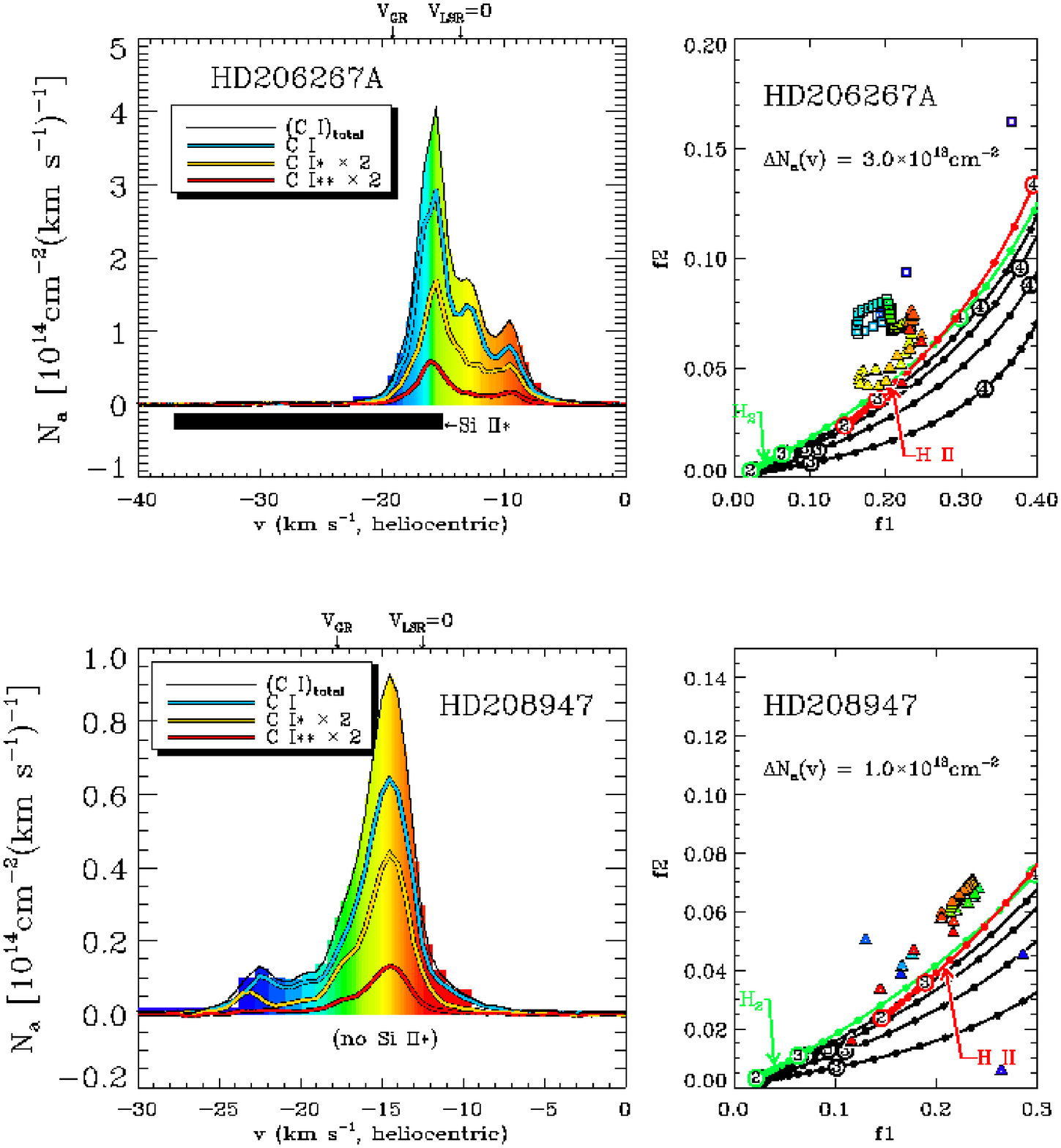}
\figurenum{7i}
\caption{continued}
\end{figure}
\clearpage
\begin{figure}
\plotone{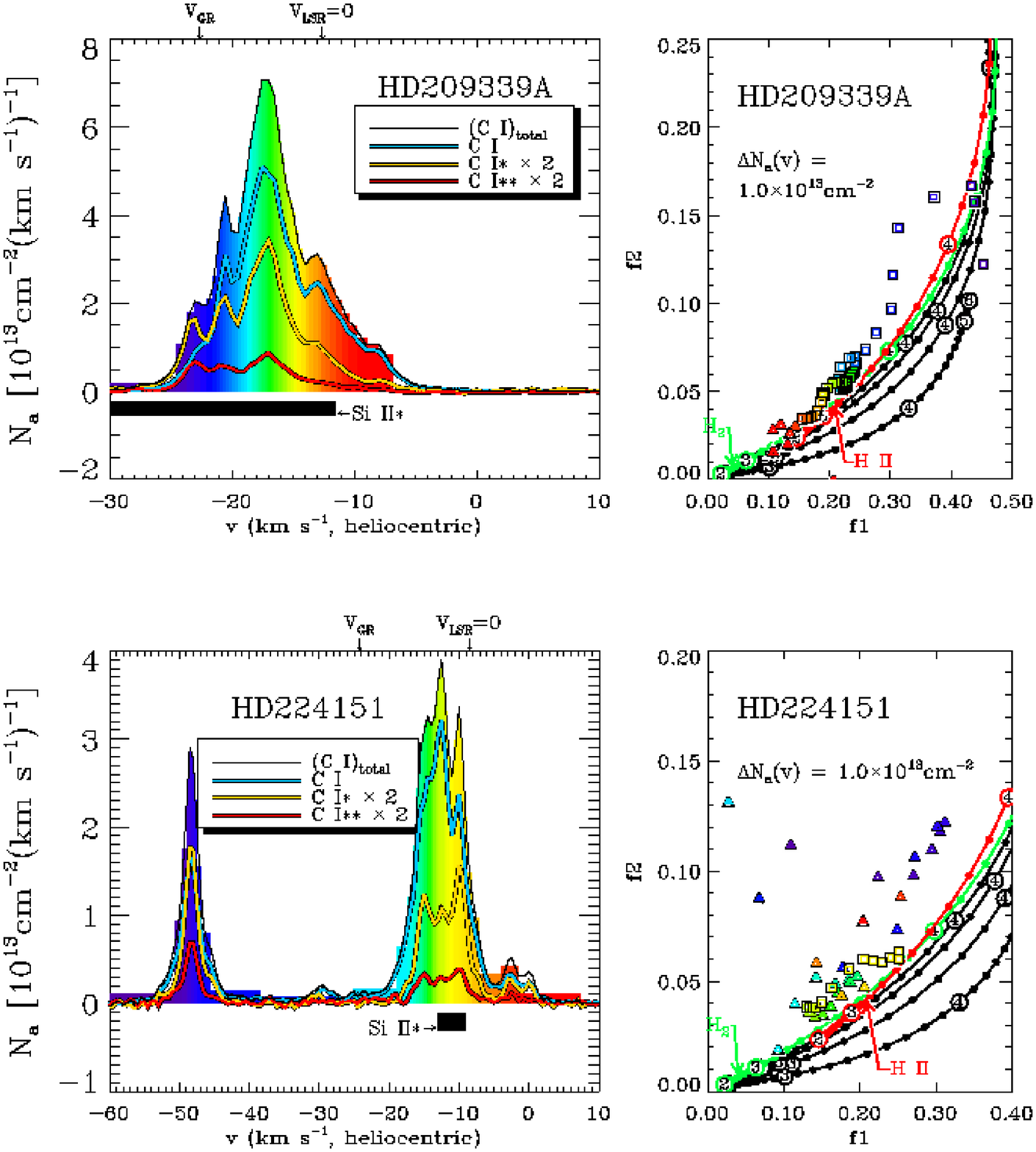}
\figurenum{7j}
\caption{continued}
\end{figure}
\clearpage

\placetable{col_dens}

\begin{deluxetable}{
r     
c     
c     
c     
c     
}
\tablecolumns{5}
\tablewidth{0pt}
\tablecaption{Integrated Column
Densities\tablenotemark{a}\label{col_dens}}
\tablehead{
\colhead{HD~~} & \colhead{$\log N$(C~I)} & \colhead{$\log N({\rm
C~I}^*)$} & \colhead{$\log N({\rm C~I}^{**})$} & \colhead{$\log N({\rm
C~I}_{\rm total})$}\\
\colhead{} & \colhead{(${\rm cm}^{-2}$)} & \colhead{(${\rm cm}^{-2}$)} &
\colhead{(${\rm cm}^{-2}$)} & \colhead{(${\rm cm}^{-2}$)}
}
\startdata
108\phantom{A}&14.810&14.320&13.926&14.972\nl
3827\phantom{A}&13.514&12.415&12.444&13.580\nl
15137\phantom{A}&14.513&13.862&13.219&14.618\nl
69106\phantom{A}&14.179&13.555&13.033&14.296\nl
88115\phantom{A}&13.956&13.161&12.766&14.044\nl
\nl
93843\phantom{A}&13.969&13.542&13.112&14.149\nl
94493\phantom{A}&14.144&13.504&12.921&14.254\nl
99857A&14.418&13.952&13.397&14.576\nl
103779\phantom{A}&14.152&13.365&12.322&14.223\nl
106343\phantom{A}&14.210&13.638&13.059&14.337\nl
\nl
109399\phantom{A}&14.155&13.584&13.049&14.285\nl
116781A&14.188&13.457&12.935&14.282\nl
120086\phantom{A}&13.076&12.472&\nodata\tablenotemark{b}&13.171\nl
122879\phantom{A}&14.309&13.682&13.042&14.420\nl
124314A&14.540&13.970&13.402&14.667\nl
\nl
203374A&14.857&14.252&13.763&14.980\nl
206267A&15.160&14.616&14.131&15.300\nl
208947\phantom{A}&14.502&13.978&13.418&14.642\nl
209339A&14.608&14.104&13.517&14.753\nl
210839\phantom{A}&14.798&14.337&14.013&14.977\nl
\nl
224151\phantom{A}&14.501&13.898&13.437&14.627\nl
\enddata
\tablenotetext{a}{$\log [\int N_a(v)dv]$ over the velocity ranges shown
in Figs.~\protect\ref{210839_f1f2} and 7.}
\tablenotetext{b}{Column density consistent with zero, to within the
measurement errors.}
\end{deluxetable}
\clearpage

\subsection{Information from O~I* and O~I**}\label{OI*}

Neutral oxygen is another element that has absorption features within
the wavelength coverage of our survey and, like neutral carbon, has
three fine-structure levels in its ground electronic state.  The
separations of the levels in energy are larger than those for C I, and
the radiative decay rates are much larger.  As a consequence, measurable
concentrations of O~I* and O~I** occur only when the pressures and
temperatures are much higher than those needed to populate C~I* and
C~I**.

All stars in our survey show absorption by telluric O~I* and O~I** (see
Appendix~\ref{telluric}) from the transitions at 1304.858 and
1306.029$\,$\AA, respectively, but only six show additional features
that we can attribute to the interstellar medium.  These features are
shown in Fig.~\ref{oisoiss}, as represented by their $N_a(v)$ profiles. 
Our derived values of $N_a(v)$ for C~I** are shown above them for
comparison.  The stars HD~93843 and HD~210839 show distinct differences
in the shapes and positions of the excited C~I and O~I features,
indicating that the regions have some velocity-dependent stratifications
in their physical conditions.

\placefigure{oisoiss}
\addtocounter{figure}{1}
\begin{figure}
\plotone{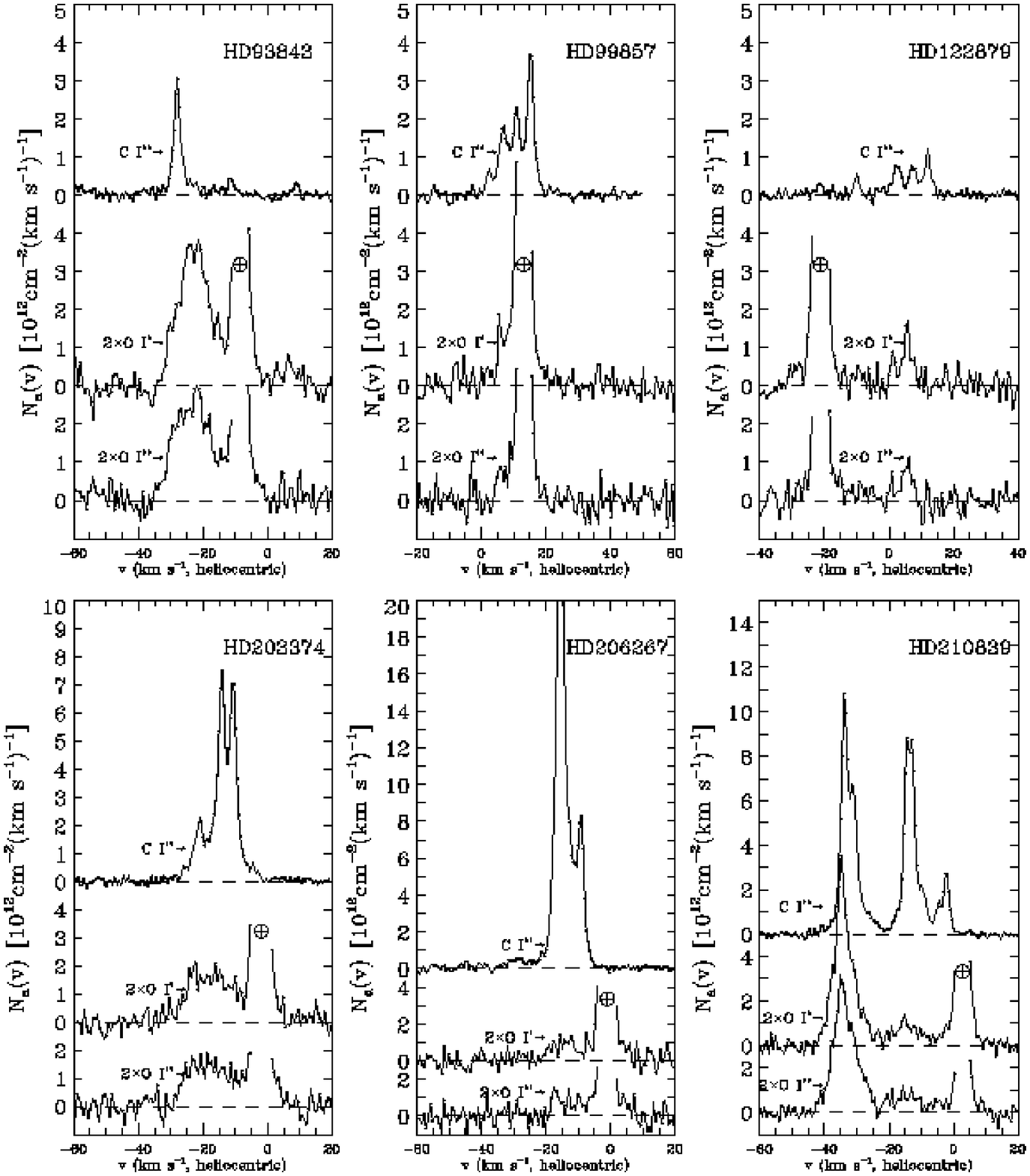}
\caption{A comparison of column densities per unit velocity for C~I**,
O~I*, and O~I**, plotted against heliocentric velocity for those stars
that showed some detectable absorption from the excited oxygen atoms. 
The locations of the telluric oxygen absorptions are indicated by the
symbol $\earth$ over the O~I* feature.  The plots for O~I* and O~I**
have their baselines offset by $-5$ and $-8$ in $y$ [unit = $10^{12}{\rm
cm}^{-2}({\rm km~s}^{-1})^{-1}$] and are expanded vertically by a factor
of 2.  Values for $N_a({\rm O I}^*)$ and $N_a({\rm O I}^{**})$ were
derived using Eq.~\protect\ref{N_a} and values of $\log f\lambda=1.804$
listed by Morton  (1991).\label{oisoiss}}
\end{figure}

In the wavelength coverage of STIS, there are only two O~I lines from
the unexcited state: one is an allowed transition at 1302.169$\,$\AA\
and the other a semi-forbidden transition at 1355.598$\,$\AA. 
Unfortunately, the former is usually very badly saturated over a large
range of velocity, while the latter is seen only when there is a large
amount of material in the line of sight.  The only case where we could
register an absorption feature at 1355$\,$\AA\ that had accompanying
discernible features of O~I* and O~I** was for HD~210839.

\subsection{Physical Conditions in the ${\bf -35\, km~s^{-1}}$ Component
toward HD~210839}\label{210839_comp}

Over all of the C~I velocity components studied in our survey, the one
centered at $-35\,{\rm km~s}^{-1}$ toward HD~210839 ($\lambda$~Cep)
seems to exhibit the clearest evidence for high pressure gas.  This
negative velocity is generally consistent with a portion of the large,
expanding shell of the Cepheus Bubble [$-28 < v_{\rm LSR} < +9\,{\rm
km~s}^{-1}$ that surrounds this and many other stars in the general
vicinity  (Patel et al. 1998)].  Evidently this velocity component is
not particularly conspicuous in CH$^+$, since its presence is not
reported by Crane, Lambert \& Sheffer  (1995), whereas they do present
information about CH$^+$ blended absorption components centered at
$v=-9.5$ and $-13.1\,{\rm km~s}^{-1}$, i.e., velocities consistent with
our stronger C~I component shown in Fig.~\ref{210839_f1f2}.

Fortunately, there is enough matter present at $-35\,{\rm km~s}^{-1}$
that we can detect a weak absorption by O~I for the transition at
1355.598$\,$\AA.  Using this line in conjunction with the excited lines
in the strong multiplet shown in Fig.~\ref{oisoiss} leads to $N({\rm
O~I})= 9.6\times 10^{16}{\rm cm}^{-2}$, $N({\rm O~I}^*)= 2.6\times
10^{13}{\rm cm}^{-2}$ and $N({\rm O~I}^{**})= 2.3\times 10^{13}{\rm
cm}^{-2}$.  For now, we consider the case where the gas is primarily
neutral, and later we will present evidence that supports this position. 
From the tables showing the expected fine-structure population ratios
for O~I presented by Keenan \& Berrington  (1988), we find that a region
with $n({\rm H~I})=60\,{\rm cm}^{-3}$ and $T\approx 1000\,$K gives a
reasonably close match to the relative populations that we derived for
this component.  The resulting pressure, $p/k=10^{4.8}{\rm cm}^{-3}$K,
is very close to the answer that we derived for the excitation of C~I at
high temperatures, as is evident in Fig.~\ref{210839_f1f2}.  This arises
from the fact that the theoretical expectations for $f1$ and $f2$ for
neutral gas with $T\sim 1000\,$K and various pressures very nearly match
the track for fully ionized gas at $T=7000\,$K (Case 3: red tracks in
the $f1,f2$ diagrams) at the corresponding identical pressures.  The
blue, box-like points in the $f1,f2$ diagram are very close to the point
representing $p/k=10^{4.8}{\rm cm}^{-3}$K on the red track.

We consider next evidence provided by the fine-structure excitation of
Si~II.  The strongest transition from Si~II* is one at 1264.7$\,$\AA,
whose strength may be compared to that of the much weaker line of Si~II
at 1808.0$\,$\AA. Because it has a positive charge, Si~II has its
fine-structure excitation dominated by collisions with electrons if the
region is not almost completely neutral.  The rate coefficient for
de-excitations is
\begin{equation}\label{downward}
\gamma_{2,1}={8.63\times 10^{-6}\Omega_{1,2}\over g_2T^{0.5}}{\rm
cm}^3{\rm s}^{-1}
\end{equation}
(Spitzer 1978, p. 73), with the reverse rate given by detailed
balancing, $\gamma_{1,2}=(g_2/g_1)\exp(-E_{1,2}/kT)\gamma_{2,1}$.  The
statistical weights of the levels are $g_1=2$ and $g_2=4$, and the
temperature equivalent for the difference in energy levels
$E_{1,2}/k=413\,$K.  Thus we find that the condition for equilibrium,
\begin{equation}\label{fsl_equilib}
n(e)\gamma_{1,2}n({\rm Si~II})=[n(e)\gamma_{2,1}+A_{2,1}]n({\rm
Si~II}^*)
\end{equation}
will lead to an equation for the electron density
\begin{equation}\label{n(e)}
n(e)={g_2A_{2,1}T^{0.5}\left[{n({\rm Si~II}^*)\over n({\rm
Si~II})}\right]\over 8.63\times 10^{-6}\Omega_{1,2}\left\{
\left({g_2\over g_1}\right) \exp\left({-E_{1,2}\over kT}\right) -
\left[{n({\rm Si~II}^*)\over n({\rm Si~II})}\right]\right\} }~,
\end{equation}
where the radiative decay probability for the upper level is
$A_{2,1}=2.17\times 10^{-4}{\rm s}^{-1}$  (Nussbaumer 1977), and the
collision strength $\Omega_{1,2}=5.6$ over a broad range of temperature 
(Keenan et al. 1985).

\begin{figure}
\plotone{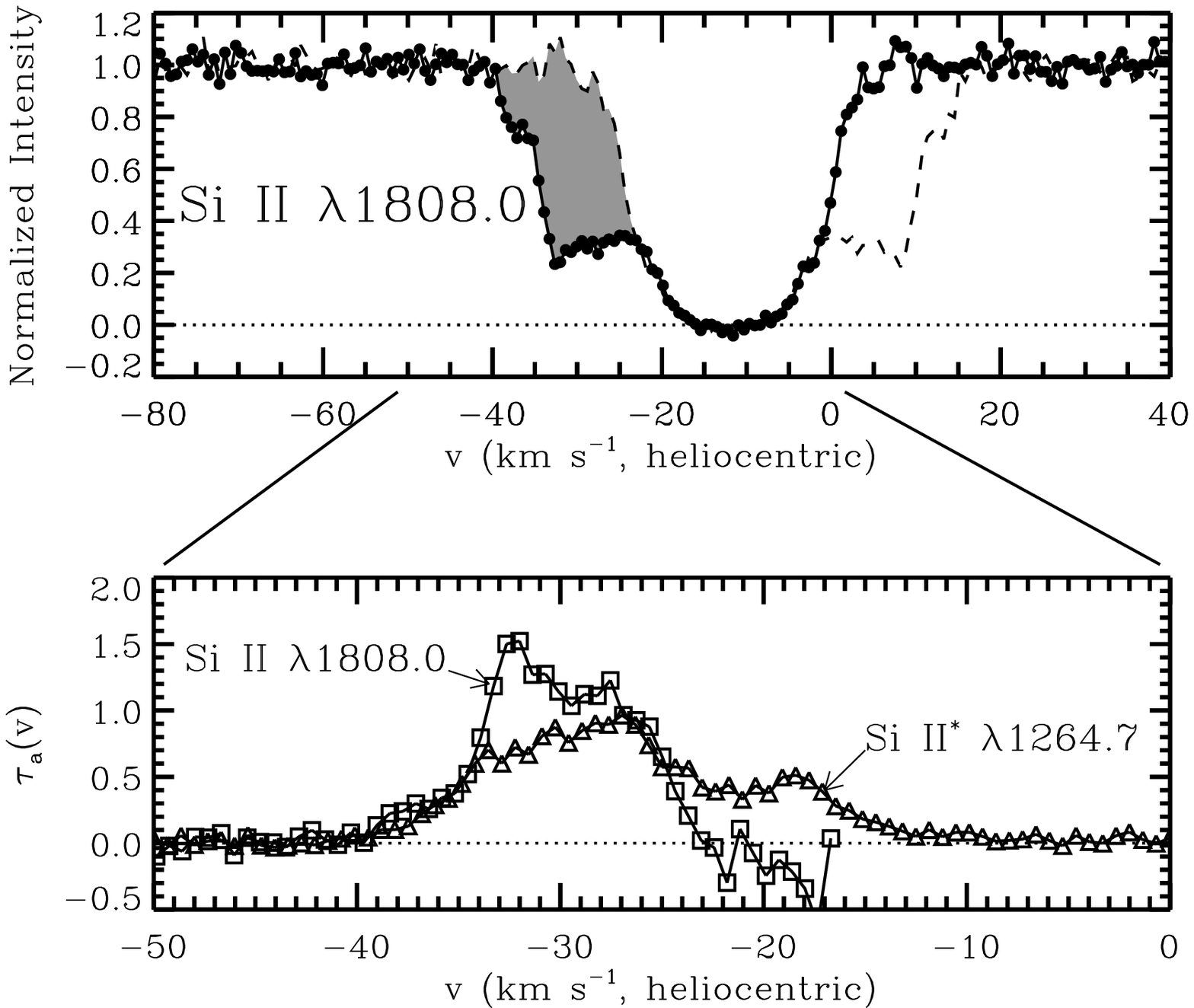}
\caption{{\it Top panel:\/} Intensity profile for the absorption by the
Si~II transition at 1808.013$\,$\AA\ toward HD~210839 ($\lambda$~Cep),
shown by the dark line with dots.  The dashed line traces a reflection
of this intensity about an axis of symmetry for the main component at
$-11.9\,{\rm km~s}^{-1}$.  The filled-in area shows the assumed
contribution for the absorption centered at $-30\,{\rm km~s}^{-1}$ that
is expected in the absence of the left-hand wing of the main feature. 
{\it Bottom panel:\/} Apparent optical depths $\tau_a(v)$ for the Si~II
and Si~II* transitions, as labeled.  The $\tau_a(v)$ for Si~II applies
only to the shaded region in the upper panel.\label{210839_si_ii}}
\end{figure}

The derivation of $N$(Si~II*) for the velocity complex near $-30\,{\rm
km~s}^{-1}$ is straightforward, but the corresponding absorption by
Si~II has strong interference from the main interstellar component at
$-18\,{\rm km~s}^{-1}$, as is shown in the top panel of
Fig.~\ref{210839_si_ii}.  On the assumption that the latter component is
perfectly symmetrical and fully resolved by the instrument, we evaluated
$\tau_a(v)$ for the high pressure component of Si~II against a continuum
level that was simply a reflection of the profile about the axis of
symmetry of the strong component at $-18\,{\rm km~s}^{-1}$.  This
absorption is depicted by the shaded region in the figure.  The derived
$\tau_a(v)$ is shown in the lower panel, overplotted on $\tau_a(v)$ for
the Si~II* feature at 1264.7$\,$\AA.  By integrating the profiles over
the velocity interval $-40$ to $-23\,{\rm km~s}^{-1}$ and using
Eq.~\ref{N_a} together with the values $\log f\lambda=0.576$  (Bergeson
\& Lawler 1993) and 3.100  (Luo, Pradhan, \& Shull 1988) for Si~II and
Si~II*, respectively, we find that $N({\rm Si~II})=1.3\times 10^{15}{\rm
cm}^{-2}$ and $N({\rm Si~II}^*)=4.4\times 10^{12}{\rm cm}^{-2}$.  These
two numbers lead to a relative population of $3.5\times 10^{-3}$ for the
excited form.  Substituting this result into Eq.~\ref{n(e)}, we find
that
\begin{equation}\label{n(e)_ans}
n(e)=1.0\exp(0.413T_3)T_3^{0.5}{\rm cm}^{-3}
\end{equation}
where $T_3$ is the temperature in units of 1000$\,$K, i.e, for $T_3=1$
we obtain $n(e)=1.5\,{\rm cm}^{-3}$.  This electron density is not high
enough to alter the excited O~I populations appreciably from those
established by neutral hydrogen collisions.  Thus, we feel that it
appropriate to identify our C~I and O~I absorptions as having come from
mostly neutral gas.  (We note also that the total integrated column
density for C~I in all three levels for this velocity component is
$1.42\times 10^{14}\,{\rm cm}^{-2}$, well in excess of what one would
predict for the contribution from an H~II region generated by one or
more O6 stars, as we will show in \S\ref{hii_regions}.)

\placefigure{210839_si_ii}

It is clear from Figs.~\ref{210839_f1f2}, \ref{oisoiss} and
\ref{210839_si_ii} that the velocities and shapes of the excited carbon,
oxygen and ionized silicon do not agree in detail, even though they are
clearly associated with a complex of gas that has a distinct identity,
by virtue of its velocity being well separated from the general gas
along the line of sight.  All profiles are asymmetrical and have a
sharper edge on the negative velocity side than on the positive velocity
one.  The velocity sequence in peak absorptions start with O~I* and
O~I** profiles at $-34.5\,{\rm km~s}^{-1}$, followed by C~I** at
$-32.5\,{\rm km~s}^{-1}$ (C~I and C~I* agree with C~I** -- see
Fig.~\ref{210839_f1f2}), and ending with Si~II* which starts to build
near the C~I and O~I maxima, but doesn't reach its peak until
$-27.5\,{\rm km~s}^{-1}$.  The Si~II* profile has an extension to nearly
$-12\,{\rm km~s}^{-1}$.  Errors in the relative positions of the peaks
are of order $1\,{\rm km~s}^{-1}$, judging from the offsets we found for
the carbon multiplets (\S\ref{vel_reg}).
\clearpage

\section{General Interpretation}\label{interpretation}
\subsection{Implications of $f2$ vs. $f1$}\label{pts}

\begin{figure}
\plotone{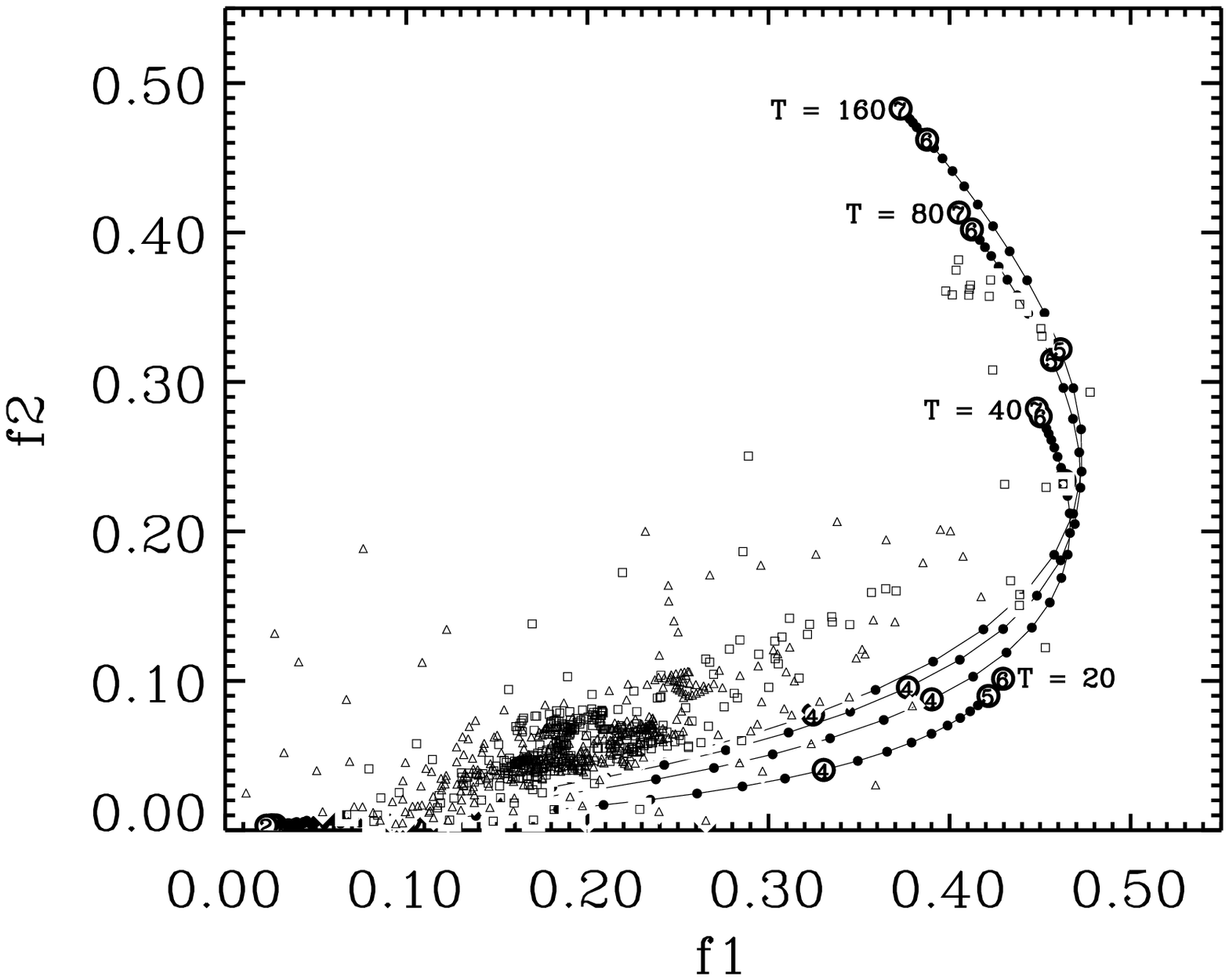}
\caption{The outcome for $f1$ and $f2$ for all stars in the survey, with
triangles representing measurements outside the velocity intervals
spanned by Si~II* absorption and squares within these intervals (see
\S\protect\ref{other_HII}).  Unlike the treatments for individual stars
shown in Figs.~7$a$, $g$, and $i$, $\Delta N_v(v)$ was never
allowed to differ from $1.0\times 10^{13}{\rm cm}^{-2}$, so that all
points represent equal amounts of C~I.  The theoretical tracks for
Cases~2 and 3 (\S\protect\ref{theoretical}) are omitted for
clarity.\label{f1f2all}}
\end{figure}

Figure~\ref{f1f2all} shows all of the C~I $f1,f2$ points for the 21
stars in the survey.  The median value for $f1$ for the entire sample is
0.196, which corresponds to $p/k=2240\,{\rm cm}^{-3}$K for a purely
atomic gas at a most favorable temperature $T=40$K.  Slightly higher
pressures would be derived for temperatures away from this value.  We
are not surprised to see a broad dispersion of the points, most of which
run parallel to the tracks that represent $10^3 < p/k < 10^4{\rm
cm}^{-3}$K.  This behavior seems to reflect chance encounters of the gas
regions with random events linked to pressurization and rarefaction in a
chaotic medium.  However, an unexpected result is that most of the
measured values for $f2$ seem to be too large for their accompanying
$f1$ values, if indeed the C~I at a particular radial velocity were to
reside in a homogeneous region corresponding to any of the three cases
for possible mixes of collision particles that we discussed in
\S\ref{theoretical} (or admixtures thereof).  This is shown by the fact
that most of the points fall above the tracks in the diagrams.  Looking
back to Figs.~\ref{210839_f1f2} and 7, we see that
particularly noteworthy manifestations of this effect show up on the
lines of sight toward HD's 108, 15137, 99857A, 106343, 109399, 116781A,
124314A, 203374A, 206267A, 208947, 210839 and 224151.  Recalling the
``center of mass'' behavior (\S\ref{genl_remarks} and
Fig.~\ref{cartoon}) for points representing a composite of different
regions on the $f1,f2$ diagram, we can reconcile this effect with the
existence of a large amount of material within an ordinary range of
pressures, i.e., $10^3 \lesssim p/k \lesssim 10^4{\rm cm}^{-3}$K, but
then accompanied by a small amount of gas at very high pressure having
the same radial velocity.  From the geometry of how the tracks wrap
around in the diagram, it is evident that for the high-pressure gas to
be able to ``pull'' the low-pressure points in a near vertical direction
away from their tracks, it must be at a pressure $p/k\gtrsim 10^5{\rm
cm}^{-3}$K.  Except for the components of C~I centered at $-33\,{\rm
km~s}^{-1}$ in the spectrum of HD~210839 and $-28\,{\rm km~s}^{-1}$ in
the spectrum of HD~93843, this gas at such extraordinary pressures seems
never to be unaccompanied by the lower pressure material.

While one might suppose that unusually large C~I** populations could
arise from the absorption of line-emission photons that are created by
H~II regions, as we discussed in \S\ref{optical_pump}, it must be
remembered that the effect works only if the radial velocity differences
are less than the characteristic line widths that average about ${\rm
FWHM}=6.7\,{\rm km~s}^{-1}$  (Huang et al. 1999).  Thus, it seems hard
to imagine that the C~I** populations could be modified over the large
ranges of radial velocity that are evident in the figures.

\placefigure{f1f2all}

It is important to recognize that while the C~I-weighted fraction of
this high pressure gas can be of order 0.1, the real fraction of gas, as
measured by $N({\rm H})$, can be much lower since a larger fraction of
the carbon is in neutral form at high pressures.  To take an extreme
example, if indeed $n({\rm C~I}_{\rm total})/n({\rm C})\propto p/k$
(from $n(e)/n(H)$ being approximately constant), we could envision that
a $10^{-5}$ fraction of the hydrogen at $p/k = 10^6\,{\rm cm}^{-3}$K
would shift a point 10\% of the distance spanned from the $f1-f2$
combination representing the remaining gas at $p/k = 10^2\,{\rm
cm}^{-3}$K.  Of course, this assumes that the duration of the
overpressurized condition is long enough for the carbon to reach a new
ionization equilibrium, which can take up to about a thousand years if
the photoionization rate is equal to the general galactic value.  (Time
and length scales are considered further in \S\ref{discussion}.)

\subsection{Simple Checks}\label{simple_check}

Since the conclusions stated above have some far-reaching implications,
it is important to verify that the phenomenon is real.  For instance,
one might have reservations about possible systematic errors that could
arise in the solutions to multiplet absorptions (\S\ref{specific_appl})
or the derivations of relative $f$-values (\S\ref{fval}).  A very simple
test that avoids these complex operations can be implemented to validate
the unusual character of our results.  Multiplet nr.~4 at 1329$\,$\AA\
has absorptions from the three levels that are well separated, although
the individual transitions from a given excited level are usually
blended.  The sums of the $f$-values out of the three levels are
identical (this property arises from LS coupling assumed in the
published results, and our rederivations agree with it).  We now
restrict our attention to sight lines that have 1329$\,$\AA\ multiplets
that are of moderate depth, so that they are neither so weak that they
are dominated by noise, nor so strong that errors from saturation are
large.  We then evaluate $\int \tau_a(v)dv$ over the entire widths of
the C~I* and C~I** profiles, knowing that the merging of the lines
within them results in our being able to measure only the total column
densities without knowing the details in the velocity structures.

Table~\ref{tau_a_comp} shows the average values of $\tau_a(v)$ for the
complex of three lines arising from C~I* centered at 1329.1$\,$\AA,
along with similar determinations for the two lines located near
1329.6$\,$\AA\ for C~I**.  Measurements of the single line out of C~I at
1328.833$\,$\AA\ are of no use, because this line is saturated when the
others are measurable.  The ratio of the average $\tau_a$'s, listed in
the last column of Table~\ref{tau_a_comp}, may be compared to the
overall averages of $f1/f2$ shown in Figs.~\ref{210839_f1f2} and
7.  The correspondence between the listed values of
$<\tau_a({\rm C~I^{**}})>/<\tau_a({\rm C~I^*})>$ and the positions of
the points in the respective $f1,f2$ diagrams is excellent for
HD~99857A, HD~224151 (both peaks) and HD~210839 (left-hand peak). 
Reasonably good matches are seen for HD~69106, HD~106343, and
HD~124314A.  HD~15137, HD~94493 and HD~122879 show matches that are
marginally acceptable.  The match for HD~116781A is bad, but the value
for $<\tau_a({\rm C~I^{**}})>$ is poorly known because the lines are so
weak in this multiplet.  Overall, our test applied for many lines of
sight indicates that our conclusion that the measurements of $f2$,
relative to those of $f1$, are indeed above the theoretical tracks in
the diagram, where for $p/k<2000\,{\rm cm}^{-3}$K we would expect that
$f2/f1$ should always be less than 0.2.

\begin{deluxetable}{
r     
c     
c     
c     
}
\tablecolumns{4}
\tablewidth{0pt}
\tablecaption{Average Optical Depths\tablenotemark{a}\label{tau_a_comp}}
\tablehead{
\colhead{HD} & \colhead{$<\tau_a({\rm C~I^*})>$} &
\colhead{$<\tau_a({\rm C~I^{**}})>$} & \colhead{$<\tau_a({\rm
C~I^{**}})>/<\tau_a({\rm C~I^*})>$}}
\startdata
15137\phantom{A}&0.846&0.170&0.201\nl
69106\phantom{A}&0.549&0.134&0.241\nl
94493\phantom{A}&0.235&0.052&0.221\nl
99857A&0.985&0.326&0.331\nl
106343\phantom{A}&0.542&0.129&0.238\nl
\nl
116781A&0.192&0.042&0.218:\nl
122879\tablenotemark{b}\phantom{A}&0.446&0.088&0.197\nl
124314A&0.740&0.172&0.232\nl
224151\tablenotemark{c}\phantom{A}&0.583&0.289&0.496\nl
224151\tablenotemark{b}\phantom{A}&0.564&0.153&0.271\nl
\nl
210839\tablenotemark{c}\phantom{A}&1.121&0.993&0.886\nl
\enddata
\tablenotetext{a}{For the following blended absorption features: C~I* =
$\lambda\lambda$ 1329.0853, 1329.1004, 1329.1233$\,$\AA; C~I** =
$\lambda\lambda$ 1329.5775, 1329.6005$\,$\AA.} 
\tablenotetext{b}{Right-hand peak only.}
\tablenotetext{c}{Left-hand peak only.}
\end{deluxetable}

A different approach for questioning the reality of the small amounts of
high pressure gas is to examine the possibility that the theoretical
tracks themselves are in error.  Could there be a problem with any of
the physical data used to derive them in \S\ref{collisional}?  It is
difficult to make a fundamental judgement on whether or not the
calculations of the collisional rate constants or radiative decay rates
have substantial enough errors to mislead us, but we can draw upon real
evidence from interstellar absorption lines that indicates that the
tracks indeed go about as low in the $f1-f2$ diagram as we have drawn
them in the figures.  An analysis of the C~I lines in the two strongest
velocity components toward HD~215733 by Fitzpatrick \& Spitzer  (1997)
shows nominal values of ($f1$, $f2$) equal to (0.140, 0.0131) and
(0.134, 0.0181) for their Components 17 and 18, respectively.  The ratio
of $f2$ to $f1$ is the critical quantity for testing the placement of
the theoretical tracks.  Fortunately, the solutions for $N({\rm
C~I^{*}})$ and $N({\rm C~I^{**}})$ in the analysis of Fitzpatrick \&
Spitzer  (1997) depend almost entirely on the data from a single
multiplet ($\lambda 1561$), thus avoiding the issue of relative errors
in $f-$values from one multiplet to the next.  This increases the
confidence in the validity of the nominal values for $f2/f1 = 0.093$ and
0.135 (but only if the LS coupling rules apply for the strengths of
lines within this multiplet).  Thus, in the vicinity of $f1\approx
0.14$, the values of $f2/f1$ are indeed consistent with the tracks.  The
agreement is satisfactory even if we take the values $f2/f1 = 0.14$ and
0.19 that arise from Fitzpatrick \& Spitzer's lower limits for $N({\rm
C~I^{*}})$ and upper limits for $N({\rm C~I^{**}})$.  By concentrating
on good quality data and choosing a pair of regions that appear to be
physically homogeneous and well separated in velocity from other C I
components, we have demonstrated that the positions of the theoretical
tracks are reasonable. (It should be evident that there is no
combination of different conditions that could produce a composite
measurement {\it below\/} the tracks in this part of the diagram).
\clearpage

\placetable{tau_a_comp}

\subsection{C~I from a Target Star's H~II Region}\label{hii_regions}

In view of the fact that each line of sight penetrates half of the
volume of at least one early-type star's Str\"omgren sphere, one might
question whether or not C~I in this region could distort our findings,
which are presumed to represent simply a composite sample of conditions
in H I regions of moderate density.  This could have a significant
impact in raising our perceived pressures, since H~II regions created by
young stars need typically about $2\times 10^6{\rm yr}$ for their
internal densities to drop to about 1/10 their initial values  (Spitzer
1978, pp. 251$-$255).  This time is significant compared to the
lifetimes of our target stars, which range from about $4\times
10^6\,{\rm yr}$ for O5 stars to $10^7\,{\rm yr}$ for B0~V stars 
(Schaller et al. 1992).

We now produce some quantitative estimates of how seriously our results
could be affected by the contamination from each target star's H~II
region.  Jenkins \& Shaya  (1979) developed a way to calculate the
amount of C~I that should be present between an early-type star and the
edge of its Str\"omgren sphere.  They found that $N$(C~I) could be as
large as
\begin{equation}\label{SS}
N({\rm C~I}_{\rm total})=2({\rm C/H})\alpha_{\rm C}\alpha_{2,{\rm
H}}^{-1}N_{0,{\rm L}} \Gamma_{0,{\rm C}}^{-1}
\end{equation}
where ${\rm C/H}=2.3\times 10^{-4}$ is the abundance of carbon relative
to hydrogen\footnote{Note that this value is larger than the one that we
adopted in \S\protect\ref{substructures} because we are not certain that
C is depleted in H~II regions.  This might lead to an overestimate for 
$N({\rm C~I}_{\rm total})$.} [adopted from the B-star abundance ratio
measured by Cunha \& Lambert  (1994)], $\alpha_{\rm C}$ is the
recombination rate of carbon with free electrons, $\alpha_{2,{\rm H}}$
is the recombination rate of hydrogen to any of its levels with $n\ge
2$, $N_{0,{\rm L}}$ is the number of Lyman-limit photons emitted (in
units of ${\rm cm}^{-2}{\rm s}^{-1}$) at the star's surface, and
$\Gamma_{0,{\rm C}}$ is the ionization rate of neutral C at the star's
surface.

\begin{figure}
\plotone{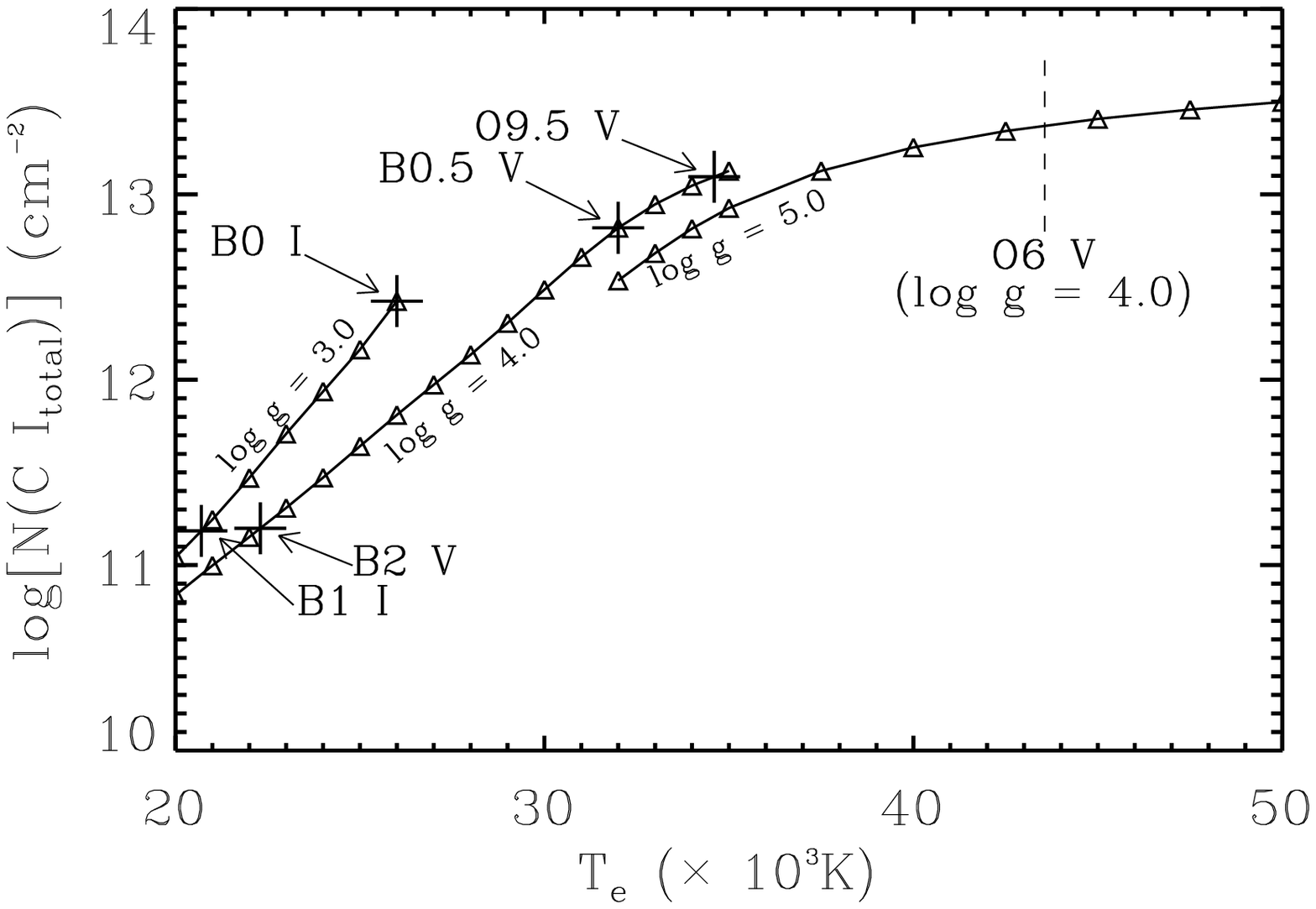}
\caption{Estimated column density of atomic carbon from the surface of a
target star to the edge of its Str\"omgren sphere, as a function of a
star's effective temperature $T_e$.  These results arise from the
application of Eq.~\protect\ref{SS}, using the fluxes from model stellar
atmospheres of Kurucz  (1993, 1997) for stars with different $T_e$ and
surface gravities.  The locations of stars of various spectral types in
this diagram are indicated, using the effective temperature scales
listed by Tokunaga  (2000) for the B-type stars and Vacca et al.  (1996)
for the O-type stars.  The O6 main-sequence stars have $\log g = 4.0$,
but model atmosphere calculations are not available for this surface
gravity at the higher values of $T_e$.\label{cisspher}}
\end{figure}

Figure~\ref{cisspher} shows our calculations of $N({\rm C~I}_{\rm
total})$ from Eq.~\ref{SS} for stars with various effective temperatures
$T_e$ and surface gravities of $10^3$, $10^4$ and $10^5\,{\rm
cm~s}^{-2}$, using the stellar atmosphere model calculations of ionizing
fluxes by Kurucz  (1993, 1997) to obtain $N_{0,{\rm L}}$ and
$\Gamma_{0,{\rm C}}$.  The recombination coefficient $\alpha_{\rm C}$ is
from the radiative and dielectronic rates given by Shull \& van
Steenberg (1982), supplemented by the additional contributions from
low-lying resonance states computed by Nussbaumer \& Storey (1986).  The
rate $\alpha_{2,{\rm H}}$ is from Spitzer  (1978, pp. 106$-$110).  Both
rates were evaluated for $T=8000\,$K.

\placefigure{cisspher}

The expression in Eq.~\ref{SS} overestimates the amount of C~I if the
Str\"omgren sphere is one of low density, since near the edge the star's
ionizing radiation is supplemented by the general interstellar ionizing
field. The results of Jenkins \& Shaya  (1979) show that this effect can
lower the estimate for $N$(C~I) by as much as 0.4~dex in some cases.

Generally, it is evident that of order $10^{13.5}{\rm cm}^{-2}$ of the
C~I might arise from the H~II region created by a very hot star, and it
follows that this region or a photodissociation front on its edge may be
responsible for some readings of high pressure.  Even so, this
perturbation in our results should be restricted to a limited span of
radial velocity in each case, one that is likely to be near the star's
radial velocity $v_*$ (see col.~3 of Table~\ref{tgt_stars}) or $v_{\rm
gal.~rot.}$ (col.~6 of Table~\ref{los}) if the star has a large peculiar
velocity, such as HD~210839.

\subsection{Si~II* as an Indicator for other H~II
Regions}\label{other_HII}

Our study has concentrated on C~I, with some occasional insights
provided by O~I. However as we discussed in \S\ref{210839_comp}, ionized
silicon provides yet one more example showing transitions from an
excited fine-structure level within the wavelength coverage of STIS.  
If Si~II* is present without an accompanying stronger absorption by
O~I*, we can be confident that the absorption arises from a region where
most of the hydrogen is ionized  (Spitzer \& Jenkins 1975).  If $p/k=
10^4\,{\rm cm}^{-3}$K in a warm ionized medium (WIM), $n({\rm
C~I}^{**})/n({\rm C}) = 2\times 10^{-4}$ if $T=8000\,$K and $\Gamma_{\rm
C}=3\times 10^{-10}\,{\rm s}^{-1}$, i.e., the normal interstellar
ionization rate with no enhancement by especially nearby stars.   Under
these same conditions, we expect that $n({\rm Si~II}^*)/n({\rm
Si~II}_{\rm total}) = 8\times 10^{-4}$; see Eq.~\ref{n(e)}.  If the Si
and C are depleted by $-1.3$~dex  (Savage \& Sembach 1996) and
$-0.4$~dex (\S\ref{substructures}), respectively, we would expect to
find roughly comparable strengths for the Si~II* line at 1264.7$\,$\AA\
and the combined effect of the two strong lines of C~I** centered at
1329.6$\,$\AA.  For $p/k > 10^4\,{\rm cm}^{-3}$K, $f2$ increases more
slowly than $n$(Si~II*)/$n({\rm Si~II}_{\rm total})$, but this is
outweighed by the linear dependence of $n({\rm C~I}_{\rm total})/n({\rm
C})$ on $n(e)$, making the Si~II* line weaker than that for C~I**.  It
is clear, in a broad sense, that we can use the 1264.738$\,$\AA\ line of
Si~II* to warn us about the presence of C~I** that may have been brought
about chiefly by gas that is fully ionized.

The black bars just below the $N_a(v)$ profiles shown in
Figs.~\ref{210839_f1f2} and 7 show the velocity ranges
over which the absorption by Si~II* at 1264.7$\,$\AA\ is greater than
10\% of the local continuum.  We find that in many cases these bars
subtend major fractions of the C~I absorbing velocity ranges.  However,
there are also several sight lines where little or no Si~II* is
detected.  

\begin{figure}[bh]
\epsscale{.6}
\plotone{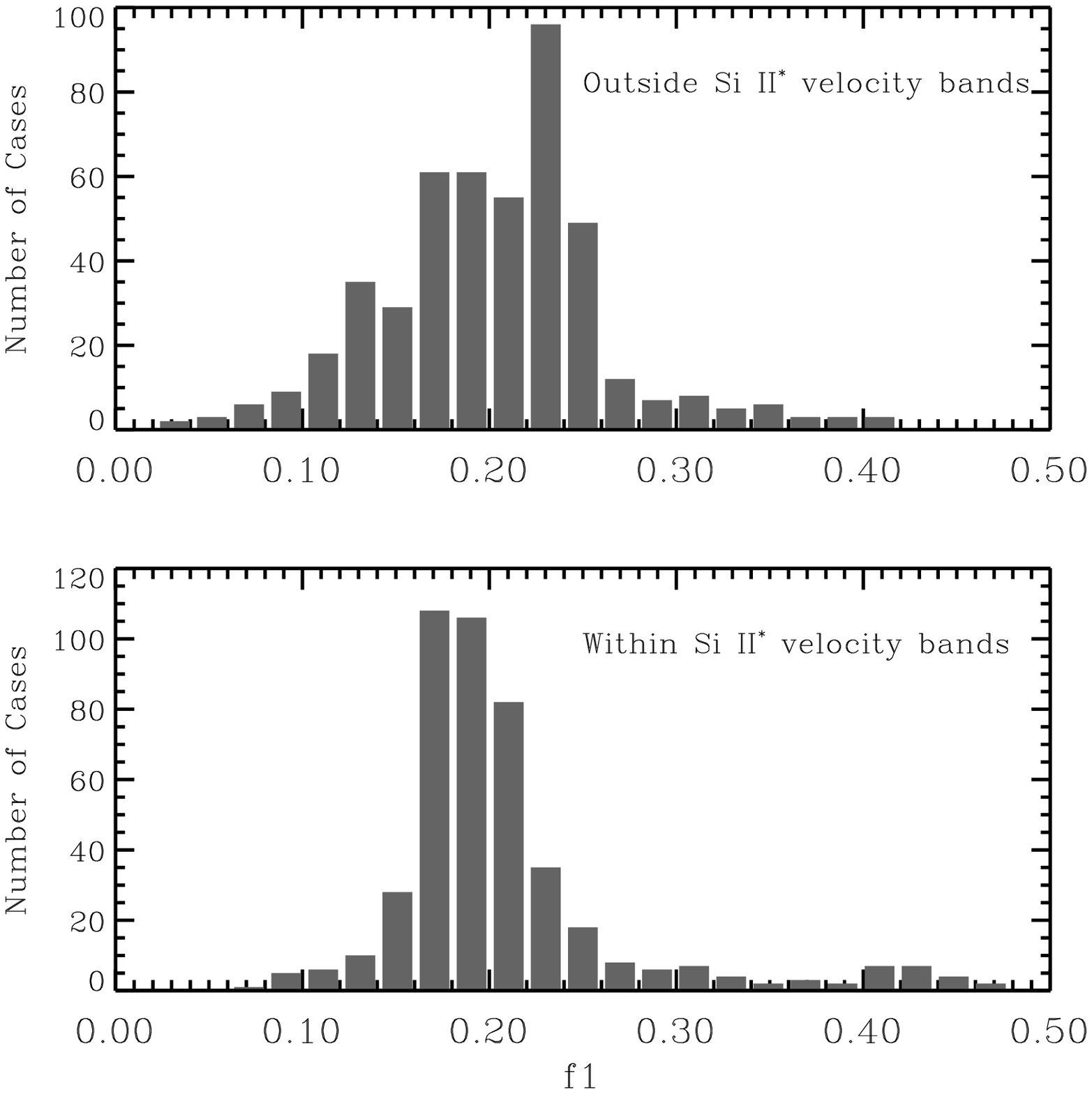}
\caption{{\it Top panel:\/} The frequency of occurrences of $f1$ for C~I
outside the velocity coverage of Si~II* ($>$10\% absorption at
1264.738$\,$\AA) compared with the analysis for points inside the Si~II*
ranges ({\it bottom panel\/}).\label{f1f2hist1}}
\end{figure}

To investigate whether or not the presence of Si~II* is associated with
any abnormalities in the C~I fine-structure ratios, we derived the
distribution functions for $f1$ that pertained to measurements both
within and outside of the Si~II* velocity spans.  The results are shown
in Fig.~\ref{f1f2hist1}.  There seems to be no significant difference
between the two, other than the fact that the measurements outside of
the Si~II* bands show a larger dispersion of values.  This outcome might
arise from the larger errors for measurements at the velocity extremes,
or alternatively, for independent kinematical reasons discussed in
\S\ref{kinematics} below.  The apparent lack of a real difference
indicates that the possible presence of H~II regions is probably not an
important factor that influences the general pressure sensed by the C~I
excitation.

\begin{figure}[th]
\plotone{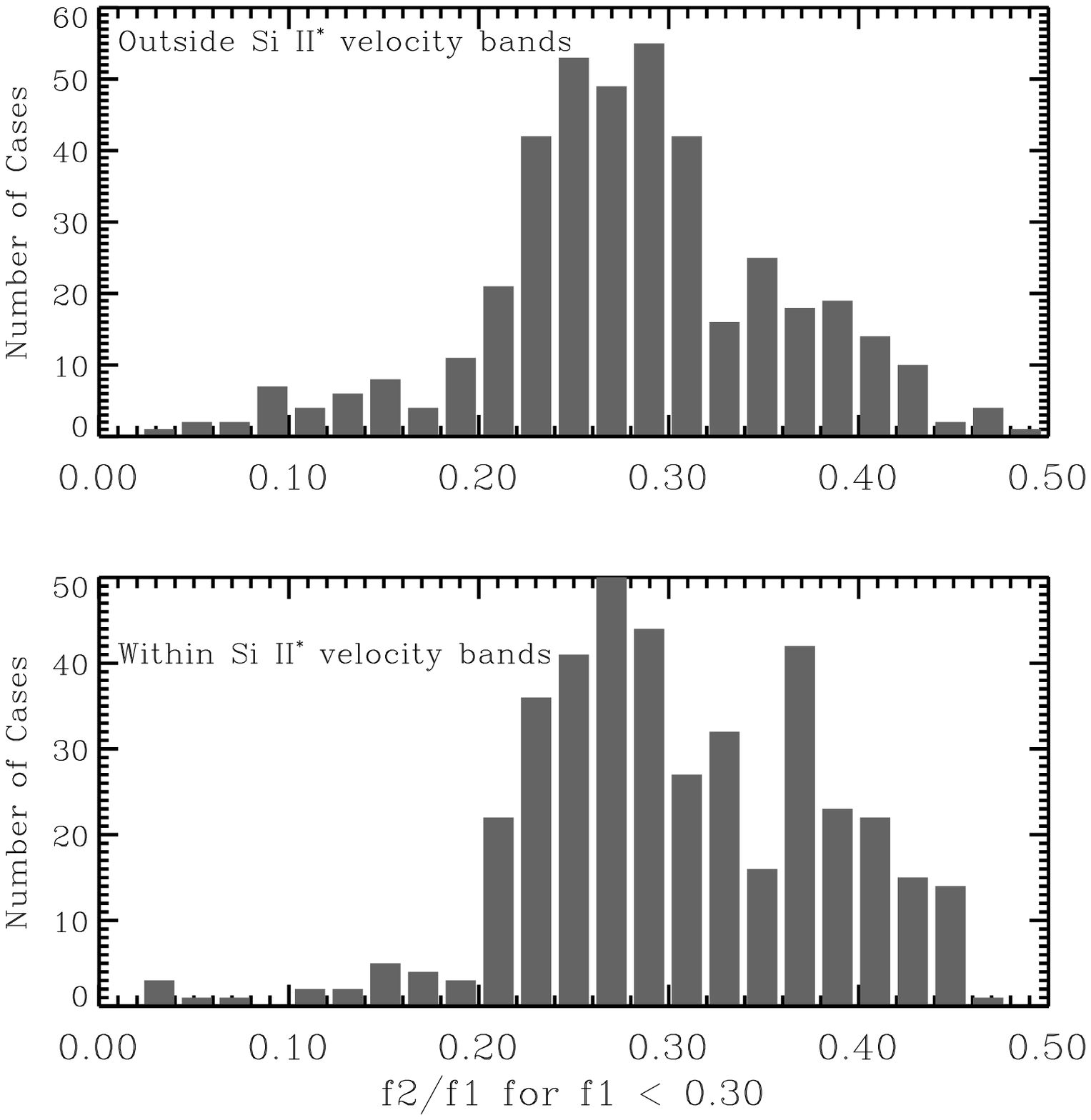}
\caption{Same as for Fig.~\protect\ref{f1f2hist1} except that the
frequencies of $f2/f1$ for $f1<0.3$ are shown for the two populations of
C~I measurements.\label{f1f2hist2}}
\end{figure}

We performed a similar comparison for $f2/f1$ for all cases where $f1 <
0.3$ to test if the presence of Si~II* had any bearing on the tendency
for points to fall above the theoretical tracks, as we pointed out in
\S\ref{pts}.  Figure~\ref{f1f2hist2} shows that C~I within the Si~II*
bands shows a slightly greater tendency to show values $0.30 < f2/f1 <
0.45$ than for determinations on the outside.  A Kolmogorov-Smirnov test
indicates that the two distributions are drawn from different
populations at the 99.9\% confidence level.  Whether or not the presence
of Si~II* is the main discriminant is still open to question, however. 
Again, the difference may simply be due to general kinematical
considerations.

\placefigure{f1f2hist1}

\placefigure{f1f2hist2}

\subsection{The Effects of Kinematics on C~I
Excitation}\label{kinematics}

Since dynamical events are probably the principal drivers of transient
changes in pressure, it worthwhile to investigate whether or not we see
evidence for changes in the C~I excitation linked to the kinematical
properties of the material.  Gas that is quiescent could be identified
as having radial velocities between the velocity of differential
galactic rotation $v_{\rm GR}$ at the position of the target star and
zero point for the Local Standard of Rest $v_{\rm LSR}=0$.  This
material is less likely to have been disturbed by some compression or
rarefaction event in the recent past.  The heliocentric velocities that
mark the adopted endpoints for such gas toward each target are indicated
in Table~\ref{los}\footnote{The endpoints are actually the maximum and
minimum values of the entries $-$Column~5 and (Column~6 $-$ Column~5) of
the table.} and Figs.~\ref{210839_f1f2} and 7.  Gas with
velocities outside this interval is most likely to represent disturbed
material.  However, we must not overlook the fact that such material may
also be found within these velocity limits, since it could be moving
transverse to the line of sight (or have a local velocity offset that
still does not transgress one of our boundaries) or represent turbulent
compression arising from oppositely directed flows whose momenta nearly
cancel each other.

To explore possible differences between moving and quiescent gases, we
start by dividing up our measurements into three velocity intervals: (1)
gases at velocities above the interval spanned by $v_{\rm GR}$ and the
zero point for $v_{\rm LSR}$, (2) gas within this interval and (3) gas
at velocities below the interval.  Interval (3) differs from (1) in one
important respect: since this interval includes includes gas that is
moving toward us with respect to gases in the general vicinity of the
star, it could represent material that is acted upon by the stellar wind
of the target star or its neighbors.  It could also signal the presence
of neutral gas situated at the foreground edge of the star's H~II region
which is accelerated by a shock driven by the suddenly overpressured,
ionized gas to the rear from our vantagepoint.  For all but one star,
HD~69106, we have the configuration that $v_{\rm GR}$ is below the zero
point for $v_{\rm LSR}$, which means that most of the time the material
moving away from the star should be separated from the more general
foreground gas.

\begin{figure}[th]
\plotone{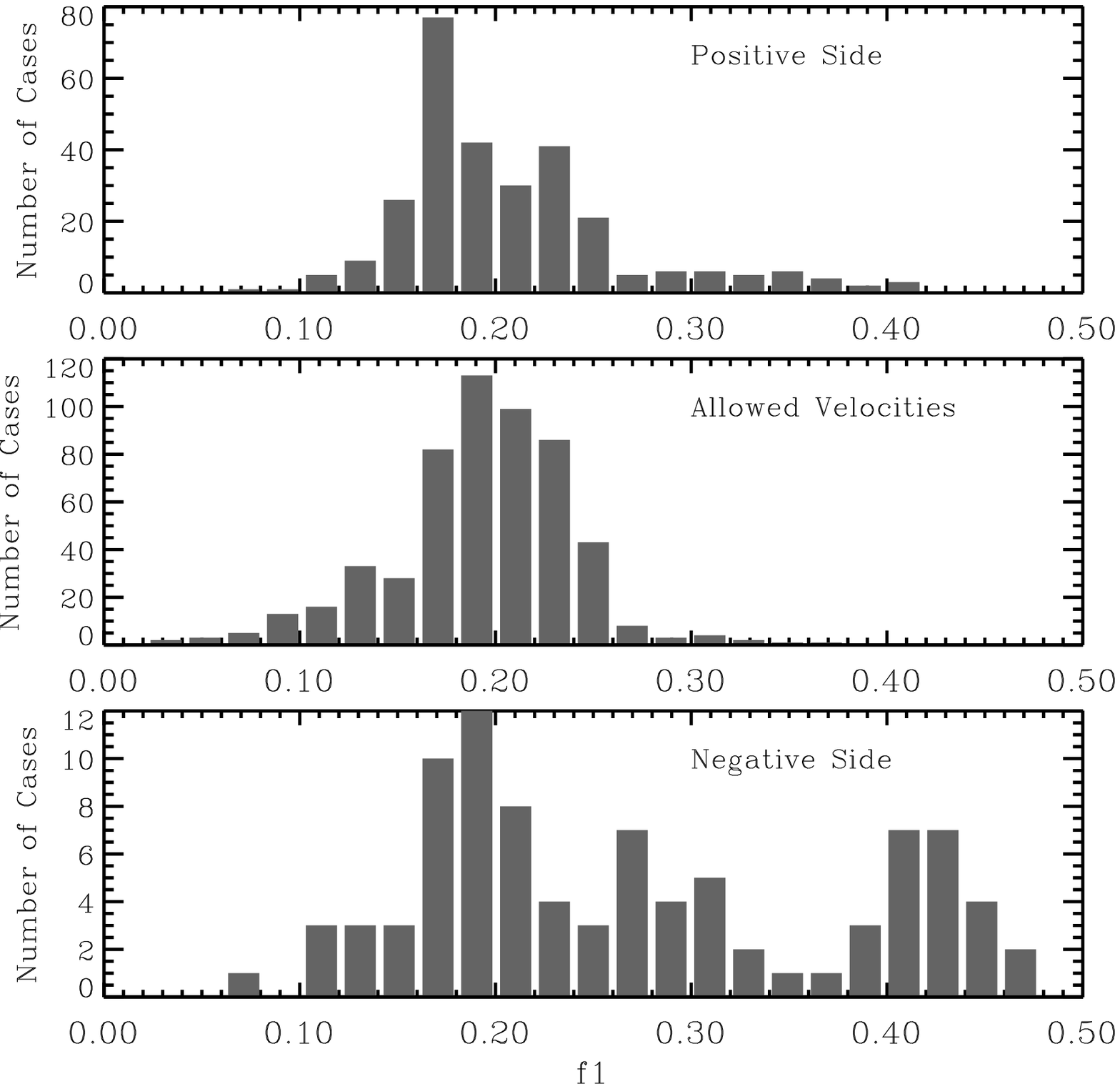}
\caption{The frequency of occurrences of $f1$ for C~I above ({\it top
panel\/}), within ({\it middle panel\/}), and below ({\it bottom
panel\/}) the ``allowed'' velocity ranges spanned by $v_{\rm GR}$ and
$v_{\rm LSR}=0$ shown for the profiles in
Figs.~\protect\ref{210839_f1f2} and
7.\label{f1f2hist3}}
\end{figure}

\begin{figure}[bh]
\plotone{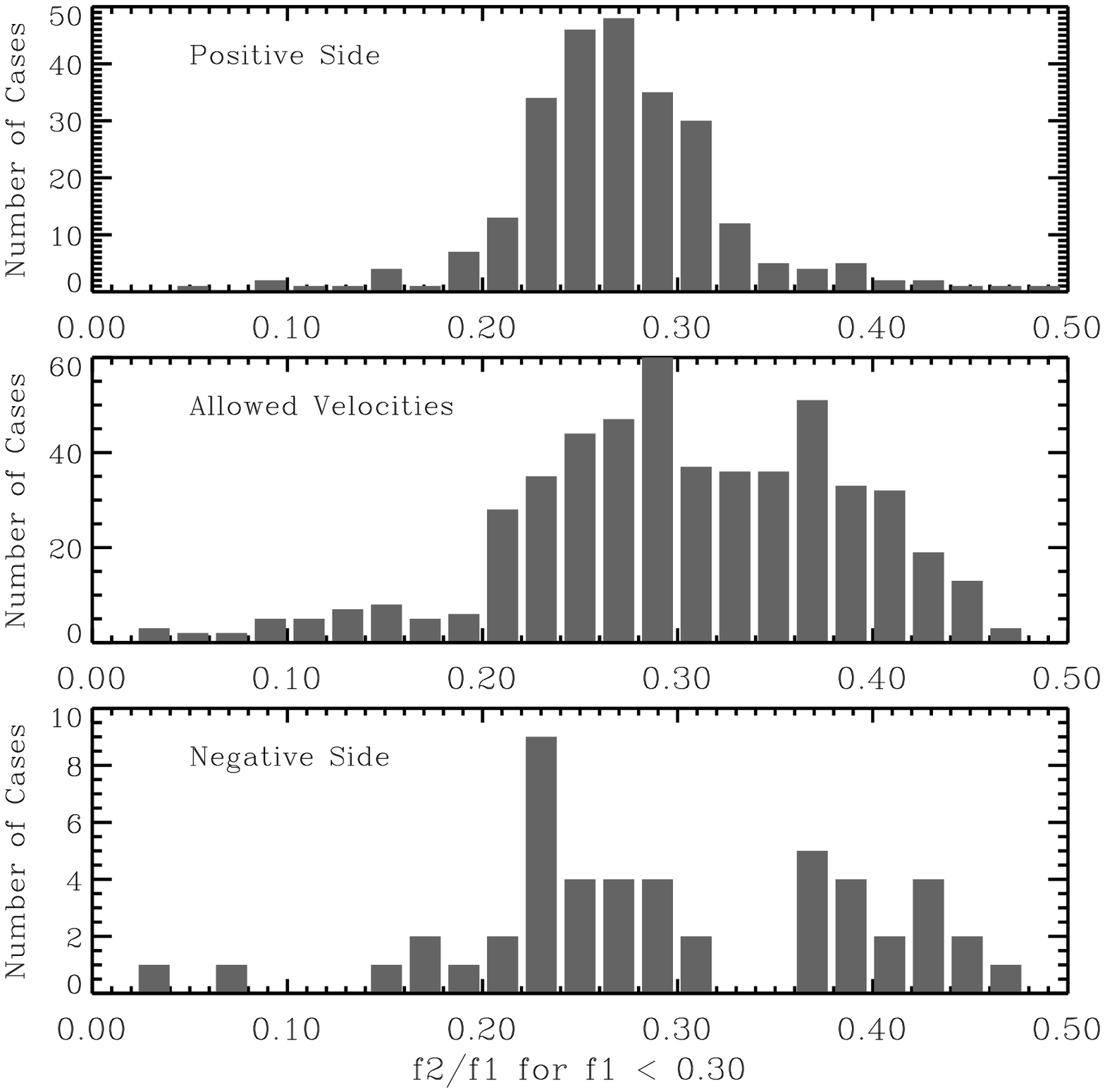}
\caption{Same as for Fig.~\protect\ref{f1f2hist3} except that the
frequencies of $f2/f1$ for $f1<0.3$ are shown for the three populations
of C~I measurements.\label{f1f2hist4}}
\end{figure}

Now that we have defined the velocity intervals, we examine how various
measurements of $f1$ and $f2$ differ from one case to the next.  The
outcomes for $f1$ and $f2/f1$ (with $f1 < 0.30$) in the three velocity
divisions are shown in Figs.~\ref{f1f2hist3} and \ref{f1f2hist4},
respectively.  From Fig.~\ref{f1f2hist3} it is evident that cases where
$f1 > 0.30$ come almost exclusively from measurements outside the range
of ``allowed velocities.''  (Note that vertical magnifications vary from
one panel to the next in the figures.) There is no question about the
reality of the samples with large $f1$ on the negative side:  the
extreme deviations toward positive $f1$ arise primarily from HD~93843
and HD~210839, and some milder excursions having $f1\sim 0.3$ come from
gas at the most negative velocities for HD's 203374A, 209339A and
224151.  It seems reasonable to identify the material that is compressed
and appears at $v\approx -22\,{\rm km~s}^{-1}$ in front of HD's 203374A
and 209339A as part of the expanding shell associated with the Cepheus
Bubble  (Patel et al. 1998).  On the positive velocity side, i.e., above
the allowed velocity range, less spectacular but nevertheless believable
results with $f1 > 0.30$ are seen for HD's 108, 99857A, 106343, 122879,
and 210839.

\placefigure{f1f2hist3}
\placefigure{f1f2hist4}

Do the deviations of $f1$ toward the low end of the distribution for the
allowed velocities, i.e., $f1 < 0.10$, arise from results that can be
trusted?  We believe that most of them, such as those those that came
from HD's 15137, 93843, 94493 [only a single contribution representing
$\Delta N_a(v)$], 103779, 116781A [only 2 $\Delta N_a(v)$
determinations], and 122879 are reliable, while those for HD's 3827,
109399, and 224151 are not.
\clearpage

To summarize, thermal pressures represented by $f1 > 0.30$,
corresponding to $p/k > 5000$ to $10^4\,{\rm cm}^{-3}\,$K (depending on
temperature), almost exclusively arise from gas outside the allowed
velocity range.  The data shown in Fig.~\ref{f1f2hist3} indicate that
about 15\% of the gas parcels moving outside the allowed velocities have
$f1 > 0.30$, whereas this applies to only 1.5\% of material within the
allowed velocities. While the most spectacular cases occur with negative
velocity gas, perhaps associated with gas driven by stellar winds from
the target star, examples for high pressures are also evident for gas
moving at extraordinary positive velocities.  For rapidly moving
material, there seem to be very few incidents showing very strong
rarefactions.  Nearly all of the cases with $f1 < 0.10$ ($p/k <
1000\,{\rm cm}^{-3}\,$K) arise from gas inside the allowed range of
velocities.
\clearpage

Table~\ref{p_dist} shows numerical values for the breakdown into broad
pressure ranges for the three kinematical divisions.  The definitions of
the pressure boundaries are approximate, due to the uncertainties in
temperature.  Also, our definition of pressure ignores the elevation of
$f1$ caused by the contribution from the high pressure matter that is
responsible for pulling $f2$ above the theoretical tracks.

\placetable{p_dist}

\begin{deluxetable}{
c     
r     
c     
c     
c     
c     
}
\footnotesize
\tablecolumns{6}
\tablewidth{0pt}
\tablecaption{Relative Fractions\tablenotemark{a}~~of Occurrences of
Different Pressures\label{p_dist}}
\tablehead{
\colhead{Velocity} &
\colhead{Number} & \colhead{$\log p/k \lesssim 10^{3.0}$} &
\colhead{$10^{3.0} \lesssim \log p/k \lesssim 10^{3.5}$} &
\colhead{$10^{3.5} \lesssim \log p/k \lesssim 10^{4.0}$} &
\colhead{$\log p/k \gtrsim 10^{4.0}$}\\
\colhead{Division} & \colhead{of Cases\tablenotemark{b}} & \colhead{$(f1
< 0.10)$} & \colhead{$(0.10 < f1 < 0.22)$} & \colhead{$(0.22 < f1 <
0.36)$} & \colhead{$(f1 > 0.36)$}
}
\startdata
Positive&290&0.001&0.658&0.310&0.031\nl
Allowed&542&0.042&0.685&0.271&0.002\nl
Negative&90&0.011&0.433&0.289&0.267\nl
All&922&0.028&0.650&0.285&0.037\nl
\enddata
\tablenotetext{a}{Fractions apply to each row in the table, not the
table as a whole.}
\tablenotetext{b}{Each occurrence applies to an accumulation of $\Delta
N_a({\rm C~I}_{\rm total})=1.0\times 10^{13}{\rm cm}^{-2}$.}
\end{deluxetable}

Turning to Fig.~\ref{f1f2hist4}, we find that our index that senses the
presence of a small amount of gas at extreme pressures ($p/k\gtrsim
10^5{\rm cm}^{-3}\,$K, see \S\ref{pts}), $f2/f1$ for $f1 < 0.3$, shows
distinct differences between the three panels.  A Kolmogorov-Smirnov
test reveals that all three cases are drawn from populations that differ
from each other at more than the 98\% level of significance.  Gas that
has no detectable contribution from high pressures, signified by $f1/f2
< 0.20$, seems to be present in all three velocity divisions: 6.6\% of
the cases in the positive range, 8.3\% in the allowed range, and 12.2\%
in the negative range.  The differences in these percentages are not
significant -- they can easily arise from the uncertainties in counting
small numbers of discrete events.  We are led to conclude that the
incidence of small contributions from high pressure gas is ubiquitous.
\clearpage

\section{Discussion}\label{discussion}

We have used C~I fine-structure excitations to indicate the distribution
of thermal pressures in the ISM.  If we exclude gas with large negative
velocities, i.e., material which might be connected in some way with
mass ejected from the target stars, we find that most of the
observations indicate pressures in the general range $10^3 \lesssim p/k
\lesssim 10^4\,{\rm cm}^{-3}\,$K, as indicated in Table~\ref{p_dist}. 
However, an unexpected and noteworthy outcome of our study is the
evidence for a small but pervasive admixture of high-pressure material
($p/k \gtrsim 10^5{\rm cm}^{-3}\,$K) within the matrix of gas exhibiting
a more normal range of thermal pressures.  Figure~\ref{f1f2all}
indicates that this phenomenon exists over most samples in the lines of
sight in our survey, with the instances showing gas without such
high-pressure contributions being the exception, instead of the rule. 
The ubiquity of small contributions from the high pressure state seems
to indicate that we are not viewing just occasional manifestations of
some isolated, sporadic events, such as the compressions following
individual shock fronts from stellar winds or nearby supernova events. 
Instead, the phenomenon seems to be a persistent but perhaps highly
convoluted pattern woven into the fabric of the general ISM.

We now address what other kinds of observations might show some evidence
that can support existence of, and give added information about, the
regions that show enhanced thermal pressures.  One obvious class is the
broad range of investigations that have revealed the presence of
small-scale enhancements in gas density, starting with the discovery of
0\farcs 1 structures in the 21-cm H~I absorption against the
extragalactic source 3C$\,$147 by Dieter, Welch \& Romney  (1976). 
Confirmations of this phenomenon came from similar studies toward other
sources  (Diamond et al. 1989; Davis, Diamond, \& Goss 1996; Faison et
al. 1998), along with observations of variations in molecular line
absorption  (Marscher, Moore, \& Bania 1993; Moore \& Marscher 1995). 
Likewise,  Frail et al.  (1994) observed 21-cm absorptions that varied
with time toward pulsars with high transverse velocities.  These
findings, along with studies of differences in visible and uv absorption
features toward visual binary stars  (Meyer \& Blades 1996; Watson \&
Meyer 1996; Lauroesch et al. 1998; Lauroesch \& Meyer 1999; Lauroesch,
Meyer, \& Blades 2000) and stars in a globular cluster  (Meyer \&
Lauroesch 1999), suggest the presence of density enhancements over size
scales ranging from tens to thousands of AU.  Indeed, one of these
absorption line observations can be directly related to a measurement of
C~I excitation in our survey.  Lauroesch \& Meyer  (1999) observed three
members of the visual multiple star HD206267 (Components A, C and D). 
They found the greatest column density variations for a velocity
component at $-17.2\,{\rm km~s}^{-1}$.  This is consistent with our
finding an upward excursion of $f2$ above the theoretical tracks at
velocities similar to this value (as depicted by the green and blue
points in the upper right panel of Fig.~7$i$).  However,
this correspondence is too fragmentary for us to claim that we have a
convincing case for a meaningful relationship between the presence of
small-scale structures and gas at greatly elevated pressures.

The extent to which the apparent changes in 21-cm and visible absorption
lines are amplified by reductions in temperature is unclear.  Both kinds
of absorption become stronger with decreasing temperature at a constant
density, scaling in proportion to $T^{-1}$ and $\sim T^{-0.7}$,
respectively.  Moreover, the very large densities ($n(H)\sim
10^5-10^6\,{\rm cm}^{-3}$) that one derives from the column density
differences divided by the transverse separations may be misleading 
(Deshpande 2000).  Heiles  (1997) pointed out the severe difficulties
that arise from the simple characterization of discrete, static blobs of
gas at such densities.  Aside from the obvious problem of how such
clouds are contained in a medium with a much lower (total) pressure,
Heiles pointed out that one should ordinarily predict the existence of
an inadmissibly large amount of H$_2$ (and accompanying extinction) in
each cloud.  [Even moderate increases of molecular content for very
small, isolated clouds that absorb at 21-cm seems to be ruled out by the
lack of corresponding CO emission  (Gibson et al. 2000)].  Heiles 
(1997) suggested that this overproduction of H$_2$ could be overcome if
the structures had large longitudinal dimensions and thus much lower
internal densities because they consisted of curved filaments or sheets
viewed near their tangent points.  He also proposed that low
temperatures ($T\sim 15\,$K) would help to resolve the pressure
imbalance.

Regardless of their geometry, if small-scale density enhancements are
indeed accompanied by significant drops in temperature, it would be hard
to relate them to our detections of small amounts of highly excited C~I
because the Boltzmann factors in the upward collisional rate constants
become very small (recall that the energies of the excited levels have
temperature equivalents $E/k = 23.6$ and 62.4$\,$K).  This is clearly
shown by the behavior of the theoretically expected $f1-f2$ tracks that
appears in lower right-hand panel of Fig.~\ref{210839_f1f2}.  For
instance, when $T=20\,$K we find that $f2$ should never exceed about
0.10, even for $p/k$ approaching $10^6\,{\rm cm}^{-3}\,$K.  We will
revisit this issue in a slightly different context further on in our
discussion.

It is plausible that our results are signaling the presence of ephemeral
enhancements of pressure caused by the inertial forces that arise from
supersonic interstellar turbulence.  This is an outgrowth of the picture
that cusp-like compressions of gas are continuously being created and
deformed by an ever-changing, stochastic swirl of converging flows. It
is a picture of the ISM that conforms with a structural description
consisting of a continuum of varying density enhancements suggested long
ago by Chandrasekhar \& M\"unch  (1952) and departs from the usual
paradigm of discrete, static clouds with well defined boundaries that
are constrained by various forms of external pressure from a surrounding
medium.  Hydrodynamical simulations of turbulence under typical
interstellar conditions indeed show that ridges of enhanced density
should arise when randomly directed flows of material collide with each
other  (V\'azquez-Semadeni, Passot, \& Pouquet 1995; Scalo et al. 1998;
Ballesteros-Paredes, V\'azquez-Semadeni, \& Scalo 1999; Elmegreen 1999),
with lifetimes that exceed the crossing time for the gas particles
through the thickness of each region  (Elmegreen 1993b).  These
structures are expected to be found over a very broad range of size
scales, reflecting the downward cascade of turbulent wavenumbers.  In
regimes of density that are higher than what we can observe, these
compressions that arise from supersonic turbulence might, along with
other instabilities, help lead to star formation  (Larson 1981; Hunter
et al. 1986; Elmegreen 1993a, b). 

As long as the size of a region is much greater than a critical length
of about $10^{-2}\,$pc (see Eq.~\ref{region_size} below), the time scale
for the gas to arrive at a thermal equilibrium after compression or
expansion is considerably shorter [$\sim 2\times 10^4\,$yr for the cold,
neutral medium  (Wolfire et al. 1995)] than the typical residence time
of atoms in the new state  (Ballesteros-Paredes, V\'azquez-Semadeni, \&
Scalo 1999).  When this requirement is satisfied, we can anticipate that
there should be a unique sequence of temperature as a function of
density governed purely by the rates of heat loss and gain by various
(sometimes indirect) means of absorbing or emitting radiation.  Thus,
over some limited range of physical parameters one may characterize
thermal equilibrium of the gas with the use of a barytropic index, or
``effective $\gamma$,'' $\gamma_{\rm eff}$, resulting in a simple
scaling of pressure with density, $p\propto \rho^{\gamma_{\rm eff}}$.

The expected lifetimes and structural properties of density enhancements
caused by turbulence depend on the average Mach number $M$ and
$\gamma_{\rm eff}$ (the latter can vary over different density regimes). 
Isothermal turbulence corresponds to $\gamma_{\rm eff}=1$, while
isobaric turbulence arises from $\gamma_{\rm eff}=0$.  As $\gamma_{\rm
eff}$ decreases below 1.0, the density peaks become taller, sharper and
farther apart, and their lifetimes increase because the effective sound
speed  decreases  (V\'azquez-Semadeni, Passot, \& Pouquet 1996; Scalo et
al. 1998).  Also, the (volume weighted) distribution of densities
evolves from a log-normal distribution with an extended power-law tail
toward low densities when $\gamma_{\rm eff} > 1$, through a purely
symmetrical log-normal distribution at $\gamma_{\rm eff} = 1$, to a
log-normal distribution with a power-law tail on the high density side
when $\gamma_{\rm eff} < 1$  (Passot \& V\'azquez-Semadeni 1998; Scalo
et al. 1998; Nordlund \& Padoan 1999).   When $\gamma_{\rm eff}$ dips
below zero between two density regimes that have positive values, $\rho$
can be double-valued, and a thermal instability  (Field 1965) will cause
the gas to bifurcate into two phases  (Field, Goldsmith, \& Habing 1969;
Shull 1987; Begelman 1990; Wolfire et al. 1995), although
Kelvin-Helmholtz instabilities and small-scale forcing of the turbulence
caused by star formation may often obliterate the distinctive bimodality
of densities and make the instability a second-order effect, according
to simulations conducted by V\'azquez-Semadeni, Gazol \& Scalo  (2000). 
Nevertheless, under the right conditions the instability can create very
small clumps of dense gas  (Burkert \& Lin 2000).

Wolfire et al.  (1995) have calculated the thermal equilibrium curves in
the representation of $\log p$ vs. $\log \rho$ for interstellar gases
under a variety of conditions, i.e., different values of column density
(which affect the low energy x-ray fluxes), abundances of the principal
coolants (C and O), far-UV radiation field strengths, and dust-to-gas
ratios.   For soft x-ray absorption equivalent to $N({\rm
H~I})=10^{20}\,{\rm cm}^{-2}$, a reasonable estimate (or lower limit)
for most of the individual regions in our lines of sight, the slope of
their curve indicates that $\gamma_{\rm eff}=0.72$ and the gas is
thermally stable for $n\gtrsim 25\,{\rm cm}^{-3}$ (and $p/k\gtrsim
10^{3.1}\,{\rm cm}^{-3}\,$K).  However, lower values of $\gamma_{\rm
eff}$ (and a shift for the thermal instability to higher pressures) can
arise at these densities when the far-UV radiation field is enhanced, as
might be the case for some portions of our lines of sight.

\begin{figure}
\epsscale{1.0}
\plotone{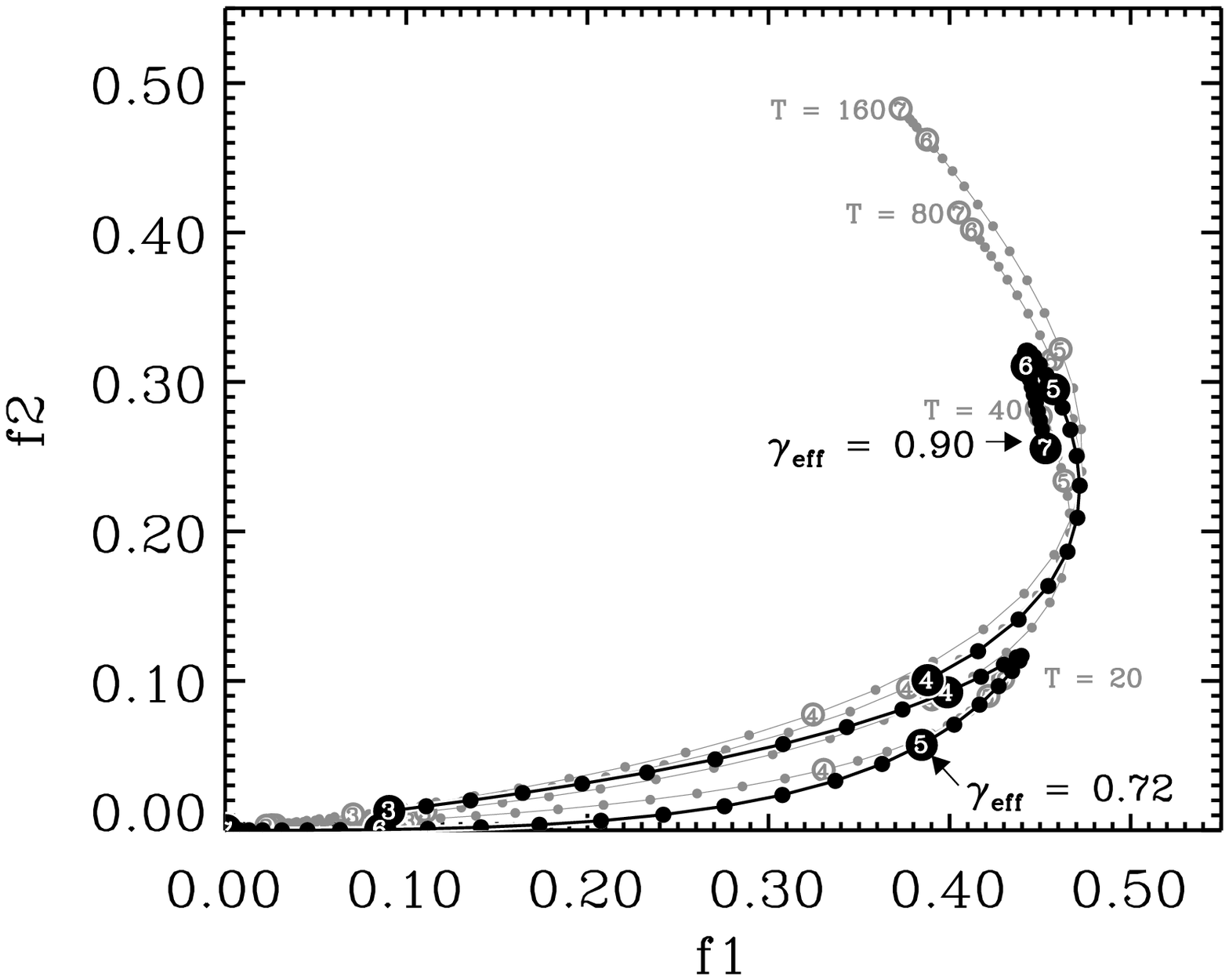}
\caption{Expected evolution of the fine-structure population ratios $f1$
and $f2$ for gas which is compressed with $\gamma_{\rm eff} = 0.72$ and
0.90, starting from initial quiescent conditions $n=10\,{\rm cm}^{-3}$,
$T=100\,$K for the thermal pressure $p/k=1000{\rm cm}^{-3}\,$K.  These
(heavy) tracks are superposed on the (lighter) tracks for atomic gas at
various specific temperatures (tracing an isothermal $\gamma_{\rm
eff}=1.00$) as an aid in comparing against the results presented in
Figs.~\protect\ref{210839_f1f2}, 7, and
\protect\ref{f1f2all}.  Numbers in the large markers indicate logarithms
for the thermal pressures (in units of ${\rm cm}^{-3}\,$K). For clarity,
the portion of the track for $\gamma_{\rm eff}=0.90$ is omitted below
$p/k=10^4 {\rm cm}^{-3}\,$K because it is so close to that for
$\gamma_{\rm eff}=0.72$.\label{gamma_eff}}
\end{figure}

If indeed the lifting of the observed $f2$ values above the lower parts
of the theoretical tracks (for homogeneous regions having $p/k <
10^4\,{\rm cm}^{-3}\,$K in Fig.~\ref{f1f2all}) is caused by small
contributions arising from density fluctuations produced by turbulence,
we can use our observations to define a lower limit on $\gamma_{\rm
eff}$.  In essence, we require that the gas responds to compression in a
manner such that its trajectory on the $f1-f2$ diagram passes through
locations that can pull the measurements of other regions at the same
velocity away from the lower portions of the tracks, in a manner that
reflects our geometrical interpretation about the effects of gas
mixtures that we presented in \S\ref{genl_remarks}. 
Figure~\ref{gamma_eff} shows the trajectories in the $f1$ and $f2$
coordinates for gas undergoing compression for two values of
$\gamma_{\rm eff}$.  For the case where $\gamma_{\rm eff} = 0.72$, it is
clear that when we start with a representative unperturbed state having
$n=10\,{\rm cm}^{-3}$ and $T=100\,$K (and thus $p/k=1000\,{\rm
cm}^{-3}\,$K), the cooling is sufficient to deactivate the C~I
excitation at higher pressures.

\placefigure{gamma_eff} 

Clearly, we need a value of $\gamma_{\rm eff}$ that is higher than 0.72
if the initial conditions stand as we have defined them.  The curve in
Fig.~\ref{gamma_eff} that represents $\gamma_{\rm eff}=0.90$ starts from
these same initial conditions, but when compressed, it undergoes less
severe cooling which in turn makes the C~I excitations appear high
enough in the diagram to satisfy our need for uplifting the points off
the low pressure portions of the tracks (rather than just dragging them
toward the right, nearly parallel to the tracks).  This modification in
$\gamma_{\rm eff}$ should be expected for smaller regions with a
characteristic size of order or less than a critical length $L_{\rm
cool}$ given by 
\begin{equation}\label{region_size}
L_{\rm cool}=4\times 10^{-3}MT^{-1/2}_{10} ~ {\rm pc}
\end{equation}
where $M$ is the Mach number of the compression and $T_{10}$ is the
temperature in units of 10K  (Ballesteros-Paredes, V\'azquez-Semadeni,
\& Scalo 1999).  This equation applies to the density regime
$10^3\lesssim n \lesssim 10^4\,{\rm cm}^{-3}$ where cooling by CO is
dominant, and it indicates the scale below which the densities start to
change more rapidly than the time needed to reach thermal equilibrium. 
It follows that as we go to smaller scales, gas accumulations with
dimensions less than $L_{\rm cool}$ will begin to respond in a manner
such that $\gamma_{\rm eff}$ deviates from the thermal equilibrium
predictions and, for the greatest disparities between the thermal
equilibrium and compression time scales, approach the adiabatic
response, i.e., $\gamma_{\rm eff}\rightarrow c_p/c_v$ where the gas
heats as it is compressed.  Fig.~\ref{gamma_eff} illustrates that once
the threshold of $\gamma_{\rm eff} \approx 0.90$ is crossed, the
temperatures at high pressures are sufficient to explain the C~I
excitations.

It is reasonable to suppose that the turbulence will cascade from large
to small scales, but for the most part only down to a critical length
$L_{\rm AD}$ where ambipolar diffusion would start to dissipate much of
the mechanical energy into heat.  Klessen, Heitsch \& Mac~Low  (2000)
give the formula for $L_{\rm AD}$,
\begin{equation}\label{AD}
L_{\rm AD}=1.3\times 10^{-4}{B\over M_{\rm A}xn^{3/2}}~{\rm pc}
\end{equation}
(see their Eq.~12) where $B$ is the magnetic field strength (in $\mu$G),
$M_{\rm A}$ is the Alfv\'en Mach number, $x$ is the fractional
ionization of the gas, and $n$ is the number density of the gas (in
${\rm cm}^{-3}$).  If we assume that nearly all of the electrons arise
from the ionization of carbon, we can set $x=1.4\times 10^{-4}$, i.e.,
the expected fractional abundance of C in the gas (see
\S\ref{substructures}).  Eq.~\ref{AD} then reduces to $L_{\rm
AD}=n^{-3/2}B/M_{\rm A}$ which equals $3\times 10^{-5}\,{\rm pc}$ if
$n=10^3\,{\rm cm}^{-3}$, a density where the elevations in pressure
start to become important for exciting C~I, and $B/M_{\rm A}$ is of
order unity.  We thus can consider a range of length scales $L_{\rm AD}
< L < L_{\rm cool}$ that spans about two orders of magnitude where
compressions can start to become adiabatic in character at the upper end
of the range and then weaken as they cascade down to the bottom end
where the fluctuations start to dissipate.

The difference between our minimum acceptable value of 0.90 for
$\gamma_{\rm eff}$ and 0.72 implied by the calculations of Wolfire et
al.  (1995) is small.  Effects that they may not have considered that
could modify the equilibrium states, including not only the slope in the
diagram for $\log \rho$ vs. $\log n$ but also the starting point
($n=10\,{\rm cm}^{-3}$, $T=100\,$K), might modify the behavior to give a
trajectory that would run high enough in the $f1-f2$ diagram.  However,
if our conclusion that the regions must be small enough to drive
$\gamma_{\rm eff}$ to values that exceed the prediction for thermal
equilibrium is correct, we add some validity to our earlier conjecture
about the commonality between our extraordinary pressures and the
evidence for small scale structures in the ISM obtained by other means. 

It would be interesting investigate further the hypothesis that the
distribution of points in Fig.~\ref{f1f2all} could arise from a
turbulent field with a (C~I-weighted) density probability function
defined from the numerical simulations for given values of the average
Mach number and $\gamma_{\rm eff}$.  However, in order to do so we must
know more about the turbulent structures than simply their density
probability functions, since the outcomes depend on how well regions
with large pressure enhancements are correlated with adjacent material
that has only a moderate elevation of pressure.  In addition, we must
know how much the one-dimensional velocities change across the typical
dimensions of these density enhancements and the degree to which
unconnected regions make contributions to a single velocity (in one
dimension), e.g., see Figs.~6 and 7 of Ballesteros-Paredes et al. 
(1999).  This information must be derived either from more complete
statistical descriptions that include the two-point density correlation
function and the coupling of density changes to velocity differences 
(Klessen 2000) or by simulations that try to duplicate how the C~I
observations would behave for random interceptions through the density
grids of some actual turbulent models.  To be realistic for our
applications, such models would need to recognize the near equality of
turbulent and magnetic energies for very small condensations  (Val\'ee
2000) and the dynamical consequences that follow  (Passot, V\'azquez-Semadeni, \& Pouquet 1995).

Finally, we briefly touch upon an entirely different interpretation for
the existence of a small amount of gas at high density but not
extraordinarily low temperatures.  In their proposal to explain the
presence of CH$^+$ and other molecular species that are difficult to
produce under standard cloud conditions, Nguyen, Hartquist \& Williams 
(2001) proposed that these molecules could arise from thin, compressed
zones at the edges of clouds.  (Note here that we are returning to the
paradigm of ``identifiable clouds immersed in an intercloud medium.'') 
These zones are pressurized by the Bernoulli effect and heated by the
dissipation of waves arising from the Kelvin-Helmholtz instability as
cloud moves through a warm, intercloud medium (or is subjected to a flow
arising from some byproduct of stellar mass loss or a supernova
explosion).  Their model for such interfaces that seems to yield the
best match to observed molecular abundance ratios has $p/k\approx
4\times 10^5\,{\rm cm}^{-3}\,$K and $T\approx 2000\,$K.  If this
proposal is correct, the interfaces that produce molecules could easily
account for the small amount of gas that shows large C~I excitations. 
As with the turbulent fluctuations, the pressurized cloud sheaths could
also be commonplace.

\section{Conclusions}\label{conclusions}

We have made a survey of interstellar C~I in its three levels of
fine-structure excitation, creating results that, by virtue of their
good velocity resolution ($1.5\,{\rm km~s}^{-1}$), are a considerable
refinement over the earlier work of Jenkins \& Shaya  (1979)  and
Jenkins et al.  (1983) using the {\it Copernicus\/} satellite.  Except
for two of the targets, HD~3827 and HD~120086, all stars are within
Galactic longitudes $\ell = 120\pm 20\arcdeg$ and $300\pm 20\arcdeg$ and
latitudes $b=0\pm 10\arcdeg$.  With distances that range from about 500
to 3700~pc, many of these stars are located in regions whose velocities
arising from differential Galactic rotation are displaced from the Local
Standard of Rest by more than $25\,{\rm km~s}^{-1}$, which allows one to
make crude identifications for the locations of the gas based on
kinematics.  This velocity separation was especially useful in allowing
us to differentiate from the foreground material a special,
highly-pressurized region that is probably very near HD~210839
(\S\ref{210839_comp}).

For each star we created column-density profiles as a function of
velocity for each level by analyzing up to 10 multiplets that contain
transitions whose strengths spanned 2 orders of magnitude.  In order to
do so, we had to solve a large system of linear equations to minimize
the overall $\chi^2$, thus enabling us to unravel the complex blends of
lines in each multiplet.  In order to obtain consistent results from one
multiplet to the next, we had to modify the transition $f-$values from
recent results published by others.  These modifications are summarized
and are compared with the published values in Table~\ref{fval_results}
(and depicted in Fig.~\ref{fvalcomp}).  A remarkable effect is that
there is a steady divergence between our values and the published ones
as the lines become weaker, with very little random scatter in this
relation.  We discussed several tests which suggest that unresolved
saturation is not artificially driving these $f-$value revisions.

While the amounts of C~I found along the lines of sight are of some
interest (the total column densities integrated over all velocities
toward each star are listed in the last column of Table~\ref{col_dens}),
we regard the ratios of C~I in the three levels as the most significant
outcome of this survey.  These ratios are expressed in terms of $f1$ and
$f2$, the fractions of carbon in the two excited levels relative to the
numbers of atoms in all three levels.  It is important to note that
validity of the numbers we derived are dependent on the correctness of
the assumption that LS coupling applies to the 1657$\,$\AA\ multiplet,
one that we chose as a standard for calibrating the strengths of
transitions in other, weaker multiplets.  Because there is a large
separation in energy of its upper electronic level from those of other
C~I multiplets, we regard the 1657$\,$\AA\ multiplet as being the least
likely to suffer perturbations that would cause deviations from the LS
coupling rules that govern the relative strengths of lines within the
multiplet.  We found it reassuring that a comparison of derived
pressures from C~I agreed with those from O~I in a special case where
both could be seen together (HD~210839, see \S\ref{210839_comp}). 

After comparing our measurements of $f1$ and $f2$ to theoretical
predictions for various physical and chemical states of the gas, we
concluded that a large percentage of the gas has thermal pressures in
the general range $10^3 < p/k < 10^4\,{\rm cm}^{-3}\,$K.  The relative
number of C~I atoms showing $p/k > 5000\,{\rm cm}^{-3}\,$K is strongly
dependent on kinematics: a figure of 15\% applies to gas that is moving
outside the range bounded by the Local Standard of Rest and the velocity
expected for differential Galactic rotation at the position of the
target, but this fraction drops by about an order of magnitude for gas
within the allowed velocity range (i.e., material that is either
quiescent or moving transverse to the line of sight).  This outcome is
consistent with the notion that rapidly moving gas parcels are likely to
experience compressions leading to pressures that are noticeably
elevated compared to undisturbed material.  We found no strong
indication that pressurized regions containing C~I were favored at
velocities where there was evidence for ionized gases, as indicated by
the presence of Si~II*.  At the opposite extreme, about 3\% of the C~I
bearing gas seems to have a gas pressure well below the total $p/k$ of
$3\times 10^4{\rm cm}^{-3}\,$K  (Boulares \& Cox 1990) caused by the
weight of material within the Galactic gravitational potential.  We
propose that this condition arises from rarefactions, probably from the
same dynamical processes that create the compressions.

For quantitative statements about the amount of material in various
pressure regimes, we emphasize that it is important to realize that the
results are weighted by the local abundances of C~I.  For regions that
have a lifetime greater than about 1000~yr, i.e., the time needed to
reach ionization equilibrium for the normal average intensity of
ionizing radiation, the C~I abundance scales in proportion to the local
density of electrons, which in turn is usually proportional to the total
density of atoms since the electrons principally come from the
photoionization of C.  One must factor in this consideration when trying
to make quantitative statements about the fractions of hydrogen at
various pressures.

Finally, it is no surprise that we occasionally find values of $f1$ and
$f2$ indicating C~I-bearing gases exclusively at $p/k \gtrsim 10^5\,{\rm
cm}^{-3}\,$K.  The clearest manifestation of this is the component at
$-35\,{\rm km~s}^{-1}$ in front of HD~210839, where there is no evidence
that this material is mixed with lower pressure gas.  In all cases, this
same material also shows excited O~I, and it probably is the result of
pressurization by the winds or expanding H~II regions created by the
target stars and their association companions.  However, somewhat
unexpectedly we also found evidence that overpressured gases may be
rather pervasive, although the amounts are small.  This conclusion
arises from our finding that most of the C~I seen at specific velocities
shows an excess in the values of $f2$ relative to expectations from
their $f1$ counterparts.  If this effect is real and not due to errors
in the assumptions about LS coupling (see above) or the collisional rate
constants used in our calculations of the expected level populations, we
conclude that a very small fraction of the gas exists at extraordinary
pressures $p/k \gtrsim 10^5\,{\rm cm}^{-3}\,$K, that is, at least 2
orders of magnitude above the overall average.  The widespread nature of
this material indicates that it arises from some generalized property of
the medium, rather than isolated, unusual events.  We suggest that
compressions created from interstellar turbulence are a likely source of
this effect.  

If turbulent motions are indeed the explanation for the compressions, we
conclude that the affected regions must be small, because we require
that the barytropic index $\gamma_{\rm eff}$ of the gas must be greater
than predictions based on thermal equilibria at different densities. 
This could happen if the regions had a compression that was closer to
adiabatic in character, as we would expect if characteristic dimensions
were less than about $10^{-2}\,$pc and there was insufficient time for
the gas to reach thermal equilibrium by various radiative processes
before it re-expanded (or collapsed further).  It is important to note
that the characteristic time for cooling $(3/2)kT/\Lambda=5\times
10^{12}n^{-1}\,{\rm cm}^{-3}{\rm s}$, if $\Lambda n^{-2}=4\times
10^{-27}\,{\rm erg~cm}^{-3}{\rm s}^{-1}$  (Dalgarno \& McCray 1972) and
$T=100\,$K, is 4 orders of magnitude shorter than the timescale for
forming an appreciable concentration of molecular hydrogen,
$R^{-1}n^{-1}$ for an H$_2$ formation rate constant $R=2\times
10^{-17}\,{\rm cm}^3{\rm s}^{-1}$  (Hollenbach \& McKee 1979) at the
same temperature.  Since our C~I evidence indicates that the lifetimes
of the compressed states are comparable to or shorter than this cooling
time, it follows that Heiles's  (1997) objection to small structures
based on excessive molecule formation, a point we raised in
\S\ref{discussion}, is not a problem.  This now eliminates an important
reason that supported Heiles's proposal that the small scale structures
must simply be curved sheets or filaments viewed edge-on.

From the considerations expressed above, we are persuaded that
small-scale, short-lived structures formed by turbulent motions in the
interstellar medium are a likely source for small amounts of
overpressurized material.  The plausibility of this interpretation is
reinforced by various types of observations that indicate the existence
of density enhancements over scales ranging upwards from about 10~pc. 
In effect, our results may serve as an independent confirmation of their
existence.  Perhaps more important, the C~I fine-structure levels can
give insights on the nature of these regions, such as the fact that they
are not extremely cold.  Also, our results can be compared to
predictions from contemporary models of interstellar turbulence, with
the objective of supplying some useful constraints on the free
parameters. 

\acknowledgements

We thank K.~Sembach for supplying IDL Legendre polynomial fitting
modules that we implemented for our continuum definitions
(\S\ref{cont_def}).  Discussions with J.~Zsarg\'o and S.~Federman were
useful in our re-examinations of the newly derived $f-$values presented
in \S\ref{fval}.  We benefitted from discussions with M.~Mac~Low and
B.~Draine on various aspects of interstellar turbulence.  We thank
C.~Bowers and F.~Roesler for their willingness to contribute some of
their personal observing time for this project under the block of time
allocated for the STIS Team Guaranteed Observing with HST.  This
research was supported by NASA Grant NAS5-30110 to Princeton University.

\appendix
\section{STIS Wavelength Resolving Power}\label{resolving_pwr}

The research discussed here clearly benefits from our having the highest
resolution possible, since the ionization equilibrium of carbon imposes
the condition that absorption features from the neutral state
preferentially arise from dense gas regions that are likely to have
small velocity dispersions.  In fact, C~I should behave much like Na~I
or K~I, species that can be detected at visible wavelengths and are
known to have narrow features  (Welty, Hobbs, \& Kulkarni 1994; Welty \&
Hobbs 2001).  It is for this reason that we placed strong emphasis on
achieving the highest wavelength resolution that STIS could provide,
leading to the selection of the E140H and E230H echelle gratings with
the narrowest possible entrance slit (\S\ref{obs}).

Along with the rigorous demands our program placed to register narrow
velocity structures, there is also an opportunity before us to use these
features as a way to learn about the wavelength resolving power of STIS
in its highest resolution mode.  To accomplish this, we searched for the
best example of an isolated, narrow C~I feature and assumed that its
intrinsic width is small enough to reveal the full potential of the
instrument to register small differences in velocity.

In principle, a study of this sort must be done with care, since two
effects can distort the conclusions.  On the one hand, a feature that is
thought to be narrow but is indeed intrinsically broad enough to degrade
the profile may lead to an underestimate of the instrument's resolving
power.  This problem grows worse as the feature is reflected in stronger
transitions where saturation sets in.  On the other hand, the selection
of the narrowest, weak features that can be seen in a spectrum may lead
to an error in the opposite direction.  The chance effects of noise can
spuriously narrow or broaden an outcome for any feature that is not seen
at extraordinarily good $S/N$, and selecting only the narrowest features
will unfairly favor the noise effects that operate in one direction
only.  To overcome these problems as best we could, we selected a
velocity component that seemed to be narrower than all the rest in our
survey.  Next, as evenhandedly as possible, we measured the widths all
manifestations of this component in the different multiplets.  We 
excluded only those cases where the lines were either too weak to be
seen or were clearly compromised by interference from other lines or
extreme saturation.  We added to the sample three of the strongest lines
of S~I.  This element, like carbon, is below its favored ionization
stage for H~I regions when it is neutral.
\clearpage

\placetable{stistbl}
\begin{deluxetable}{
c     
c     
c     
c     
}
\tablecolumns{4}
\tablewidth{0pt}
\tablecaption{Line Widths\tablenotemark{a}~~ for a Component at
$-2\,{\rm km~s}^{-1}$ in the Spectrum of HD~210839\label{stistbl}}
\tablehead{
\colhead{$\lambda$} & \colhead{Origin} & \colhead{Equivalent Width} &
\colhead{Profile FWHM}\\
\colhead{(\AA)} & \colhead{} & \colhead{$W$ (Hi-Res Pixels)} &
\colhead{(Hi-Res Pixels)}
}
\startdata
1329.123&C~I*&1.300&2.055\nl
1329.578&C~I**&1.078&2.493\nl
1329.600&C~I**&0.434&2.490\nl
1279.890&C~I*&1.465&2.350\nl
1280.333&C~I**&0.482&1.712\nl
\nl
1280.404&C~I*&0.726&2.410\nl
1280.597&C~I*&1.092&2.245\nl
1279.056&C~I*&0.261&0.990\nl
1277.723&C~I**&1.378&2.213\nl
1277.550&C~I**&1.613&2.758\nl
\nl
1277.723&C~I**&0.751&3.231\nl
1277.723&C~I**&0.629&2.975\nl
1276.483&C~I&1.018&1.484\nl
1276.750&C~I*&0.509&2.109\nl
1260.927&C~I*&1.753&2.322\nl
\nl
1261.122&C~I*&1.216&1.919\nl
1261.552&C~I**&0.681&1.899\nl
1316.542&S~I&0.322&1.999\nl
1296.174&S~I&0.192&1.152\nl
1295.653&S~I&0.744&1.821\nl
\enddata
\tablenotetext{a}{Least-squares fits of the observed profiles to those
of inverted, single Gaussian profiles of the form
$I=I_0\{1-D\exp[-(x-x_0)^2/2\sigma^2]\}$, where each profile's center
$x_0$, depth $D$, and standard deviation $\sigma$ are free parameters
that arise from a minimum $\chi^2$.  The equivalent widths $W$ (column
3) equal $(2\pi)^{1/2}\sigma D$.  The widths listed in column 4 are
given in terms of the profiles' full width at half maxima ${\rm FWHM} =
2^{3/2}(\ln 2)^{1/2}\sigma$.  The data listed here are plotted in
Fig.~\protect\ref{stisres}.}
\end{deluxetable}

\begin{figure}[t]
\epsscale{.5}
\plotone{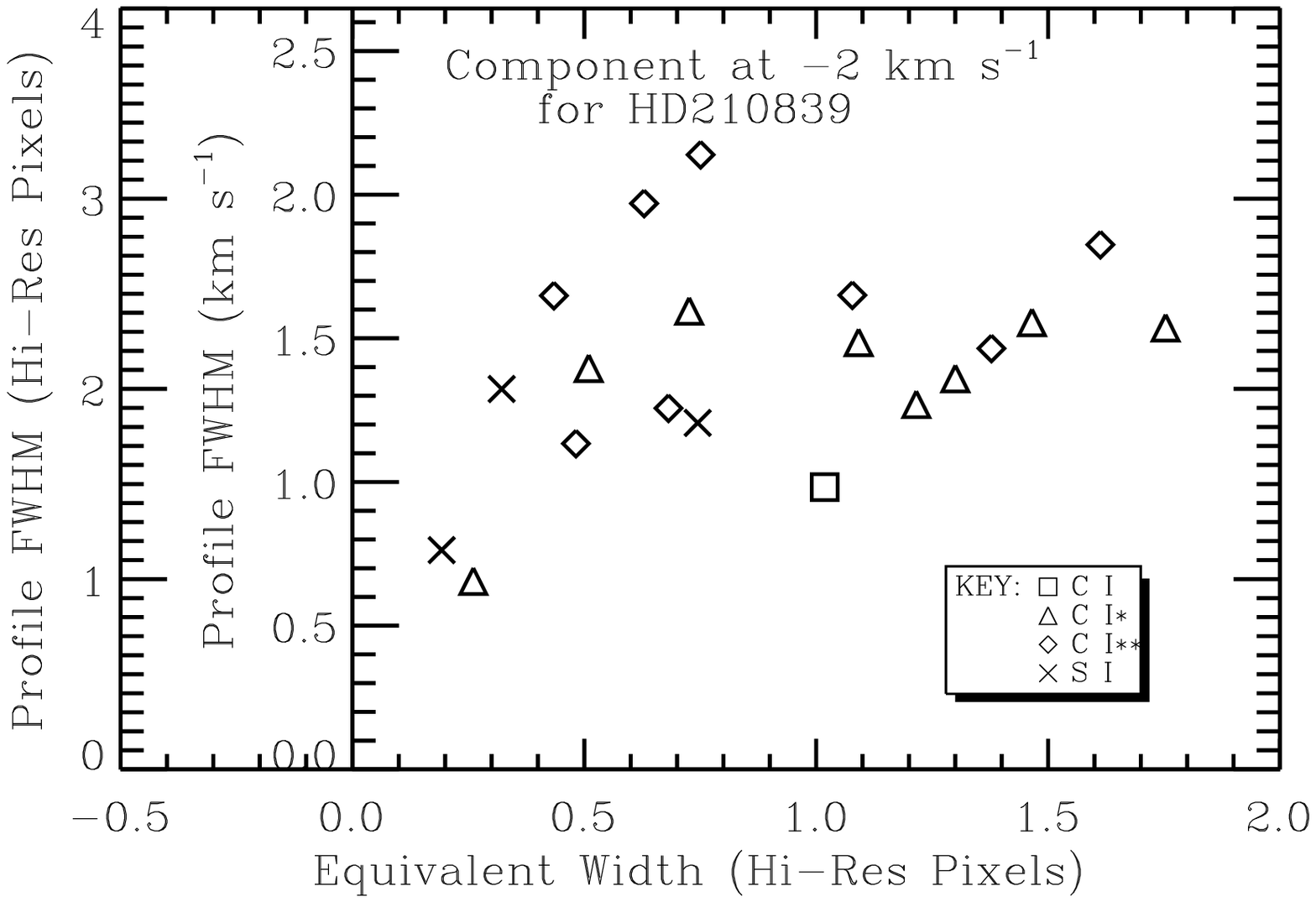}
\caption{Widths (FWHM) of best-fit, inverted Gaussians that match the
intensity profiles recorded by STIS operating in the E140H mode with its
narrowest ($0\farcs 1 \times 0\farcs 03$) entrance slit.  Measurements
of different transitions arising from C~I, C~I*, C~I** and S~I, as
indicated by plot symbols, apply to an interstellar velocity component
at $-2\,{\rm km~s}^{-1}$ in the spectrum of HD~210839.  These widths
(ordinates) are plotted against line strengths (abscissae), expressed as
equivalent widths in units of a Hi-Res pixel of the far-UV MAMA
detector.\label{stisres}}
\end{figure}

Table~\ref{stistbl} shows the results for many measurements of our
selected component: a feature at $-2\,{\rm km~s}^{-1}$ in the spectrum
of HD~210839 recorded with the E140H grating.  We chose to measure the
widths of individual transitions, rather than the width of the component
expressed in our derived composite $N_a(v)$, so that the outcome would
not be compromised by deficiencies in our velocity registrations
(\S\ref{vel_reg}), errors in the scale factors for \AA/pixel along a
single diffraction order, or small offsets in the published laboratory
wavelengths from the true separations of transitions within a multiplet.
The widths given in the table represent the FWHM for a best (minimum
$\chi^2$) fit of the intensity to a single Gaussian profile that is
inverted.

To obtain a better understanding about whether or not the results
depended on line strength, we plotted the apparent widths of the
profiles as a function of their equivalent widths (expressed in units of
a Hi-Res pixel).  This plot is presented in Fig.~\ref{stisres}.  The
strong lines do not seem to be significantly wider than the weak ones,
which indicates that saturation is probably not an important factor in
broadening these particular lines.  As expected, the weak lines show
more scatter caused by the influence of random noise.

\placefigure{stisres}

From the average location of points shown in Fig.~\ref{stisres}, it
seems reasonable to conclude that the resolution of the E140H mode of
STIS, as configured for our survey, is $R=200,000$ (for the profile's
FWHM) which equals $1.5\,{\rm km~s}^{-1}$.  This assumes, of course,
that the profile has an intrinsic width that is much smaller than
$1.5\,{\rm km~s}^{-1}$.

Unfortunately, we were unable to perform a good measurement of the
resolution of STIS operating with the E230H grating.  The only
absorption line in Multiplet~1 (the only multiplet covered by our longer
wavelength setting) that was weak enough to be unsaturated for the
$-2\,{\rm km~s}^{-1}$ component was the one arising from the weak
transition from C~I** at 1658.1$\,$\AA.  A satisfactory measurement of
this line could not be made because the amplitude of the noise was about
half as large as the line depth. 

\section{Telluric Profiles of O~I* and O~I**}\label{telluric}

Initially, when we attempted to determine the resolution of STIS we
thought that features at 1304.858$\,$\AA\ and 1306.039$\,$\AA\ arising
from excited oxygen atoms in the Earth's upper atmosphere would yield
appropriate signals for this task.  These features are especially strong
for our observing program because most of our targets are in the CVZ
(\S\ref{tgt_select}), a configuration that typically produces large
zenith angles for the telescope's viewing axis.  They rarely suffer from
interference from interstellar O~I* and O~I** (\S\ref{OI*}).  Moreover,
the fact that both lines are near the ends of the echelle diffraction
orders meant that we could sense them twice at very different locations
on the MAMA detector.  This made the lines especially attractive, since
we could obtain double the normal number of measurements and look for
possible changes in resolution along one of the image dimensions.

After we had examined the C~I data, it was clear that the telluric
profiles are indeed resolved by STIS, or at least partially so, and thus
inappropriate for measuring the resolving power of STIS.  Nevertheless,
the outcomes are of possible use in understanding the properties of
Earth's atmosphere above the altitude of HST's orbit.

\begin{figure}[t]
\plotone{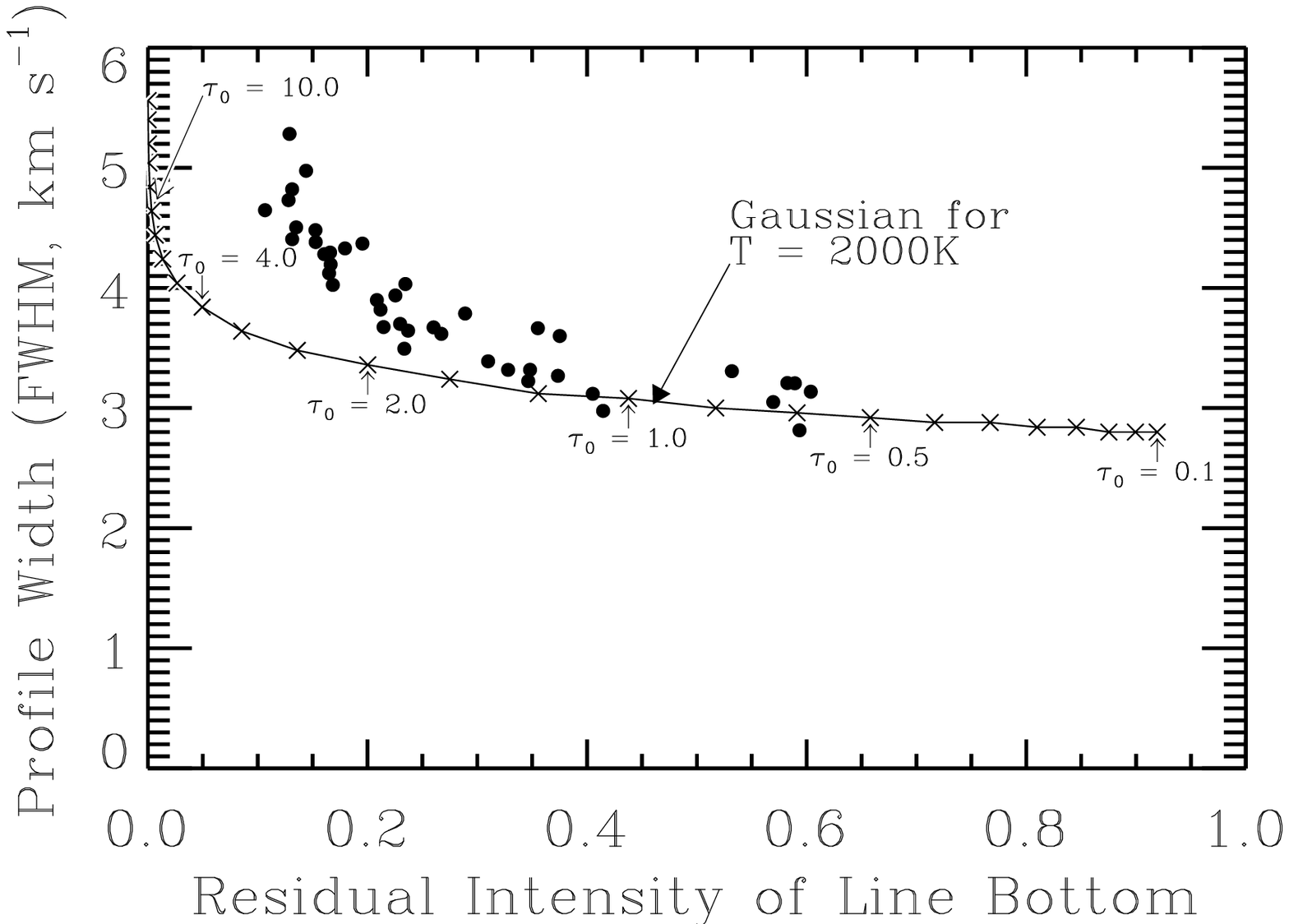}
\caption{Measured widths (FWHM) of the telluric O~I* and O~I** profiles 
for the transitions at 1304.858$\,$\AA\ and 1306.039$\,$\AA,
respectively, plotted against their minimum intensities.  As the lines
become stronger (going from right to left in the plot), they broaden. 
While this behavior is expected as the lines saturate, they do so faster
than that expected for a pure Gaussian velocity profile seen at the STIS
resolution, as indicated by the line with $\times$'s and markers for the
central optical depths $\tau_0$.\label{oitellur}}
\end{figure}

Figure~\ref{oitellur} shows the recorded widths of the telluric O~I* and
O~I** features, plotted as a function of their central intensities.  We
did not employ the Gaussian fitting procedures discussed
\S\ref{resolving_pwr} because distortions that arise when the lines
become saturated lead to poor fits.  For comparison, the expected
behavior for lines of different strength arising from oxygen atoms with
no macroscopic velocities but with a Doppler broadening corresponding to
$T=2000\,$K is also shown (line with $\times$'s).  These expected
profiles are Voigt profiles convolved with the STIS line-spread function
(assuming it's Gaussian), whose determination was described in
\S\ref{resolving_pwr}.

\placefigure{oitellur}

Weak profiles are consistent with a pure Doppler broadening at
$T=2000\,$K, an amount that exceeds generally accepted value of around
1000$\,$K for the temperature of the Earth's exosphere, such as that
given in the MSIS$-$86 model atmosphere  (Meier 1991).  As the profiles
grow stronger (due in large part to higher zenith angles of the
observations) they broaden faster than the expected behavior of a simple
Voigt profile.  Evidently, the wings of the profile suggest the presence
of some bulk macroscopic motions in the upper atmosphere of the Earth.

\end{document}